\documentclass[aps,preprint,onecolumn]{revtex4-2}
\usepackage[version=3]{mhchem}
\usepackage{mathptmx}
\usepackage{subcaption}
\usepackage{hyperref}
\usepackage{float}
\usepackage{xcolor}
\usepackage[normalem]{ulem}
\usepackage{mathtools}
\usepackage{amsmath}
\usepackage{amssymb}
\renewcommand{\figurename}{Figure}

\begin{document}

\title[Funneling and spin-orbit coupling in TMDCs nanotubes and wrinkles]{Funneling and spin-orbit coupling in transition-metal dichalcogenide nanotubes and wrinkles}

\author{M. Daqiqshirazi and T. Brumme}
\address{Bergstrasse 66c, Theoretical chemistry, Technische Universität Dresden, Dresden, Germany.}

\email{thomas.brumme@tu-dresden.de}



\begin{abstract}

Strain engineering provides a powerful means to tune the properties of two\--di\-men\-sion\-al materials. 
Accordingly, numerous studies have investigated the effect of bi- and uniaxial strain. Yet, the strain fields in many systems such as nanotubes and nanoscale wrinkles are intrinsically inhomogeneous and the consequences of this symmetry breaking are much less studied. Understanding how this affects the electronic properties is crucial especially since wrinkling is a powerful method to apply strain to two\--di\-men\-sion\-al materials in a controlled manner. In this paper, we employ density functional theory to understand the correlation between the atomic and the electronic structure in nanoscale wrinkles and nanotubes of the prototypical transition metal dichalcogenide \ce{WSe2}.
Our research shows that the symmetry breaking in these structures leads to strong Rashba-like splitting of the bands at the $\Gamma$ point and they thus may be utilized in future tunable spintronic devices. The inhomogeneous strain reduces the band gap and leads to a localization of the band edges in the highest-curvature region, thus funneling excitons there. Moreover, we show how wrinkles can be modeled as nanotubes with the same curvature and when this comparison breaks down and further inhomogenities have to be taken into account.

\end{abstract}




\maketitle

\clearpage

\section{Introduction}

Two-dimensional (2D) materials have been the focus of a myriad of researches in the last decade due to their fascinating properties. After the successful synthesis of graphene \cite{novoselov2004electric} other materials such as hexagonal Boron nitride (h-BN) \cite{song2010large}, transition metal dichalcogenides (TMDCs) \cite{ayari2007realization,kuc2011influence,lorenz2014mos2,xu2021two} and black phosphorous \cite{woomer2015phosphorene} joined the class of 2D materials quite quickly. Many researchers investigated the intriguing properties of this new class of materials such as their extraordinary strength and high deformation before rupture, high mobility, and ease of property alteration \cite{miro2014atlas,ghorbani2016single,akinwande2017review,brumme2015first,dai2019strain,hong2017recent,falin2021mechanical}. 2D materials proved to be useful for many applications such as nanoelectronics, spintronics, and catalysis \cite{ahn20202d,lu20202d,li2021carbon,anju2020biomedical}.

Still, researchers are trying to expand the applicability of these materials by methods such as alloying \cite{lin2021controllable}, introduction of defects \cite{ghorbani2022two}, creation of van der Waals heterostructures or by applying external pressure and fields \cite{ma2021tunable, daghigh2021multiscale}. Another method offering a reversible and non-destructive route to modulate the properties of 2D materials is strain engineering.\cite{du2021strain,plechinger2015control,ghorbani2013electromechanics, ghorbani2013strain,heine2015transition,peng2020graphenestrain,jiang2020straintronics,digiorgio2022NEMS} Uniaxial and biaxial strains have been studied extensively \cite{haastrup2018computational,roldan2015strain} and there are standard techniques which can be used already during the synthetization of 2D materials \cite{dai2019strain}. Often the application of in-plane stress will lead to the formation of wrinkles and there are established methods to produce these wrinkles \cite{lee2022surface} which can even be used to determine the mechanical properties of the layered material \cite{schweikart2009controlled, knapp2021controlling, yu2019tackling}. Furthermore, the changes of the electronic structure in these wrinkles leads to funneling, \textit{i.e.}, a preferential emission of light from certain spatial position along the wrinkle.\cite{peng2020graphenestrain,lee2020switchable,koo2021tip,shao2022probing,jiang2022analysis,digiorgio2022NEMS} However, the local strain in such samples is far from being homogeneous and the comparison with calculations of homogeneously strained systems might be misleading. Understanding this (local) inhomogeneous strain in 2D materials requires further research due to the vast opportunities for future applications such as polarized single photon emitters \cite{parto2021defect,wang2021highly} and flexible optoelectronics \cite{du2021strain}.

In order to understand the influence of the inhomogeneous strain in 2D systems one can study idealized model systems which have a similar strain state but are easier to control from an experimental point of view or need less approximations in the respective theoretical description.
Nanotubes (NTs) are structures where strain fields can play an important role \cite{tenne1992polyhedral,sinha2021mos2} and even though those are not strictly inhomogeneous,
since the strain can be defined by a constant curvature, they represent such a simple model system which is different from the uni- or biaxially strained 2D layers and closer to the wrinkled systems in experiments. 

The investigation of the electronic properties of inhomogeneously strained materials requires methods such as density functional theory (DFT) or density functional based tight binding (DFTB) which can be computationally demanding for large 
systems. Fortunately, for NTs, researchers have developed a method to reduce the size of the curved systems -- cyclic DFT \cite{ghosh2019symmetry,nepal2019first,kumar2020bending} employs the helical boundary condition in nanotubes in order to reduce the cost of the calculation and the method was successfully used to determine the bending modulus of various 2D systems. Yet, to the best of our knowledge, no one investigated the similarities and differences of NTs and wrinkles even if this understanding is crucial as in larger wrinkles (or other system where the variation of strain is due to a variation of the local curvature) it is computationally impossible to model the system entirely. NTs could then be used to further simplify the calculation by the use of cyclic DFT.

The sheer size of inhomogeneously strained structures causes many theoretical investigations to neglect relativistic effects beyond scalar-relativistic limits in order to enhance computational speed. Including spin-orbit coupling (SOC) in systems with heavy elements (like TMDCs) is however very important to understand many fascinating physical effects occurring in 2D materials such as the Hall effect at room temperature \cite{safeer2020spin}, the Rashba effect \cite{manchon2015new, zibouche2014transition} and spin-valley coupling \cite{klein2022electrical}.

The Rashba effect, \textit{i.e.}, the emergence of spin splitting in momentum direction, \cite{manchon2015new,chen2021spin} occurs due to inversion symmetry breaking in the presence of SOC. Such splitting was initially observed in zinc blends, wurtzite, and in systems under an external electric field. Another type of system in which this splitting is observed, is Janus-type 2D materials due to the resulting internal electric field perpendicular to the structure. Yet, the internal electric field and the subsequent splitting are quite small ($\alpha \approx \mathcal{O}(10~{meV \text{\AA}})$). Cheng \textit{et al.} \cite{cheng2013spin} studied such polar TMDCs (\ce{WSSe} as an example) proving their stability and recommended them to be used for Datta-Das spin field effect transistors. This behavior can be even more interesting if the splitting manifests adjustability without the need of an external electric field. Yao \textit{et al.} \cite{yao2017manipulation} utilized biaxial strain to manipulate the Rashba splitting in Janus heterostructures and concluded that the change of orbital overlaps increases the splitting. 
Since many symmetries including inversion symmetry are broken in curved systems such as NTs and wrinkles, one could also imagine the emergence of similar phenomena especially since such a curvature-induced SOC has already been shown for carbon NTs.\cite{CNTSOC1,CNTSOC2}

Having mentioned the importance of inhomogeneous strain in 2D structures, in this paper, we investigate \ce{WSe2} in the form of NTs and wrinkles theoretically. We concentrate on these systems specifically since we also want to estimate if and when NTs can be used to model large scale wrinkles in order to ease the computational load using helical boundary conditions. The similarities between NTs and wrinkles are expounded and strain associated phenomena in the band structures are described. We explore NTs and wrinkles in ranges larger than any previous investigation to the best of our knowledge with an \textit{ab initio} method including SOC effects which provides electronic insight about these materials for better future applications.

\section{Results and discussion}
\begin{figure}[p]
\includegraphics[width=1.0\columnwidth]{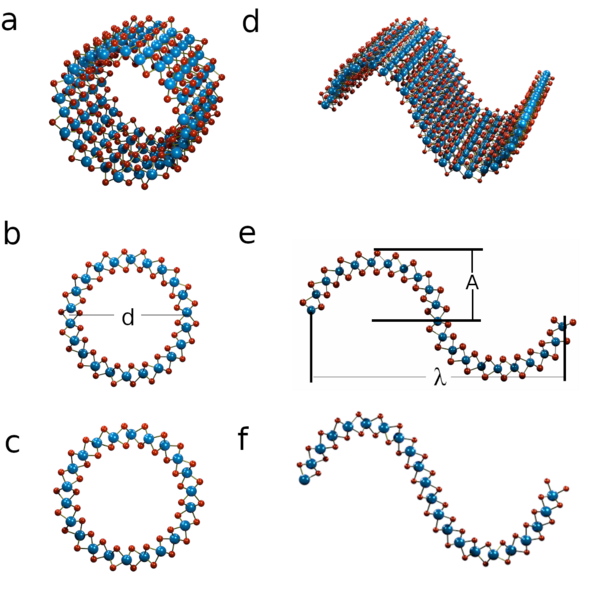}%
\caption{\ce{WSe2} structures investigated in this study. (a), (b) nanotube 3D and side view, (d), (e) wrinkle 3D and side view (periodic boundary condition on back and front). (c) and (f) side view of the relaxed nanotube and wrinkle, respectively. 
The structural parameters such as the nanotube diameter, $\mathrm{d}$, and the amplitude or wavelength of the wrinkle, $\mathrm{A}$ or $\lambda$, are indicated in (b) and (e).
Wrinkles with an initially elliptical profile ($\lambda = 4A $) relax into a structure with smoother areas between peaks and valleys.
 \label{figure_method_creation}}
\end{figure}

In the following, we will investigate how inhomogeneous strain affects wrinkles and NTs of monolayer \ce{WSe2} -- a prototypical example of the TMDCs. In order to compare wrinkles and NTs, the initial wrinkled structure was created with a circular profile as shown in Figure \ref{figure_method_creation} with a wavelength to amplitude ratio of $\lambda/A = 4$ and using NTs as input. 
We will discuss the deformation energy and changes in the band dispersion and we will explain the origin of funneling in nanoscale wrinkles as well as a Rashba-like splitting that occurs in curved TMDC structures. We will further explain in detail the similarities of electronic structure of NTs and wrinkles.

\subsection{Deformation energy/band gaps \label{section_wrinkle_profile}} 

The relaxation of the initial structure is very different for NTs and wrinkles. While the former only change the diameter (\textit{cf.}, Figures \ref{figure_method_creation}b and \ref{figure_method_creation}c), the latter relax into structures which do not resemble the initial nanotube-like profile anymore (Figures \ref{figure_method_creation}e and \ref{figure_method_creation}f), especially in the region close to the middle plane of the unit cell (inflection point).  In this region, the wrinkled structure resembles a flat monolayer and is thus under less local strain in comparison to the corresponding nanotube. However, the peak of the wrinkles deforms stronger, leading to areas with higher curvature. The curvature in NTs on the other hand is constant since they always remain circular (Figure \ref{figure_method_creation}c).

\begin{figure}[t]
    \includegraphics[width=1.\columnwidth]{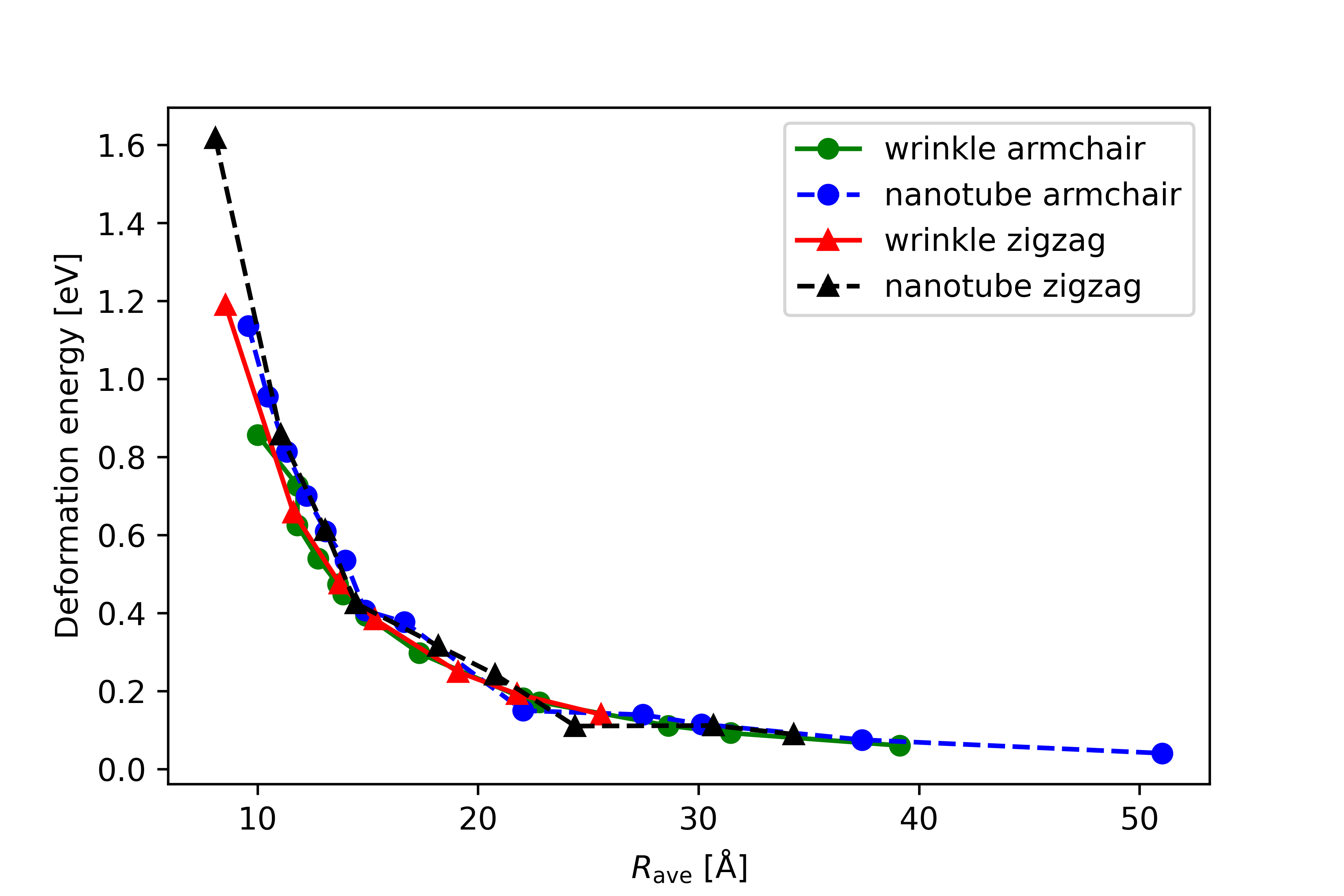}
    \caption{Variation of the deformation energy with increasing average radius of curvature, $R_\mathrm{ave}$, \textit{i.e.}, with increasing wavelength and diameter for wrinkles and nanotubes, respectively.}
    \label{fig_deformation_energy_R_ave}
\end{figure}
In order to allow for a better comparison between NTs and wrinkles, we will introduce two new measures -- the average and the maximum radius of curvature, $R_\mathrm{ave}$ and $R_\mathrm{max}$ -- by fitting a spline to the positions of the tungsten atoms and calculating its curvature. The comparison of the deformation energies, $E_\mathrm{def}$, as function of the average radius of curvature in Figure \ref{fig_deformation_energy_R_ave} shows that there is a small difference between NTs and wrinkles for the smaller systems (\textit{i.e.}, with larger average curvature) while the difference between armchair and zigzag is negligible.
The deformation energy, $E_{\mathrm{def}}$, has been calculated as follows:
\begin{equation}
    E_{\mathrm{def}}=\frac {E_{\mathrm{sys}}}{N_\mathrm{u.c.}}-E_{\mathrm{mono}},
\end{equation}
where $E_{\mathrm{sys}}$ is the energy of the wrinkle/nanotube, $N_\mathrm{u.c.}$ is the number of formula units, and $E_\mathrm{mono}$ is the energy of the flat monolayer.
This indicates that the local relaxation at the inflection point of the wrinkles and the corresponding increasing curvature at the maxima can be understood as a redistribution of the strain which leads to a total energy gain. Figure \ref{fig_deformation_energy_R_ave} furthermore shows that this energy gain (per formula unit) becomes very small for $R_\mathrm{ave}\gtrsim 25\text{\AA}$.

\begin{figure}[b]
 \centering
 \includegraphics[width=1.\columnwidth]{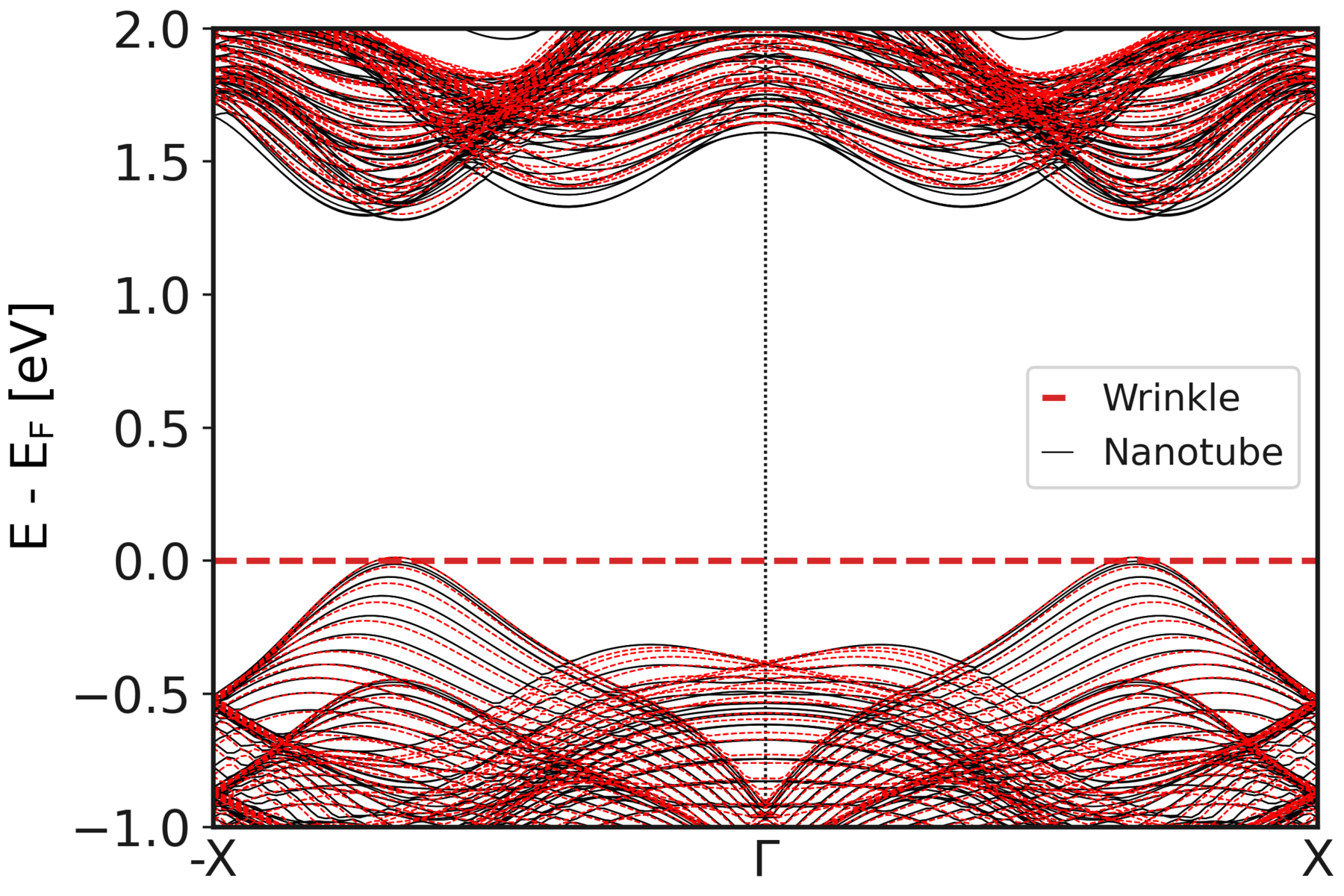}
 \caption{Comparison of the band structures for the (24,24) nanotube and the corresponding wrinkle. The smallest direct band gap is the high-symmetry $\mathrm{K}$ point of the monolayer which is mapped to the $\Gamma-\mathrm{X}$ line (\textit{cf.}, element-projected band structures in the SI \cite{ThisPaperSI}, Figure S3, and the section ``Brillouin zone of wrinkles/nanotubes and spin texture'').}
    \label{fig_nano_wrinkle2424}
\end{figure}

\begin{figure}[b]
    \includegraphics[width=1.\columnwidth]{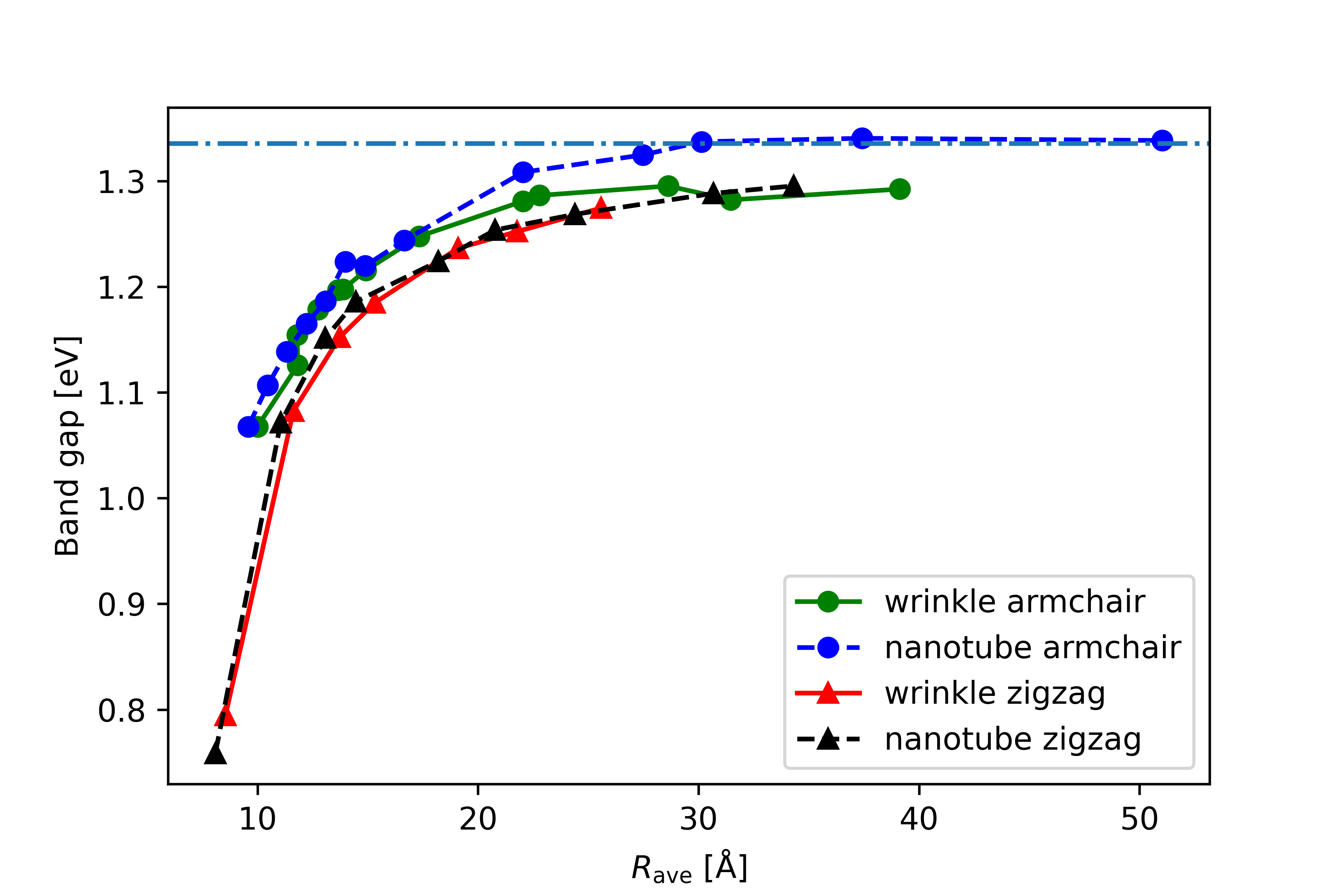}
    \caption{Variation of the global direct band gap, $E_\mathrm{gap}$, due to the strain field formed in wrinkles and nanotubes of \ce{WSe2}. The dash-dotted line indicates the direct band gap of the flat monolayer.}
    \label{fig_bandgap_allStructures}
\end{figure}

Interestingly, the profile of the relaxed wrinkles differs from the sinusoidal wave which is assumed in continuum mechanics and which follows from the harmonic approximation. The deviation from the sinusoidal shape is more prominent for shorter wavelengths and vanishes for wrinkles with larger wavelength which are more similar to monolayers. In fact, the long-range behavior is expected and can already be predicted by analyzing the average curvature of differently wrinkled curves (see section ``Curvature analysis'' in the supporting information (SI) for more details \cite{ThisPaperSI}). In brief, for $\lambda/A\gtrapprox3.5$ the profile tends to be sinusoidal, while for smaller $\lambda/A$ an elliptical or circular profile is preferred. This deviation from the harmonic solution is important for analyzing the strain fields using electron microscopy images for which one requires an assumption about the shape of the wrinkle (\textit{cf.}, Refs.~\citenum{xie2021quantitative,xie2022nonlinear}).
Our structures with $\lambda\approx4A$ after relaxation are in fact better fitted by two sine functions with one having an almost three times larger wavelength. \\
Such wrinkle profiles, having periodically wrinkled areas with peaks and valleys have already been observed in wrinkling experiments on polymeric substrate \cite{daghigh2021multiscale,knapp2021controlling}. Yet, in other wrinkling experiments \cite{wang2021strained}, single wrinkles with only peaks connected with areas of lower strain have been found. In order to keep the discussions in this paper general, we focus on fully relaxed wrinkles which correspond to the expected relaxed freestanding wrinkle profile. The investigation of substrate-induced effects is an interesting topic which is however beyond the current investigation. \\
The relaxation in the wrinkles leads to differences in the electronic structure compared to the nanotube-like structures used for the initial geometry. Yet, the comparison of the band dispersion for NTs and wrinkles with approximately the same average curvature (especially for large wavelength or diameter) also reveals similarities especially for the conduction and valence band (CB and VB). Figure \ref{fig_nano_wrinkle2424} shows the band structures for a (24,24) nanotube (d=44 $\r{A}$ ) and the corresponding wrinkle ($\lambda=88 \text{\AA}$).
Both the valence-band maximum (VBM) and conduction-band minimum (CBM) show large contributions from tungsten atoms (\textit{cf}. Figure S3 for the monolayer in the rectangular unit cell) indicating that those are the high-symmetry $\mathrm{K}$ points of the primitive unit cell which are mapped to the $\Gamma-\mathrm{X}$ line (more details about the backfolding can be found in the SI \cite{ThisPaperSI}, section ``Brillouin zone of wrinkles/NTs and spin texture'').
Not only the global band gap, $E_\mathrm{gap}$, is comparable for this specific example but also the dispersion of the first few VBs and CBs and accordingly derived physical quantities such as mobility and conductivity. This is a promising outcome as it suggests that for global variables of large-scale wrinkles a similar nanotube can be used to reduce the computational cost. In Figure \ref{fig_bandgap_allStructures} we compare the direct band gap as function of the average curvature for all investigated systems.

The global direct band gap of most wrinkles is slightly smaller than those of the NTs with similar average curvature and in both systems the band gap approaches the value of the monolayer, $E_\mathrm{gap}^{ml}\approx 1.34\:\mathrm{eV}$, for small average curvature, \textit{i.e.}, large radius of curvature -- please note that the calculated band gap for the monolayer can vary depending on the settings\cite{tran2022}. Thus, NTs can possibly be used to model the global band gap changes in large wrinkles (or similar systems under inhomogeneous strain) if the average curvature of the wrinkles is taken into account. It is worth to reemphasize that the \textit{ab-initio} modeling of inhomogeneous strain in wrinkles is computationally very demanding due to the symmetry breaking, the presence of many heavy atoms, and spin-orbit coupling and that this similarity can be utilized to simplify the theoretical modeling. One can then subsequently utilize helical boundary conditions to further reduce the computational cost. However, there are also some differences for the bands close to the VBM/CBM -- in wrinkles we find more bands with similar dispersion but small differences in the maximum/minimum energy and this can explain the funneling found in wrinkled systems as explained in the following.

\subsection{Funneling} \label{section_result_funneling}

Funneling is the phenomena of absorption and emission of light from different spatial positions along the wrinkle.
The directional guiding of the excitons can be achieved by a spatial modification of, \textit{e.g.}, the dielectric screening\cite{peimyoo2020dielectric} or the band gap of the material due to external strain.\cite{lee2020switchable}
This phenomena attracted the attention of numerous scientists, and was subject to several researches. \cite{lee2020switchable,koo2021tip,dirnberger2021,shao2022probing,peng2020graphenestrain,digiorgio2022NEMS} It has for example been shown that the photovoltaic behavior of 2D materials \cite{harats2020limits} and light emitting diodes \cite{jiang2021energy} can significantly be enhanced. Furthermore, it was proposed that highly directional exciton transport promises not only compelling advantages for exciton-based applications but that it could also be interesting for reaching truly 1D regimes to study quantum transport phenomena of correlated many-body states.\cite{dirnberger2021} From the experimental point of view, photoluminescence microscopy is routinely used to study these systems although it can be quite challenging at the nanoscale; \cite{koo2021tip} other techniques such as time-resolved transient absorption microscopy\cite{yuan2020twist} are possible too but also need additional input from theory to interpret the results. However, in conjunction with theoretical estimates of the band structure changes due to external strain\cite{koo2021tipinduced,cho2021,koo2022dynamicalcontrol} or different stacking regions\cite{cho2021,yuan2023strong} a quantitative description of experimentally observed shifts of excitonic peaks is possible.

\begin{figure}[t]
 \centering
 \includegraphics[width=1.\columnwidth]{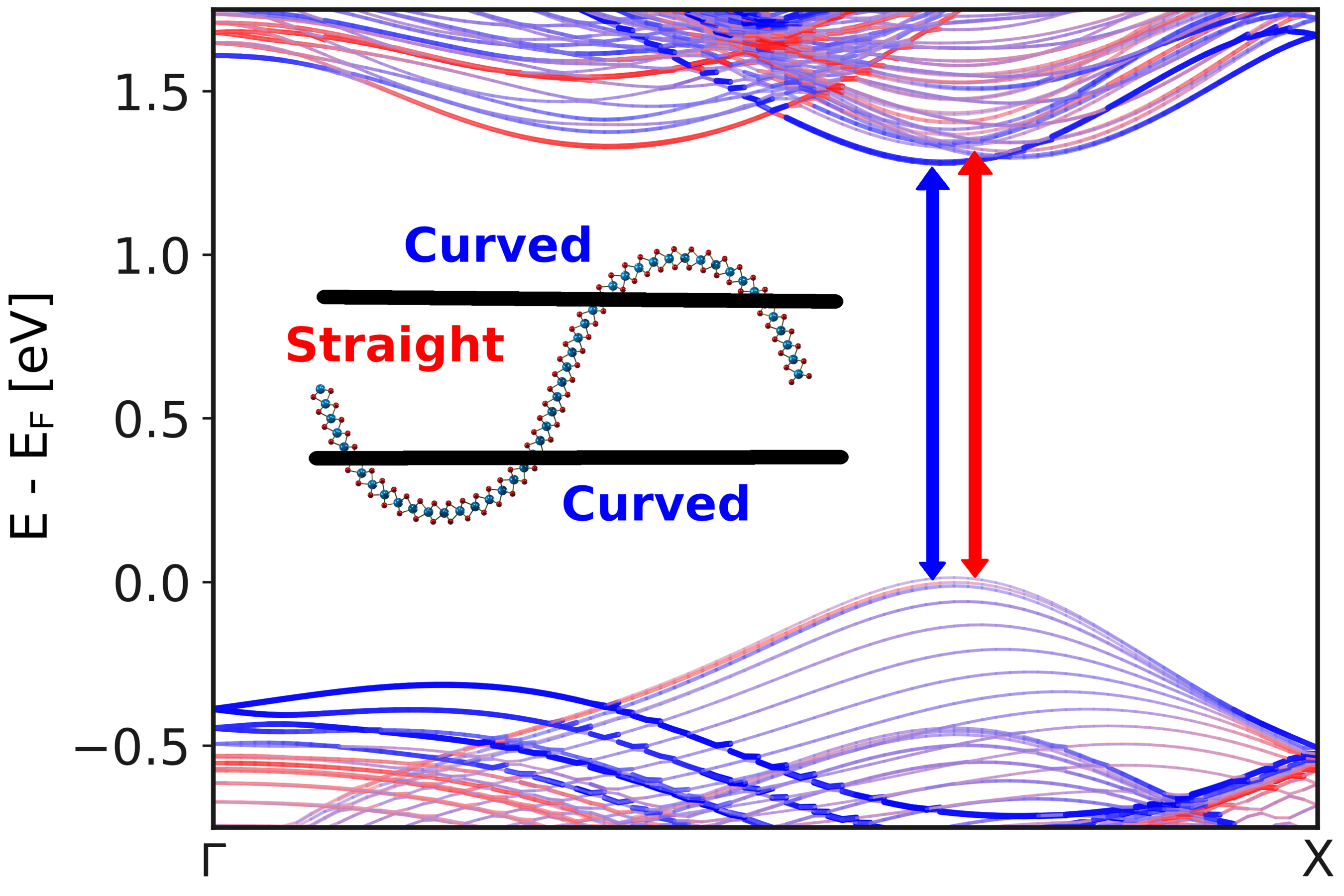}
 \caption{Contribution of curved and straight sections of the (24,24) wrinkle to its band structure, the smallest local band gaps from each region are also indicated by arrows. The inset depicts the region division.}
 \label{figure_regionContribution}
\end{figure}

In order to understand the difference between all the bands close to the VBM and CBM and relate this to experimental observations, we projected the band structure on different atoms along the wrinkle. 
Figure \ref{figure_regionContribution} shows that the VBM is spread all over the wrinkle while the VB-1 and VB-2 (the two bands below the VB) close to the VBM are more localized in the straight and the curved region, respectively. The CBM on the other hand is localized at the top of the wrinkle in the regions with large local strain while the higher CB also show contributions from the straight regions. Since the different minima of the CBs have a larger difference in energy than the VBs, the lowest band gap is found in the regions with large local curvature. Figures S14a and S14b in the SI \cite{ThisPaperSI} show the variation of the local density of states ({LDOS}) along the (24,24) wrinkle close to VBM and CBM, respectively, and confirm that the energy levels close to the band gap are more localized close to the peaks and valleys of the wrinkle. Yet, since the band extrema also have small differences in the momentum direction which complicate the situation, further studies using, \textit{e.g.}, the Bethe-Salpeter equation to describe the excitonic states are needed which are -- at the moment -- however only possible for the smallest systems of our study.\cite{jiang2022} Nevertheless, as shown previously, the changes of the local band gap give a very good estimate of the shifts of excitonic peaks.\cite{koo2021tipinduced,cho2021,koo2022dynamicalcontrol}

\begin{figure}[b]
 \centering
 \includegraphics[width=0.8\columnwidth]{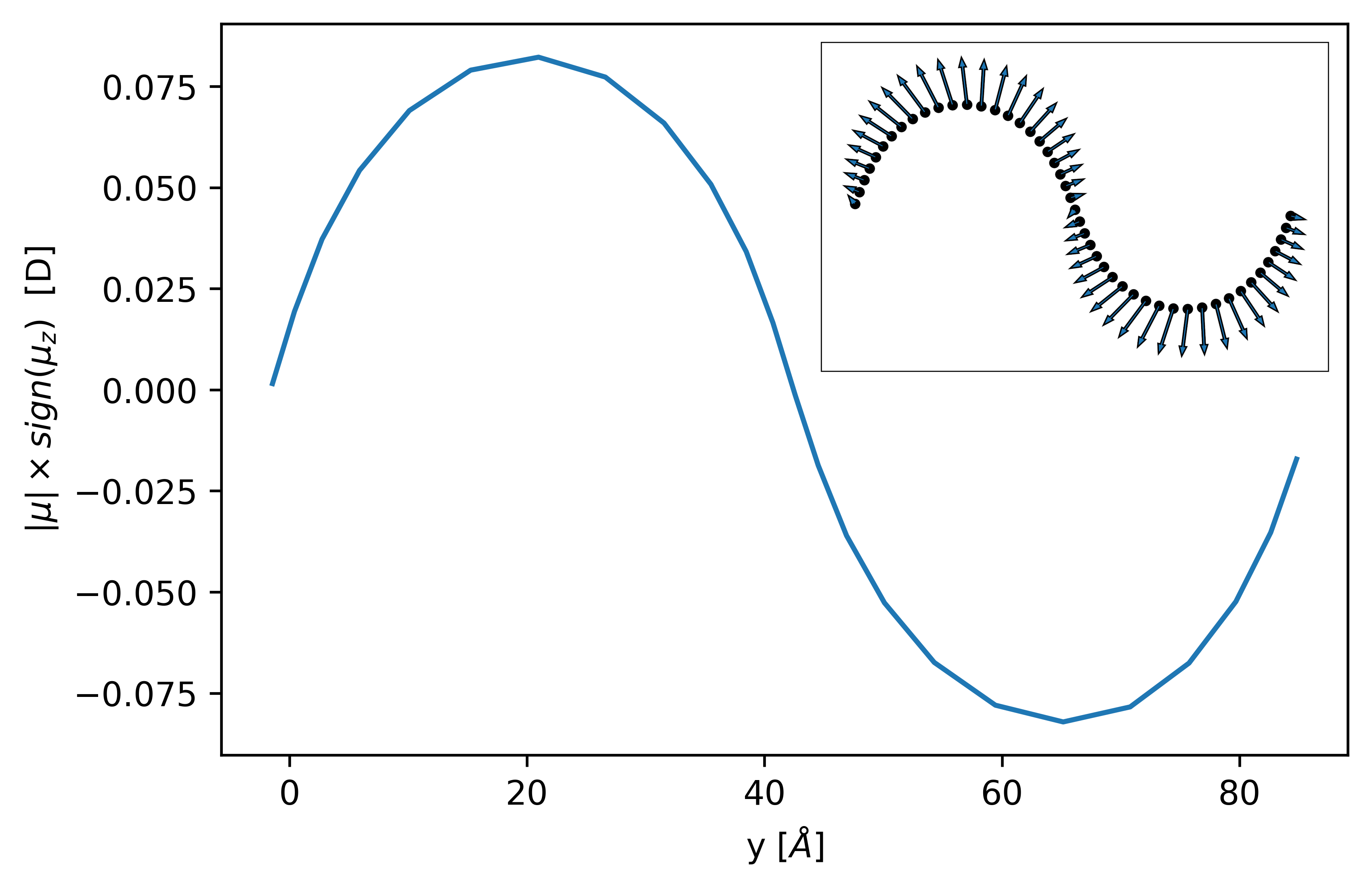}
 \caption{Magnitude of the Hirshfeld dipole vectors\cite{hirshfeld1977} of the tungsten atoms along the (24,24) wrinkle -- in order to better visualize the position along the wrinkle, the magnitude has the same sign as the z component, $\mu_z$. The inset sketches the rotation of the dipole vectors along the wrinkle (black balls are the W positions). \label{figure_dipole_hirshfeld_wrinkle1515}}
\end{figure}

Another important effect which will influence the dynamics of excitons, are the internal electric fields which are induced by the curvature \cite{shi2018flexoelectricity}. In order to estimate the local electric field which result from the local curvature, we use the dipoles as calculated with the Hirshfeld partitioning scheme \cite{hirshfeld1977}. Figure \ref{figure_dipole_hirshfeld_wrinkle1515} shows the magnitude and direction of the tungsten dipoles for the (24,24) wrinkle. One can clearly see that the dipoles at the peaks and valleys of the wrinkle are larger -- due to this inhomogeneity there will be an effective electric field which could be another force driving the excitons to the regions with higher curvature. This inhomogeneity can also be seen in the contour plot of the total electrostatic potential, Figure S13. Yet, in experiments the screening by a substrate might be important as well and we will leave this fascinating topic for future work since this is beyond the current investigation.

Most interestingly, the internal electric field due to the dipoles leads to a Rashba-like SOC splitting as can be seen in Figures \ref{fig_nano_wrinkle2424} and S15 for the CB states crossing at the $\Gamma$ point.

\subsection{SOC splitting} \label{section_result_SOC_splitting}

\begin{figure}
  \centering
  \includegraphics[width=0.49\columnwidth]{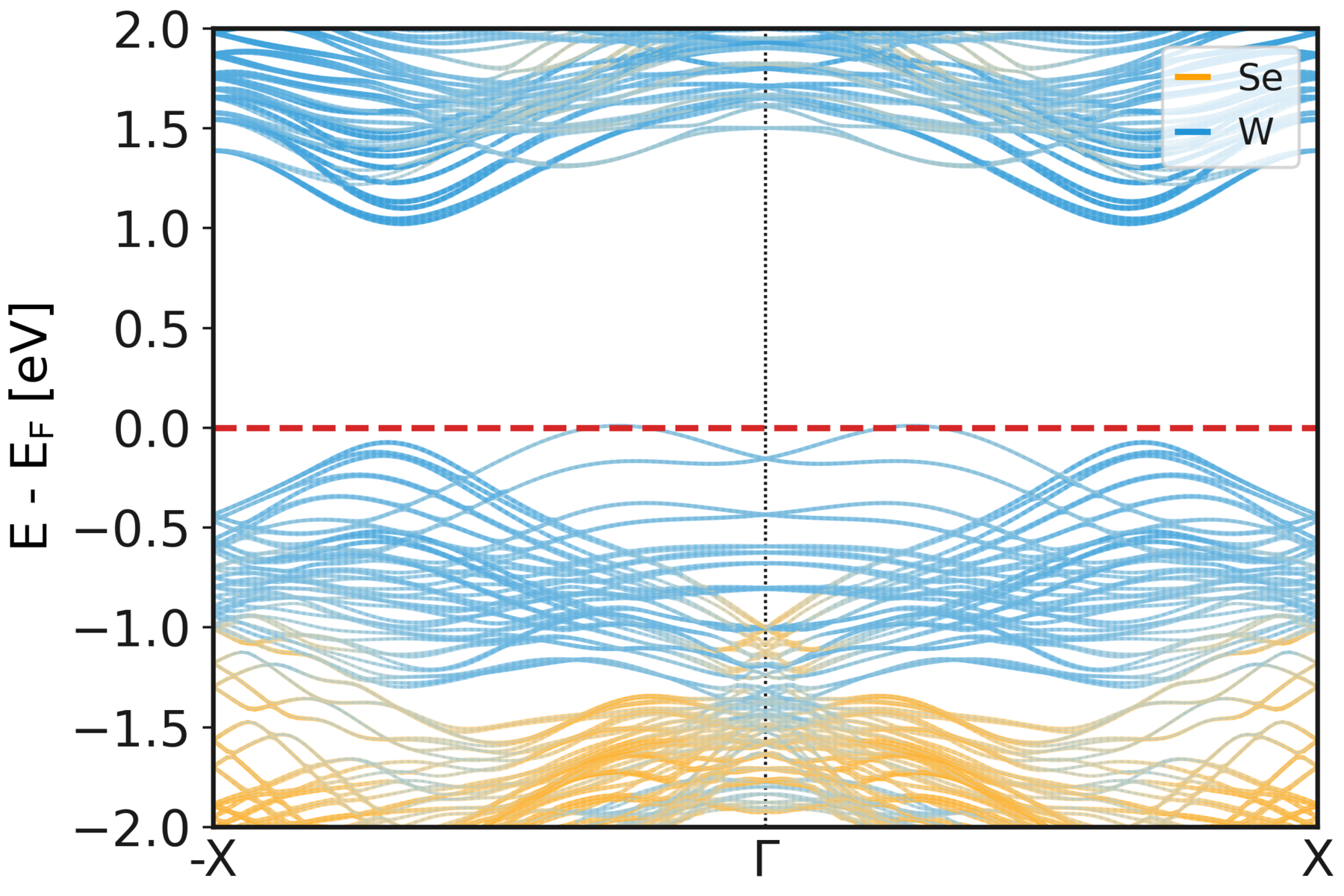}
  \includegraphics[width=0.49\columnwidth]{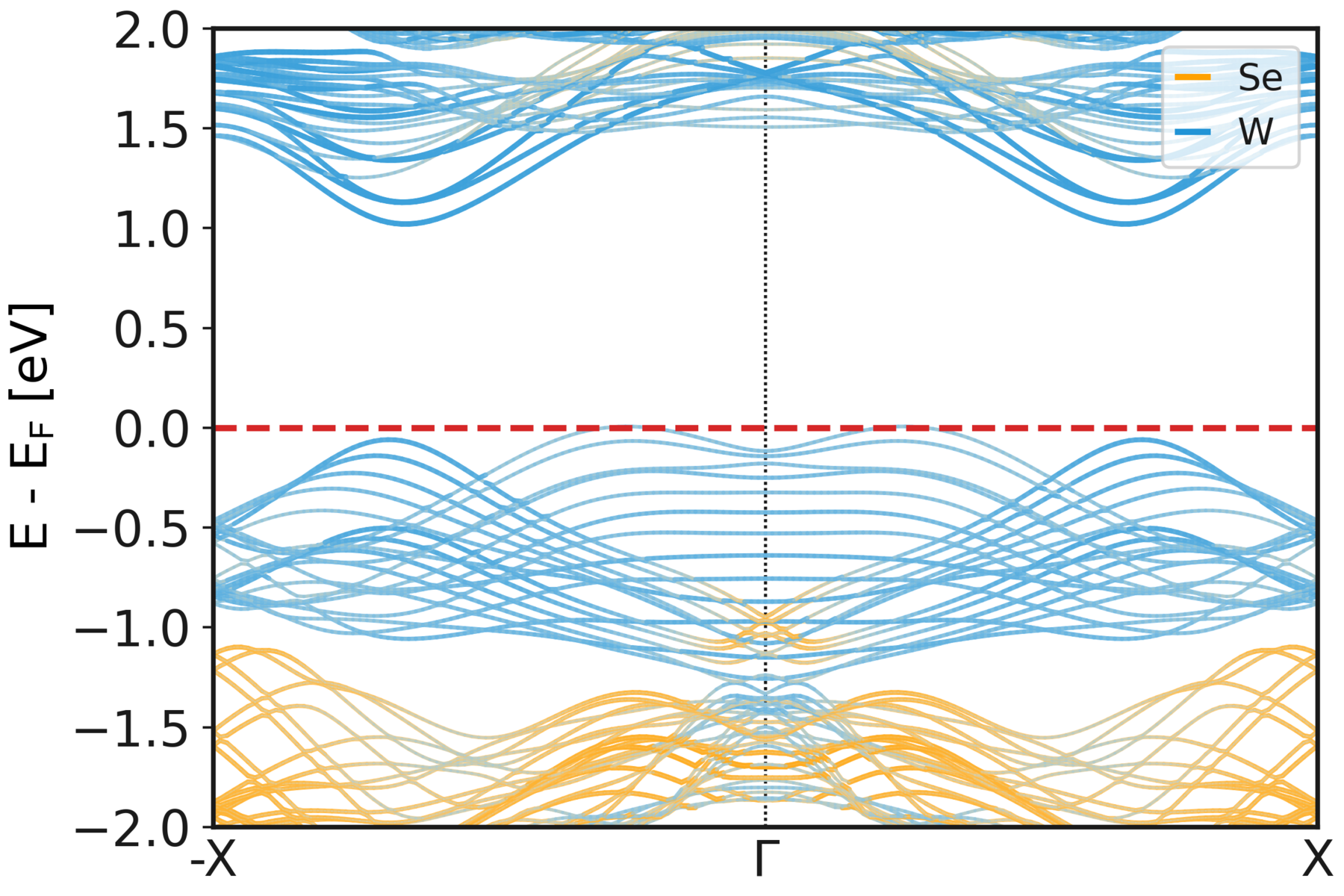}
  \caption{Band structure of the (11,11) wrinkle (left) and nanotube (right). The Rashba-like splitting in the momentum direction is clearly visible for the VB in the vicinity of the $\Gamma$ point and is -- for this system with a small $\lambda/A$ ratio -- even above the former VBM at the $\mathrm{K}$ point.}
  \label{fig_results_wrinkle1111}
\end{figure}

The splitting of the bands close to the $\Gamma$ point resembles the Rashba spin-orbit splitting found in quantum wells or Janus-type TMDCs \cite{gan2022chemical, yin2018hydrogen}. Comparing band structures with and without SOC (shown in the SI \cite{ThisPaperSI}, Figure S16), we can directly see the effect of SOC on the band dispersion. We observed this Rashba-like splitting (\textit{i.e.}, splitting of the band energies in momentum direction \cite{manchon2015new}) in all investigated systems and furthermore found in very small NTs an apparent avoided crossing of the SOC-split states which might be either due to the interaction between atoms of opposite sides or an artifact due to possible strain-induced changes of the hexagonal symmetry (\textit{cf.}, Figures \ref{fig_results_wrinkle1111} and S7).

The Rashba splitting in nanoscale wrinkles and NTs occurs due to the symmetry breaking caused by the inhomogenous strain field and the resulting electric dipoles perpendicular to the wrinkle and nanotube. Hence, in presence of spin-orbit coupling two degenerate spin bands split into two separate bands in momentum direction. These SOC-split states are mainly localized at the top (\textit{i.e.}, highest curvature part) of the wrinkle (\textit{cf.}, Figure \ref{figure_regionContribution}).

Figure \ref{fig_results_wrinkle1111} depicts the band structure of the armchair (11,11) wrinkle and nanotube highlighting also the changes with increasing curvature if compared to Figure \ref{fig_nano_wrinkle2424} -- the splitting not only increases in momentum direction but the Rashba-split states also move up in energy such that they eventually become the VBM (see also the Figures S17--S20). One main difference of the wrinkle with respect to the nanotube is the larger splitting between the uppermost VBs and the lower bands which might be due to the higher curvature at the top of the wrinkles and the more diverse local strain state in wrinkles.

Another important difference can be found in the spin texture of the highest valence bands and lowest conduction bands as shown in the section ``Brillouin zone of wrinkles/NTs and spin texture'' in the SI \cite{ThisPaperSI}.
While the NTs always show twofold degenerate bands coming from the K and K’ point of the 2D material, the degeneracy is slightly lifted in the wrinkle probably due to the different strain states along one period of the wrinkle – in fact, a slight asymmetry is also visible in the contour plot of the total electrostatic potential shown in Figure \ref{fig_tot_pot}. This also leads to a different spin texture (compared to the NT) in which the VB of the wrinkle does not automatically have the opposite spin expectation value of the band just below (VB-1). The largest contribution of $\langle\sigma_i\rangle$ for the Rashba-split states close to $\Gamma$ is always coming from $\langle\sigma_y\rangle$ and $\langle\sigma_z\rangle$, \textit{i.e.}, the directions perpendicular to $\mathbf{k}$. Figures S5 and S7 furthermore show that for the wrinkle the lowest CB has the opposite spin polarization of the VB at the K point (the extrema closer to the X point); the inhomogeneous strain is not large enough to change the pattern which is also observed in the monolayer.\cite{zhang2015}

\begin{figure}
    \centering
    \includegraphics[width=1.0\columnwidth]{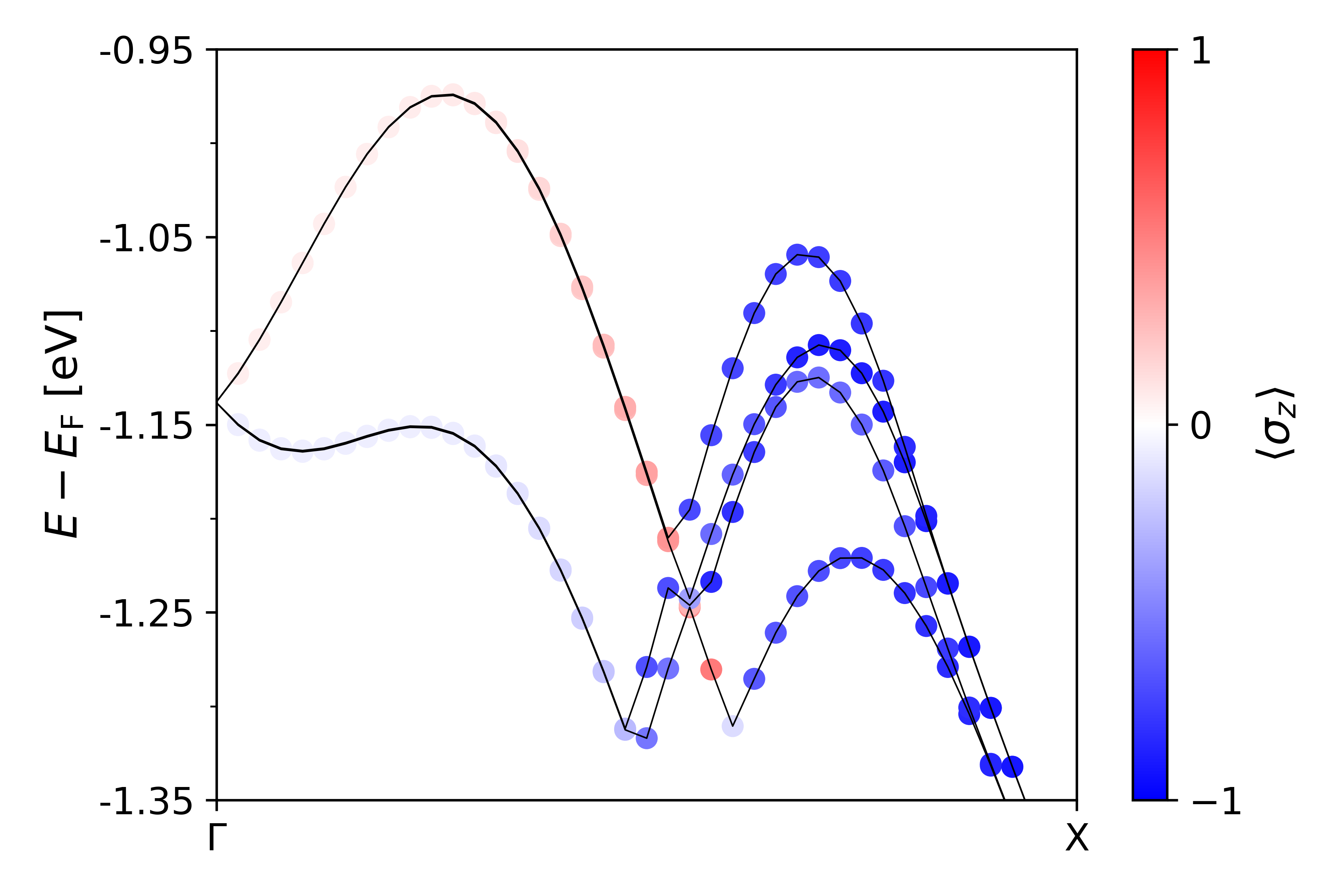}
    \caption{The expectation value $\langle\sigma_z\rangle$ for the (11,11) wrinkle is shown by the colored dots for the four highest valence bands. Note that only two bands are visible since the bands are doubly-degenerate due to the folding to the 1D Brillouin zone.}
    \label{fig_spin_texture}
\end{figure}

Figure \ref{fig_spin_texture} depicts the expectation value $\langle\sigma_z\rangle$ for the four highest valence bands of the (11,11) wrinkle. Note that close to $\Gamma$ only two bands are visible since the bands are doubly-degenerate due to the folding to the 1D Brillouin zone. We thus show the average expectation value of the two degenerate states $(\langle\sigma_z\rangle_1+\langle\sigma_z\rangle_2)/2$. The expectation values $\langle\sigma_x\rangle$ and $\langle\sigma_y\rangle$ are either zero or the two degenerate states have opposite signs. A complete discussion about the individual spin states can be found in the SI \cite{ThisPaperSI}, Figures. S5–S8.

In order to examine the strength of the Rashba-like splitting the Rashba coupling parameter\cite{kuc2015electronic}, $\alpha_R$, 
has been calculated for the armchair systems using
\begin{equation}
    \alpha_{R}=\frac{2E_R}{k_R},
\end{equation}
where $E_R$ and $k_R$ are the Rashba energy and the shift of the bands in the momentum direction, respectively. Unfortunately, the different back folding of the bands in the zigzag structures leads to the primitive unit cell's $\mathrm{K}$ point being mapped to $\Gamma$ thus obscuring the SOC-split states. This prevents an easy and correct fitting of the Rashba model to the band structure even if the band structures in the SI \cite{ThisPaperSI} clearly show the same splitting.

\begin{figure}[H]
    \includegraphics[width=1.\columnwidth]{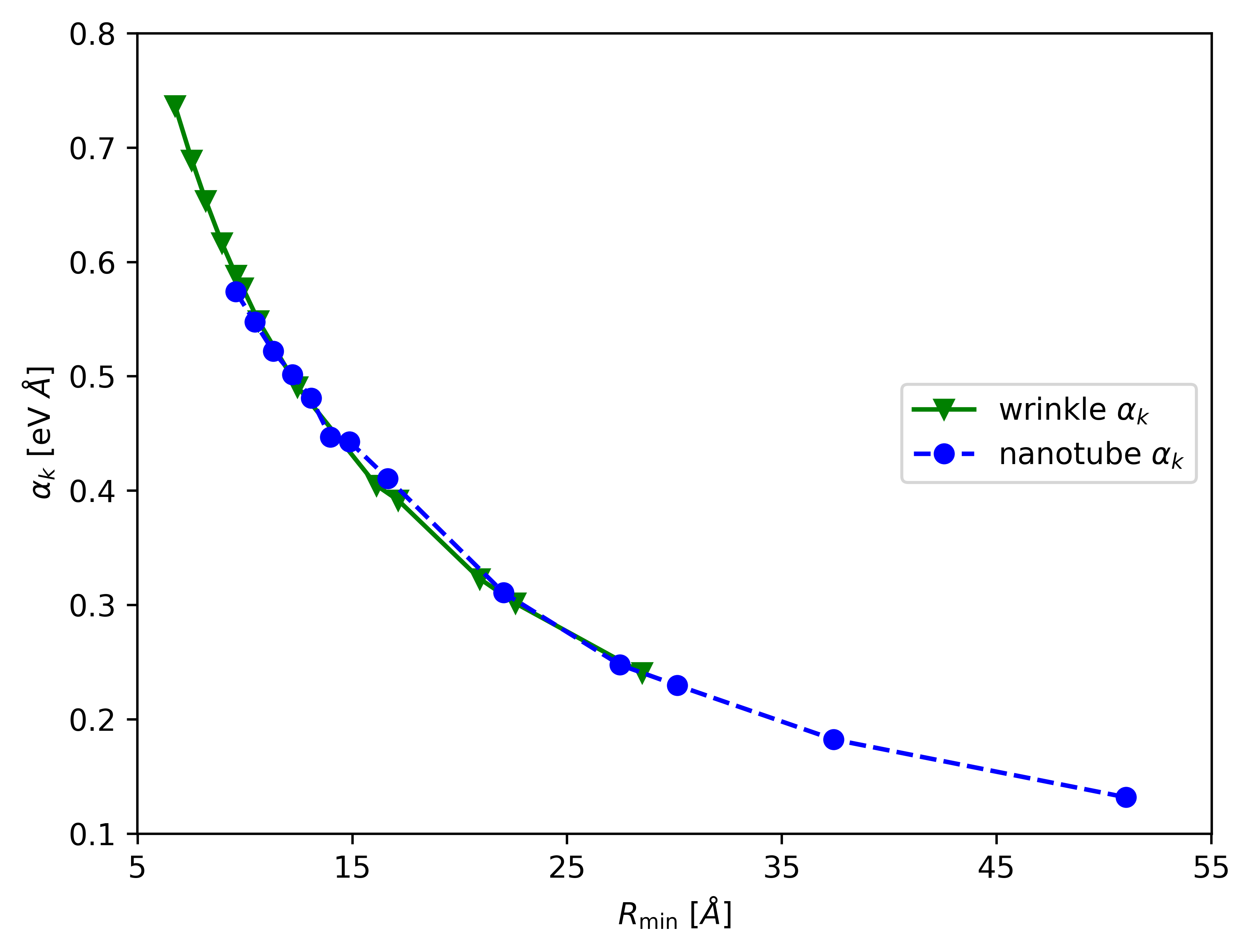}
    \caption{The Rashba coupling parameter in armchair \ce{WSe2} nanotubes and wrinkles. The splitting reduces as the $R_{min}$ increases, \textit{i.e.}, as the local strain due to the curvature decreases.} 
    \label{fig_rashba_parameters}
\end{figure}

The Rashba coupling parameter shown in Figure \ref{fig_rashba_parameters} decreases as the wrinkle wavelength or nanotube diameter increases. This trend can be explained by the decreasing strain difference between the outer and the inner chalcogen layer and the corresponding smaller induced electric field. It is also evident that band gap and band dispersion of large NTs and wrinkles are converging to the corresponding flat monolayer (\textit{cf.} Figures S3 and S19).
Furthermore, the Rashba coupling parameter is only comparable between wrinkles and NTs if it is shown with respect to the minimum radius of curvature, $R_\mathrm{min}$, i.e. the highest curvature. This is once more due to the wrinkles having higher curvature at their peaks. The Rashba coupling parameter in our structures are relatively high and almost half the size of elemental surfaces \cite{bihlmayer2015focus} and one order of magnitude larger than in Janus type TMDCs \cite{cheng2013spin,liu2022janus}.

Furthermore, the coupling parameter in wrinkles/NTs with small $R_\mathrm{min}$ can also be comparable to the one found in heterostructures including \ce{BiSb}\cite{singh2017} even if one probably needs to include higher order terms to properly describe the large splitting.\cite{gupta2021rashbahetero}
It should be noted that the largest splittings are those easiest to reach experimentally since the bands move above the monolayer VBM at $\mathrm{K}$.
This is also a possibility for tuning the electronic structure and the SOC-induced splitting for the application in spintronic devices: using (mono-)layers of TMDCs deposited on elastic substrates which can be used to induce different wrinkle morphologies as shown in, e.g., Refs.~\citenum{giambastiani2022strain, lee2022wrinkled}.

It should be noted however that the exact shape of the wrinkle can be different from our research, e.g., due to substrate effects; the conclusions in above sections still hold as they are due to strain effects and the subsequent symmetry breaking. Yet slight quantitative differences are expected and require further investigations. We expect that nanotubes can still be used to model the curvature effects in wrinkled systems and that the variation of the local band gap leads to exciton funneling. Furthermore, the interaction with substrates and additional external fields might lead to even higher Rashba-like splittings.


\section{Conclusion}
We investigated NTs and 2D wrinkles of \ce{WSe2} theoretically and analyzed the influence of the induced inhomogeneous strain on their electronic properties but the following conclusions should generally apply to all TMDCs. We found that the inhomogeneous strain causes symmetry breaking in these structures which leads to a Rashba-like splitting of the valence band at $\Gamma$. Therefore, these structures -- particularly the smaller wavelengths -- could be promising candidates for spintronic applications. In fact, spin-polarized STM using an additional graphene layer as electrode\cite{qiu2021visualizing,nieken2022direct} should be able to measure the Rashba-like splitting of the VBM. 
We believe that this is a general feature of wrinkled 2D TMDCs due to the non-uniform strain and our study will thus pave the way for the employment of a wide range of materials in spintronic devices.
This bears witness to the important role of SOC in the physics of 2D TMDCs nanoscale wrinkles and NTs. Moreover, wrinkling should be regarded as a method for introducing out of plane dipoles in the 2D system which might be useful in other contexts. Investigations of bilayers and hetrostructures made of dichalcogenides could even result in appearance of more fascinating phenomena.

Furthermore, nanoscale TMDC wrinkles do not follow a sine wave profile as suggested by continuum mechanics and which has been widely utilized in strain analysis. The profiles in our study are basically composed of a superposition of two sine waves. Thus, the curvature varies smoother in the profiles maxima and minima. This suggests that the classical formulation of such structures conceals important physical properties and conclusions based upon might be misleading, \textit{e.g.}, in the calculation of strain fields from the surface topology in nanoscale wrinkles. 
Moreover, we attribute the funneling of excitons to the localization of states in different spatial locations due to the presence of the inhomogeneous strain. Yet, more advanced calculations are needed to evaluate the influence of the, \textit{e.g.}, induced dipoles or substrate and an additional investigation of the wrinkled samples via STM \cite{qiu2021visualizing,nieken2022direct} could furthermore help to understand the band alignment better.

Additionally, we demonstrated that NTs can be used to approximate the wrinkles for reduction of computational costs, nonetheless one needs to take care of the differences that exist.


\section{Methods}

NTs and wrinkles of a monolayer of \ce{WSe2} having two edge types: armchair and zigzag were investigated using an all-electron method based on DFT as implemented in FHI-aims code \cite{blum2009ab}. We utilized the Perdew–Burke–Ern\-zer\-hof (PBE) exchange-correlation functional \cite{perdew1998perdew} together with the Tkatchenko-Scheffler dispersion correction \cite{tkatchenko2009accurate} and the non-self-consistent SOC implementation \cite{huhn2017one}. Additionally, no symmetry was imposed on any of the structures in this calculations. In order to compare wrinkles and NTs, the initial wrinkled structure was created with an elliptical profile as shown in Figure \ref{figure_method_creation} with a wavelength to amplitude ratio of $\lambda/A = 4$ and using NTs as input. The unit cell was fixed only in the direction of the wrinkles with $\lambda$ to retain the compression. All structures have been relaxed utilizing the Broyden-Fletcher-Godfarb-Shanno method to reach forces below $1\: \mathrm{meV}/\text{\AA}$. Subsequently, the Mulliken projected band structures with and without spin-orbit coupling were calculated. NTs are labeled with the rolling vector (m,n) and for wrinkles the similar notation is used such that the (m,n) wrinkle is similar to the (m,n) nanotube. The geometries as well as the band structures for all systems which have been investigated within this work were uploaded to the NOMAD repository, Ref.~\citenum{nomadupload}

\section*{Conflicts of interest}
There are no conflicts of interest to declare.

\medskip
\begin{acknowledgments}
The authors would like to thank Prof. T. Heine, Dr. A. Kuc, R. Kempt and F. Arnold for fruitful discussions. 
This project was financially supported by the SFB 1415, Project ID No. 417590517. We would like to acknowledge the Center for Information Service and High Performance Computing [Zentrum für Informationsdienste und Hochleistungsrechnen (ZIH)] at TU Dresden. Also, the authors gratefully acknowledge the Gauss Centre for Supercomputing e.V. (www.gauss-centre.eu) for funding this project by providing computing time through the John von Neumann Institute for Computing (NIC) on the GCS Supercomputer JUWELS \cite{alvarez2021juwels} at Jülich Supercomputing Centre (JSC).
\end{acknowledgments}
\medskip

\begin{figure}
\textbf{Table of Contents}\\
\medskip
  \includegraphics[width=1.0\columnwidth]{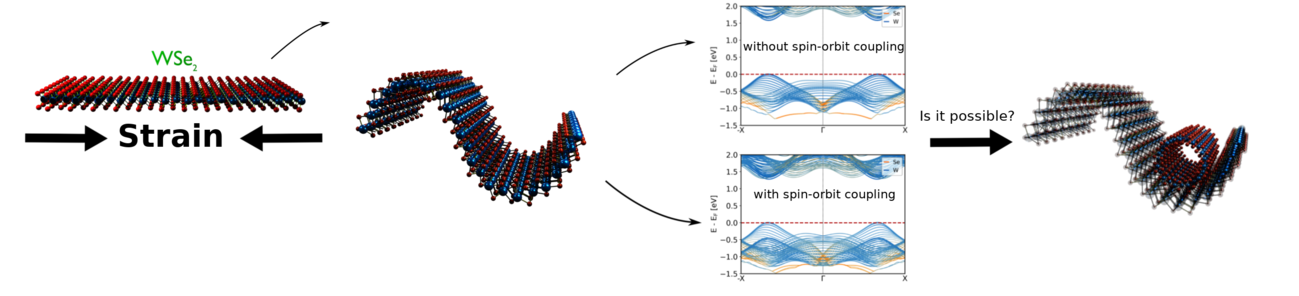}
  \medskip
  \caption*{\textbf{Strain to introduce Rashba-like splitting:} By employing an \textit{ab-initio} method (Density functional theory) a Rashba-like splitting is observed in the electronic band structure of strained two-dimensional transition metal dichalcogenides in the forms of wrinkles and nanotubes. Additionally, The assumption of modeling wrinkles as nanotubes is investigated for the electronic studies of these structures.}
\end{figure}
\clearpage

\newpage
\section*{References}
\bibliography{manuscript}

\newpage
\appendix
\setcounter{figure}{0}
\renewcommand{\figurename}{Figure}
\renewcommand{\thefigure}{S\arabic{figure}}
\section{Supporting Information}

\section{Curvature analysis}
\begin{figure}
\centering
\includegraphics[width=0.4\linewidth]{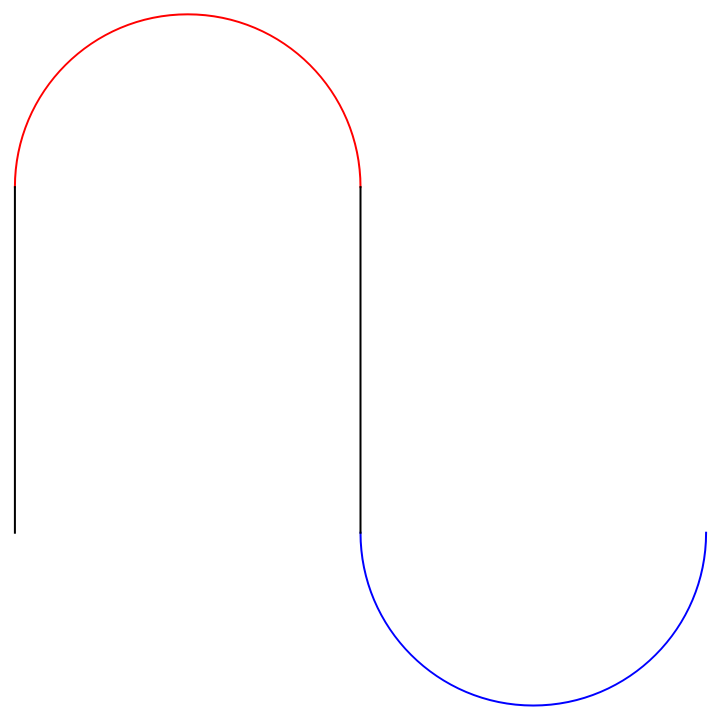}
\includegraphics[width=0.4\linewidth]{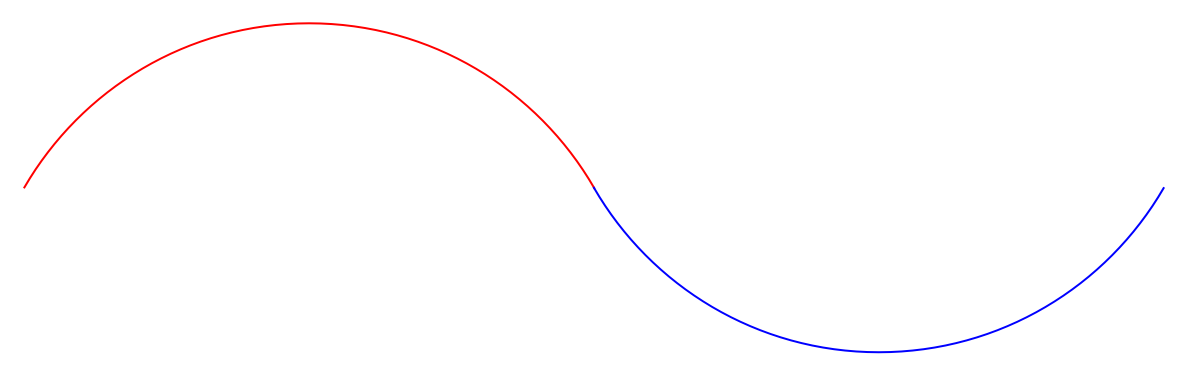}
\caption{Circular profile for $\lambda<4A$ (left), and $\lambda\geq4A$ (right).\label{fig_circle}}
\end{figure}

\begin{figure}
\centering
\includegraphics[width=0.8\linewidth]{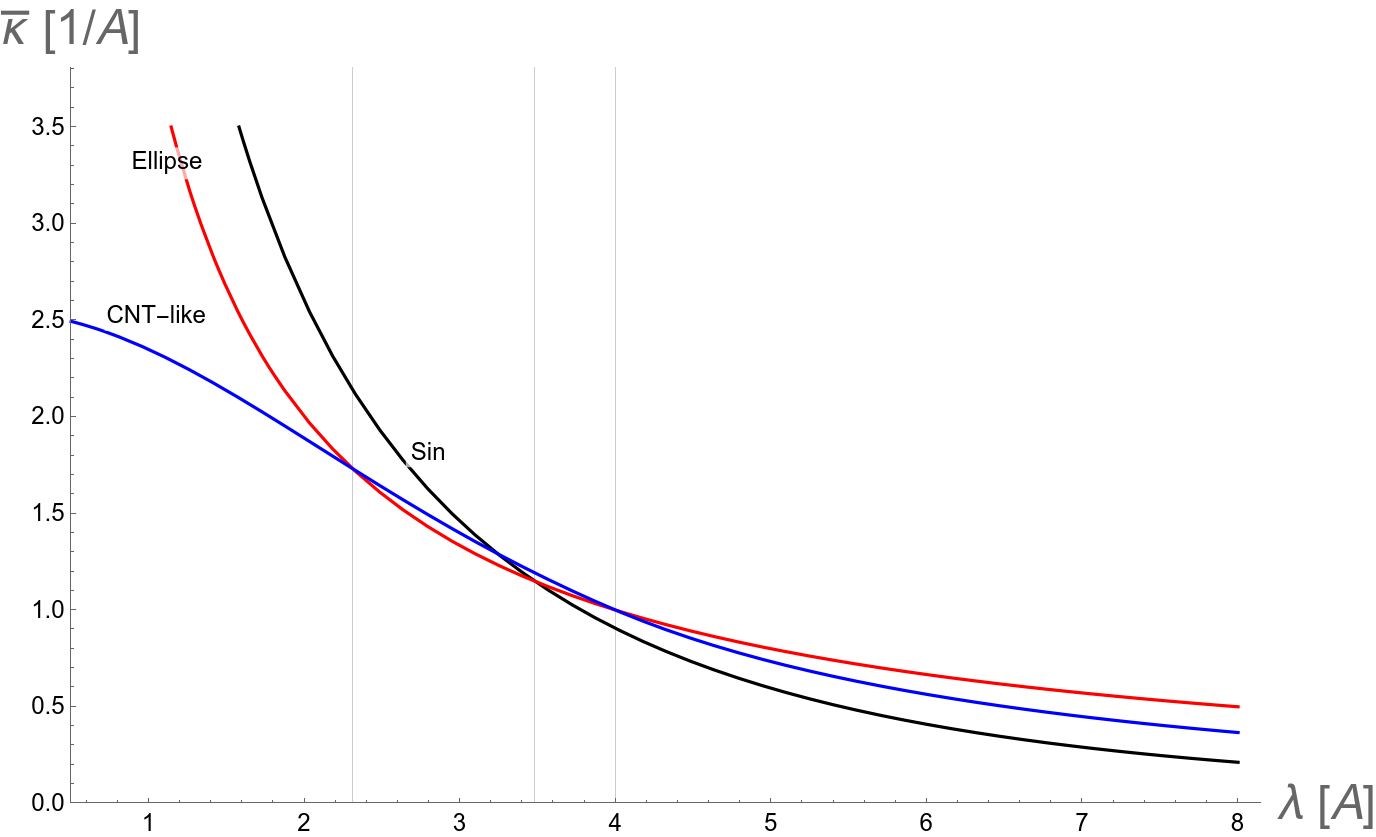}
\caption{Comparison of the average curvature of different profiles as function of the wavelength $\lambda$ in units of the amplitude $A$.\label{figure_profileComparison}}
\end{figure}

Curved systems made from a flat two dimensional (2D) material have an energy which is higher by the amount of the energy that is required to create the curvature. Therefore, in a first approximation, the 2D material tends to acquire the profile that has the smallest average curvature. Accordingly, to estimate possible starting structures which are also comparable to the nanotubes, we analyze the average curvature of three different wrinkle profiles: sine wave, circular (similar to a nanotube cut) and elliptical.
A sine wave can be described by the parametric equation
\begin{equation}
    y= A \sin(t), x=\frac{t}{2\pi}\lambda,
\end{equation}
leading to a curvature of
\begin{equation}
    \kappa = \frac{4 \pi ^2 A}{\lambda  \sqrt{4 \pi ^2 A^2+\lambda ^2}},
\end{equation}
where $\kappa$, $\lambda$ and $A$ are the curvature, wavelength and amplitude, respectively.
The average curvature $\overline{\kappa}$ can be estimated using the integral of the curvature over the first quarter of one period (due to symmetry) divided by the arc length of this part, resulting in
\begin{align}
    \overline{\kappa} = \frac{8 \pi ^3 A}{\lambda ^2 \sqrt{4 \pi ^2 A^2+\lambda ^2}\:K\!\!\left(-\frac{4 A^2 \pi ^2}{\lambda ^2}\right)},
\end{align}
with $K(z)$ being the complete elliptic integral of the first kind.
Similarly, for an ellipse with one axis being a quarter wavelength and the other axis being the amplitude, we have
\begin{equation}
    y=A \cos(t);x=\frac{\lambda}{4} \sin(t)
\end{equation}
resulting in an average curvature of
\begin{align}
    \overline{\kappa} = \frac{16\:K\!\!\left(1-\frac{\lambda ^2}{16 A^2}\right)}{\lambda ^2 \:K\!\!\left(1-\frac{16 A^2}{\lambda ^2}\right)}.
\end{align}
The circular profile we construct with a half circle connected to another half by straight lines as long as $\lambda\leq4A$, see also the left-hand panel of Figure \ref{fig_circle}. The average curvature for this profile is then a combination of the constant curvature $1/R=4/\lambda$ of the half circles and a zero curvature for the straight parts leading to an average curvature of
\begin{equation}
    \overline{\kappa}=\frac{8 \pi }{\lambda  \left(2 \left(2 A-\frac{\lambda }{2}\right)+\frac{\pi  \lambda }{2}\right)}.
\end{equation}
For larger circular profiles with $\lambda>4A$ we have essentially two circular arcs which are connected to each other (right-hand panel of Figure \ref{fig_circle}), resulting in
\begin{equation}
    \overline{\kappa} = \frac{2 \left| \sin ^{-1}\left(\frac{8 A \lambda }{16 A^2+\lambda ^2}\right)\right| }{\frac{\lambda ^2}{16 A}+A}.
\end{equation}
Figure \ref{figure_profileComparison} shows the average curvature as a function of the wrinkle wavelength in units of the amplitude. The figure clearly shows that for different $\lambda/A$ values, different profiles are preferred. In the long wavelength limit, the harmonic, \textit{i.e.}, sine-like profile is the lowest curvature solution. The elliptical and circular profile cross at $\lambda=4A$ as one would expect since the ellipse reduces to a circle with our definition. For $2.309A\lesssim\lambda\lesssim3.483A$ the elliptical profile is preferred and for even smaller wavelength the circular arcs shown in the left panel of Figure \ref{fig_circle} become the lowest curvature solution.
However, in several test with different starting geometries for the wrinkles with $\lambda=4A$ we found that the relaxed structure does not follow a simple sine wave (or elliptical) profile but that the profiles are composed of a superposition of two sine waves leading to a smoother variation of the curvature in the profiles maxima and minima.

The average and maximum curvature derived from the relaxed positions of the tungsten atoms (\textit{cf.}, main text for details) for all investigated systems are summarized in Tables.~\ref{table_wrinkle_curvature} and \ref{table_nanotube_curvature}.
\begin{table}
\begin{tabular}{c|c|c}
Name    &   Average curvature [1/\AA]  & Max curvature  [1/\AA] \\
\hline
(10,10) &   0.1000	& 0.1481 \\ 
(11,11) &	0.0057	& 0.1327 \\
(12,12) &	0.0847	& 0.1222 \\ 
(14,14) &	0.0733	& 0.1042 \\ 
(15,15)	&   0.0721  & 0.1009 \\
(16,16)	&   0.0670	& 0.0941\\ 
(18,18)	&   0.0577	& 0.0803 \\ 
(24,24)	&   0.0439	& 0.0583\\
(30,30)	&   0.0349	& 0.0477\\
(33,33)&	0.0318	& 0.0442\\
(41,41)	&   0.0256	& 0.0349\\
(14,0)	&   0.1170	& 0.1796\\
(20,0)	&   0.0860	& 0.1246\\
(24,0)	&   0.0728	& 0.1038\\
(27,0)	&   0.0653	& 0.0922\\
(34,0)	&   0.0524	& 0.0733\\
(39,0)	&   0.0459  & 0.0639\\
(46,0)	&   0.0391	& 0.0542
\end{tabular}
\caption{Average and maximum curvature for all investigated wrinkles.}
\label{table_wrinkle_curvature}
\end{table}
\begin{table}
\begin{tabular}{c|c}
Name & Curvature [1/\AA] \\
\hline
(10,10) & 0.1044 \\
(11,11) & 0.0955 \\
(12,12) & 0.0883 \\
(13,13) & 0.0818 \\
(14,14) & 0.0764 \\
(15,15) & 0.0715 \\
(16,16) & 0.0672 \\
(18,18) & 0.0600 \\
(24,24) & 0.0453 \\
(30,30) & 0.0364 \\
(33,33) & 0.0332 \\
(41,41) & 0.0267 \\
(56,56) & 0.0196 \\
(14,0)  & 0.1236 \\
(20,0)  & 0.0905 \\
(24,0)  & 0.0765 \\
(27,0)  & 0.0685 \\
(34,0)  & 0.0550 \\
(39,0)  & 0.0481 \\
(46,0)  & 0.0410 \\
(58,0)  & 0.0326 \\
(65,0)  & 0.0291
\end{tabular}
\caption{Curvature for all investigated NTs. \label{table_nanotube_curvature}}
\end{table}

\newpage
\begin{figure}
      \includegraphics[width=1.\linewidth]{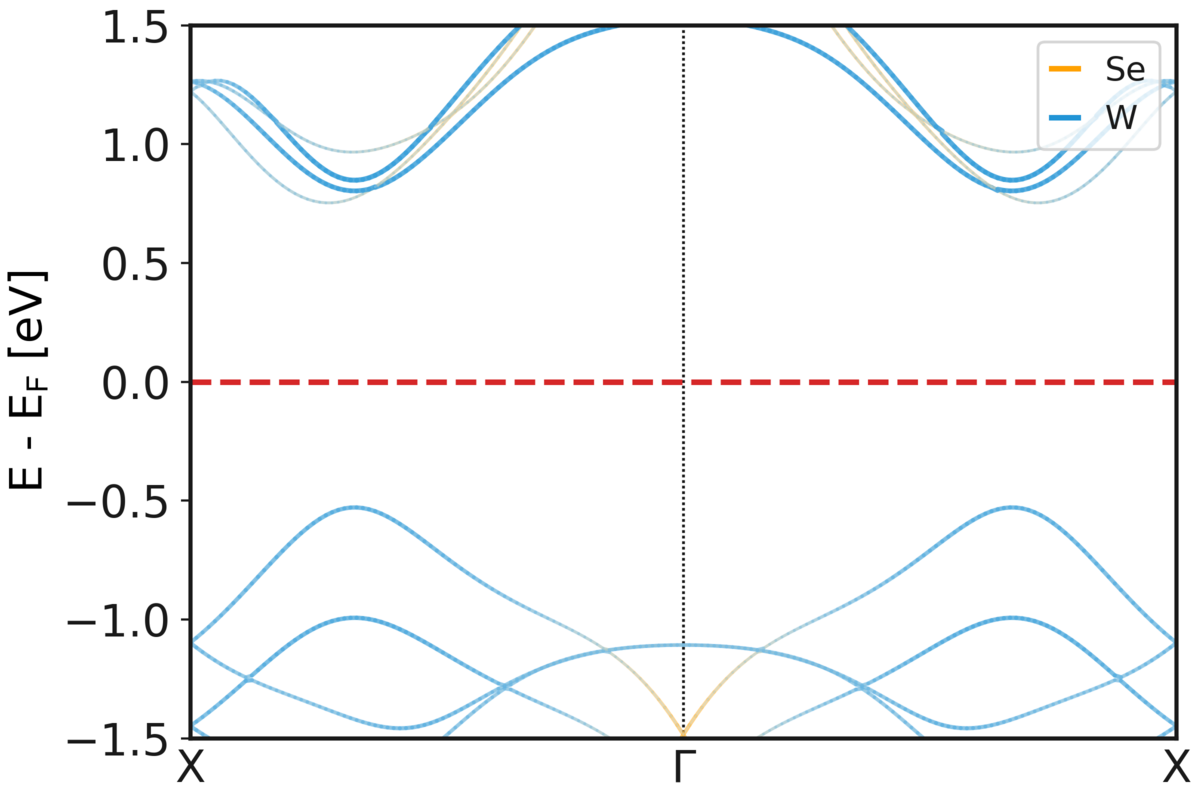}
      \caption{Projected band structure of monolayer \ce{WSe2} in the rectangular unit cell described by the supercell matrix $\begin{pmatrix}1&0&0\\-1&2&0\\0&0&1\end{pmatrix}$. The direct band gap at the hexagonal unit cells K point is mapped onto the $\Gamma-\mathrm{X}$ as well as the minimum which is between K and $\Gamma$ which can be seen in this band structure as the light-blue band which is the CBM.}
\end{figure}

\newpage
\section{Brillouin zone of wrinkles/nanotubes and spin texture}
In order to understand the differences in the band structure and the spin texture near the $\Gamma$ point for wrinkles and nanotubes, we want to briefly summarize, how the Brillouin zone (BZ) for 1D tubes can be derived from a 2D monolayer with hexagonal symmetry.\cite{saito}
\begin{figure}
\centering
\includegraphics[width=0.7\linewidth]{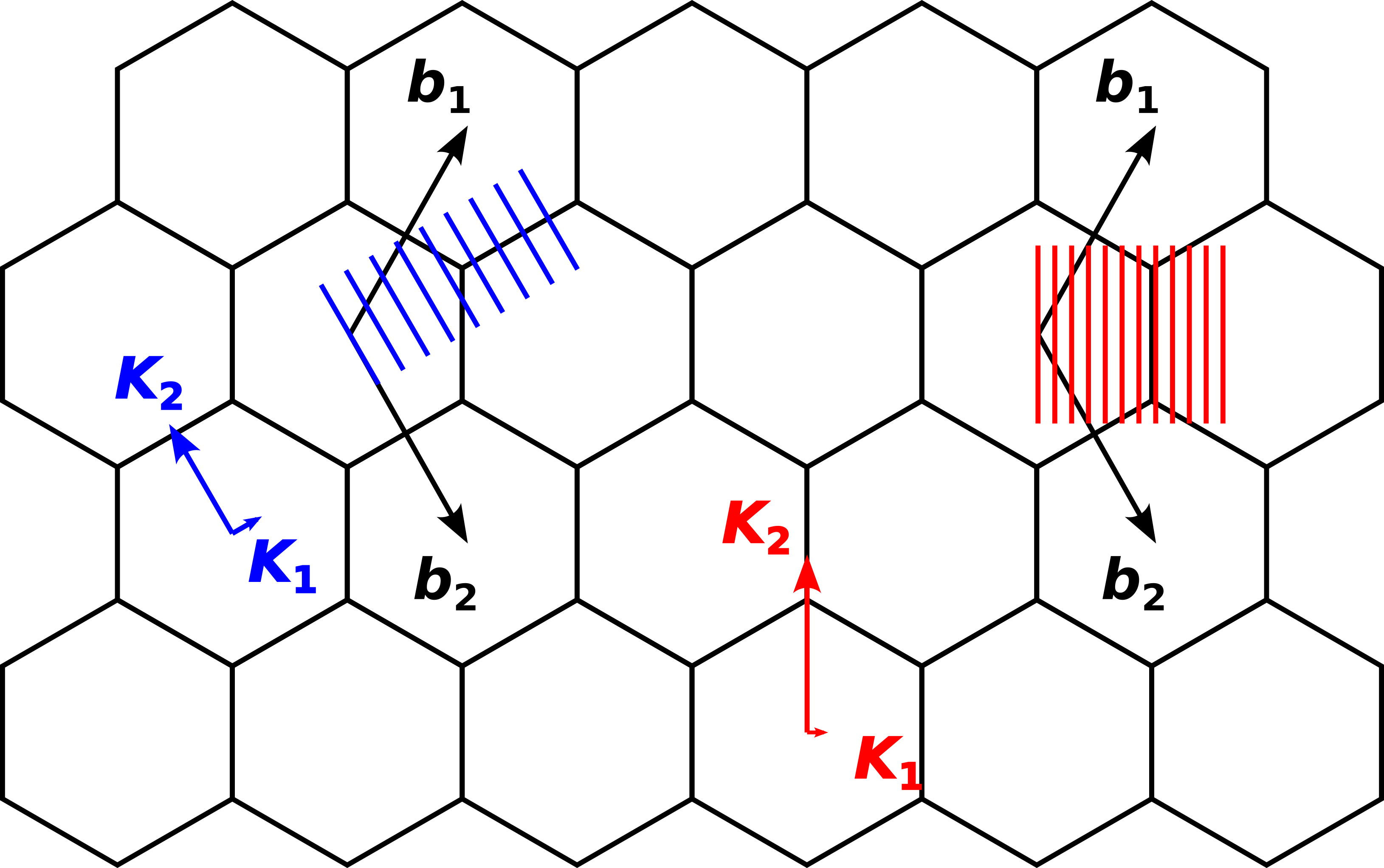}
\caption{2D-to-1D band structure mapping for zigzag (blue) and armchair (red) nanotubes. The reciprocal lattice vectors of, \textit{e.g.}, monolayer \ce{WSe2} are indicated by $\mathbf{b}_1$ and $\mathbf{b}_2$. $\mathbf{K}_1$ and $\mathbf{K}_2$ are the reciprocal lattice vectors of $\mathbf{C}_h$ and $\mathbf{T}$ which define the geometry of the nanotube.\cite{saito} The blue/red line segments correspond to the unique crystal momentums that satisfy the circumferential periodic boundary conditions of a zigzag/armchair nanotube. The separation between two segments equals $|\mathbf{K}_1|$, and the length of each line segment is $|\mathbf{K}_2|$. For zigzag and armchair tubes the number of line segments is $2n$. The 1D band structure of a nanotube can be obtained by cutting the band structure of the 2D monolayer along the colored line segments and mapped to 1D along the direction of $\mathbf{K}_2$. This mapping will be changed slightly once the distortion of the hexagonal lattice due to the curvature-induced strain is taken into account.\label{fig_bz}}
\end{figure}

From a structural perspective, a nanotube of a hexagonal system such as graphene or transition-metal dichalcogenides can be derived from the monolayer which is rolled up to form a tubular structure with a finite radial curvature. The direction of rolling and the circumference of the resultant tube are
determined by the chiral vector $\mathbf{C}_h=n \mathbf{a}_1 + m \mathbf{a}_2$, where $\mathbf{a}_1$ and $\mathbf{a}_2$ are the 2D primitive lattice vectors of the corresponding monolayer.\cite{saito} The two integer indices $n$ and $m$ form a pair $(n, m)$ that specifies the chirality of the tube, with $(n, n)$
corresponding to armchair tubes, $(n \neq 0, m = 0)$ to zigzag tubes, and other cases of $(n \neq m \neq 0, m \neq 0)$ being chiral tubes. The translational vector $\mathbf{T}$, which is parallel to the tube axis and normal to $\mathbf{C}_h$, reflects the 1D translational symmetry of the nanotube and is defined as:
\begin{align}
 \mathbf{T}&=t_1\mathbf{a}_1 + t_2\mathbf{a}_2,\quad\text{with}\\
 t_1&=(2m+n)/d_R, \quad t_2=-(2n+m)/d_R,\quad d_R=\gcd(2n + m,2m + n),
\end{align}
where $\gcd(i,k)$ denotes the greatest common divisor of the 2 integers $i$ and $k$.

In this 1D system, the Bloch wavefunction must be the same after a distance corresponding to the chiral vector $\mathbf{C}_h$, that is, $\Psi_\mathbf{k}(\mathbf{r} + \mathbf{C}_h) = \Psi_\mathbf{k}(\mathbf{r})$, where $\mathbf{k}$ is a 2D crystal momentum. This means that the wave vector associated with the circumferential direction $\mathbf{C}_h$ becomes quantized. The energy bands of the 1D system thus consist of a set of one-dimensional energy dispersion relations which are cross sections of those of the 2D system as indicated by the colored lines in Figure~\ref{fig_bz}. For more details we refer the interested reader to Ref.~\citenum{saito}.

For $\lambda\gg A$ and large diameter tubes, the main difference between tubes and wrinkles is the boundary condition -- while the tubular structure leads to a quantized wave vector which is associated with the circumferential direction $\mathbf{C}_h$, the wrinkle does not have this restriction. The creation of a small amplitude/large wavelength wrinkle is like the formation of a supercell with a perturbation which has the periodicity defined by the wavelength $\lambda$. Thus, the corresponding 1D bandstructure does not consist of ``cuts'' of the 2D bandstructure but is -- for small perturbations -- a backfolded version of the full 2D bandstructure in which the ``perturbation'' due to the curvature leads to a splitting of the bands. Yet, in our structures with $\lambda\approx4A$ the deformation cannot be considered a small perturbation and results in the localization of states in specific regions (\textit{cf.}, Figure \ref{figure_regionContribution} in the main text). The band structure is thus a combination of 1D localized systems which are perturbed by the small interactions between them and is very similar to the band structure of the nanotubes with the same curvature as the maximum curvature in the wrinkles.

\begin{figure}
\centering
\includegraphics[width=\linewidth]{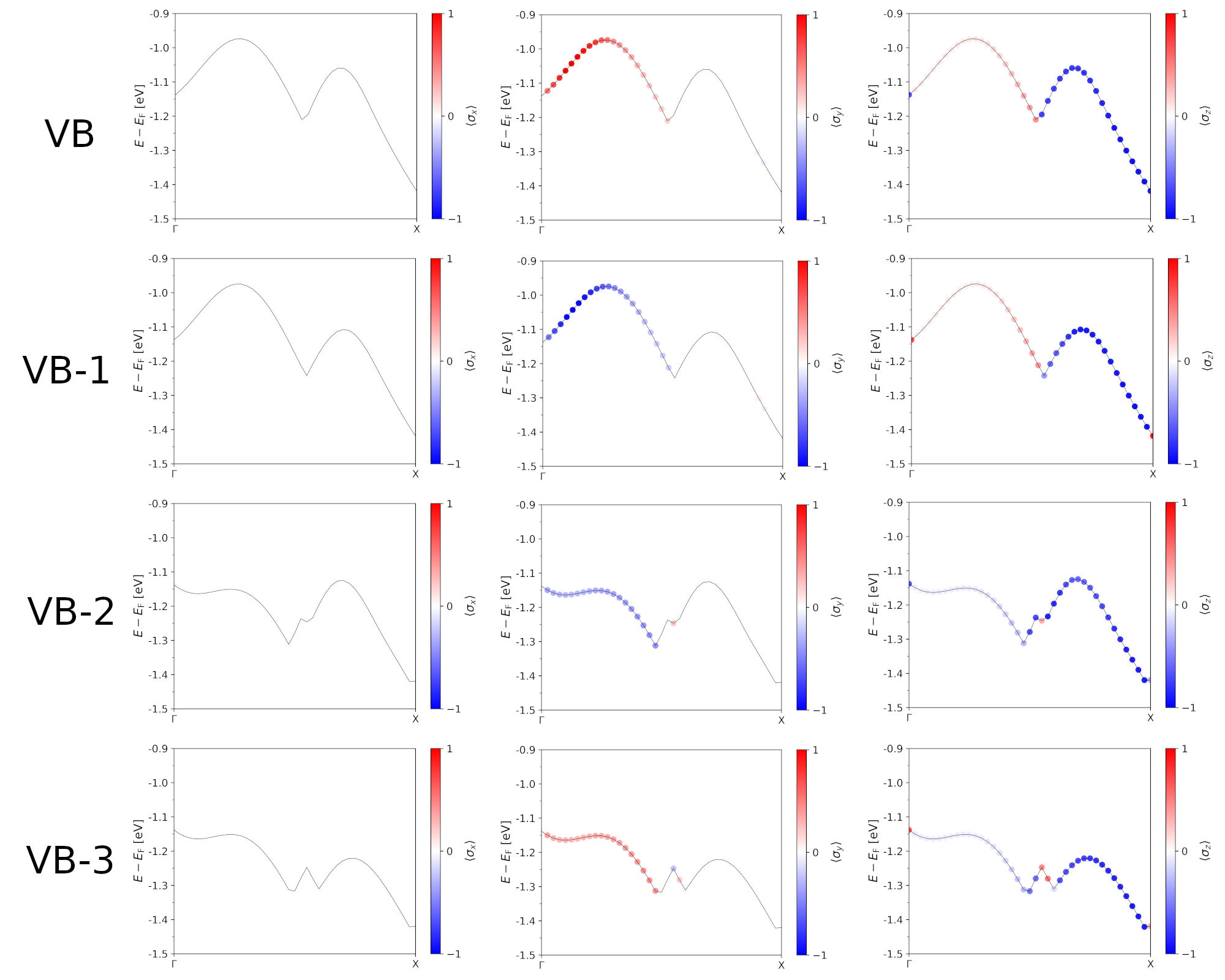}
\caption{Expectation values of the Pauli matrices $\langle\sigma_i\rangle$ for the four uppermost valence bands (VB) for the (11,11) wrinkle as calculated with the FHI-aims code.\label{fig_spinwrinkleVB}}
\end{figure}

\begin{figure}
\centering
\includegraphics[width=\linewidth]{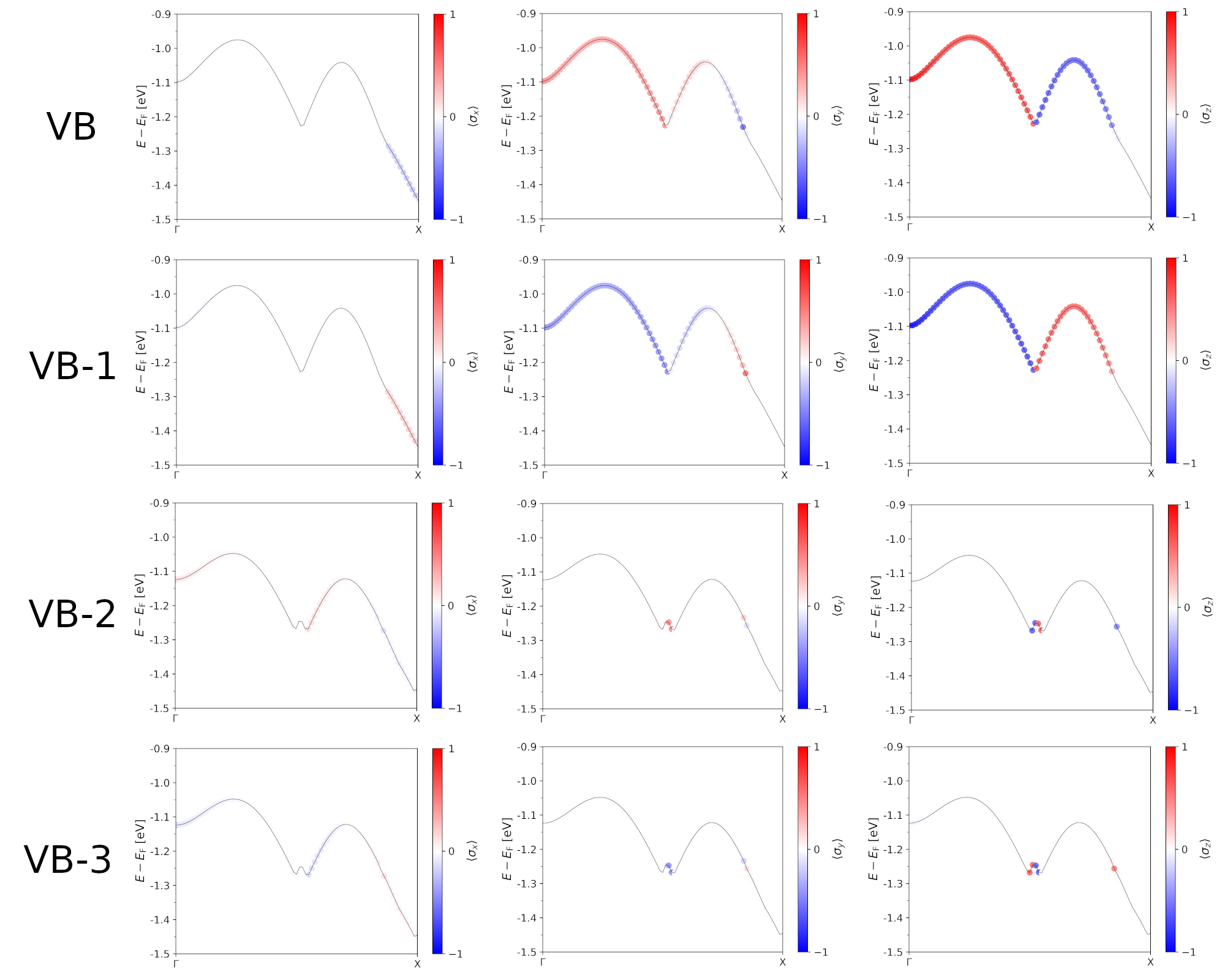}
\caption{Expectation values of the Pauli matrices $\langle\sigma_i\rangle$ for the four uppermost valence bands (VB) for the (11,11) NT as calculated with the FHI-aims code.\label{fig_spintubeVB}}
\end{figure}

\begin{figure}
\centering
\includegraphics[width=\linewidth]{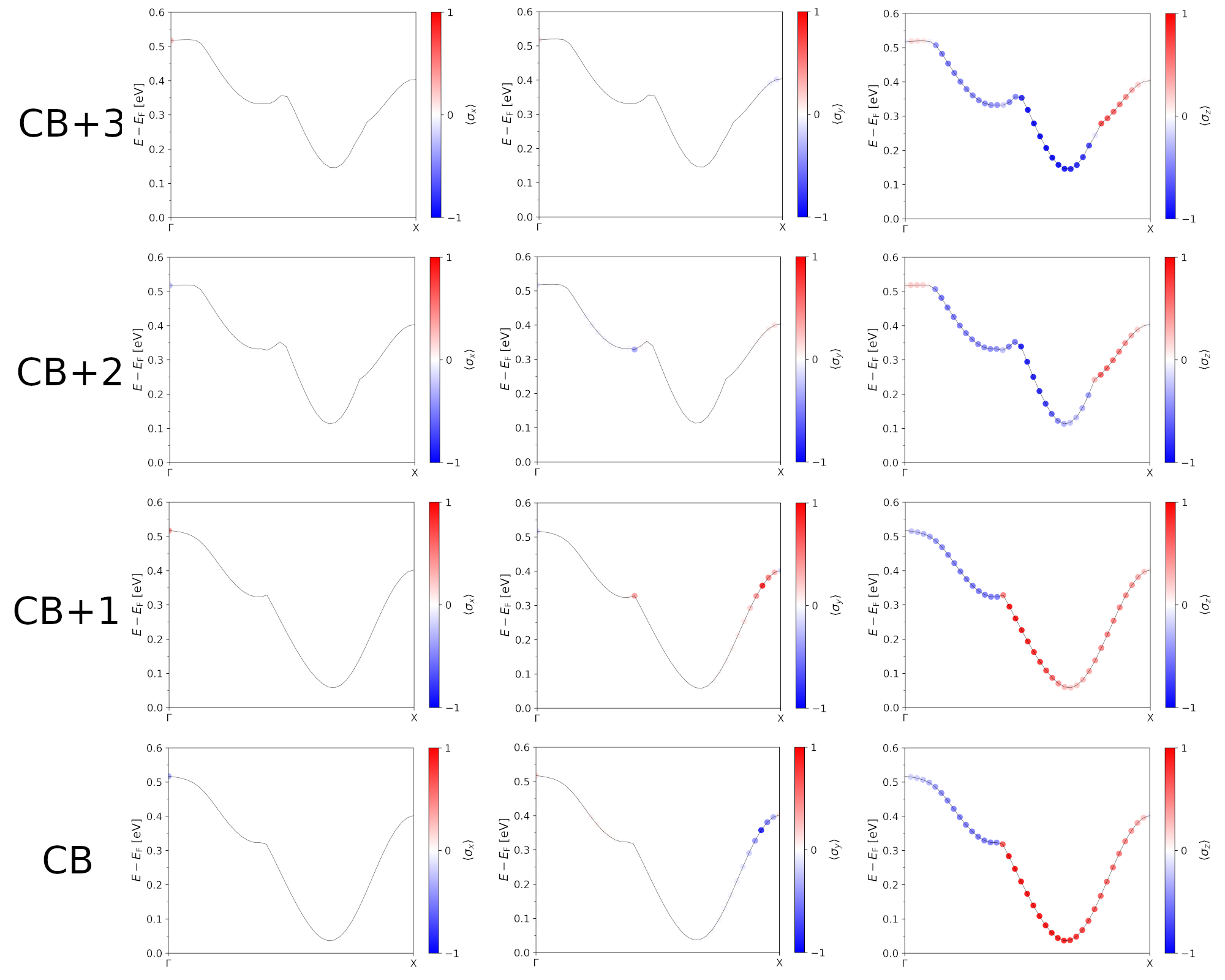}
\caption{Expectation values of the Pauli matrices $\langle\sigma_i\rangle$ for the four lowest conduction bands (CB) for the (11,11) wrinkle as calculated with the FHI-aims code.\label{fig_spinwrinkleCB}}
\end{figure}

\begin{figure}
\centering
\includegraphics[width=\linewidth]{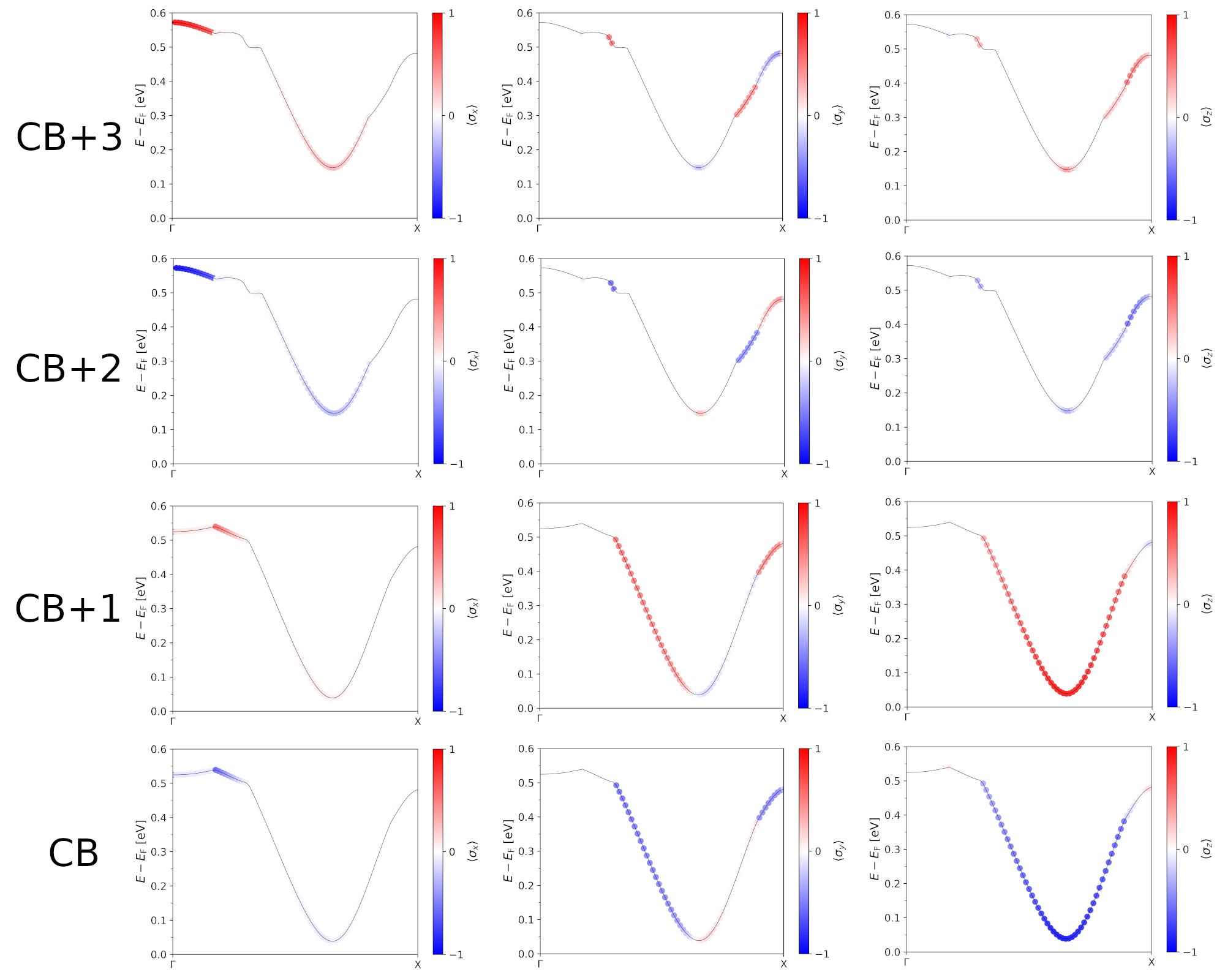}
\caption{Expectation values of the Pauli matrices $\langle\sigma_i\rangle$ for the four lowest conduction bands (CB) for the (11,11) NT as calculated with the FHI-aims code.\label{fig_spintubeCB}}
\end{figure}

Interestingly, there are however differences in the spin texture of the valence and conduction bands. Figures \ref{fig_spinwrinkleVB}--\ref{fig_spintubeCB} show the expectation values of the Pauli matrices for the uppermost 4 valence bands (VB) for an armchair (11,11) nanotube and the corresponding wrinkle of the same size and with wavelength equals four times the radius of the nanotube.

\begin{figure}
\centering
\includegraphics[width=\linewidth]{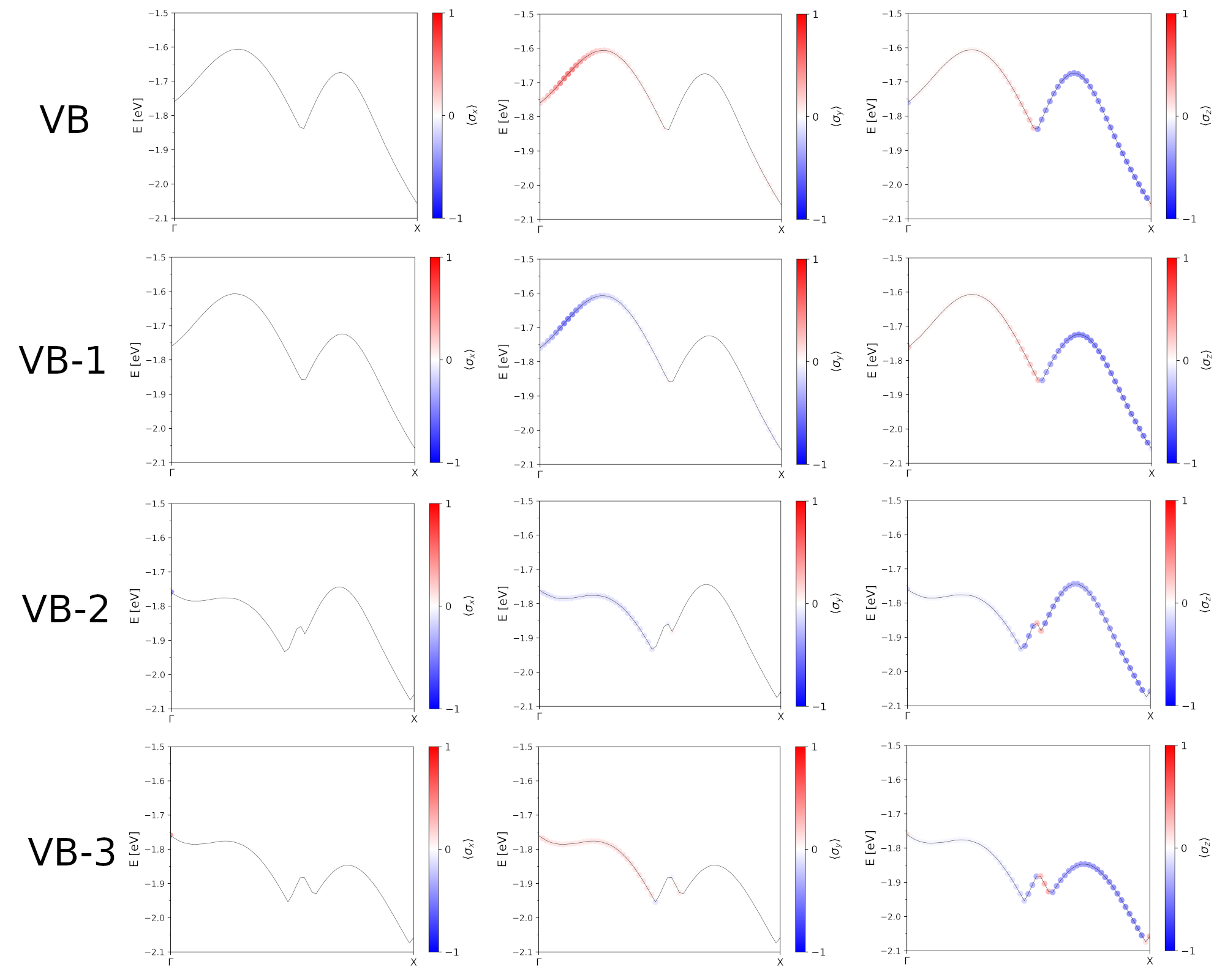}
\caption{Expectation values of the Pauli matrices $\langle\sigma_i\rangle$ for the four uppermost valence bands (VB) for the (11,11) wrinkle as calculated with {\sc Quantum ESPRESSO}.\label{fig_spinwrinkleVB_QE}}
\end{figure}

\begin{figure}
\centering
\includegraphics[width=\linewidth]{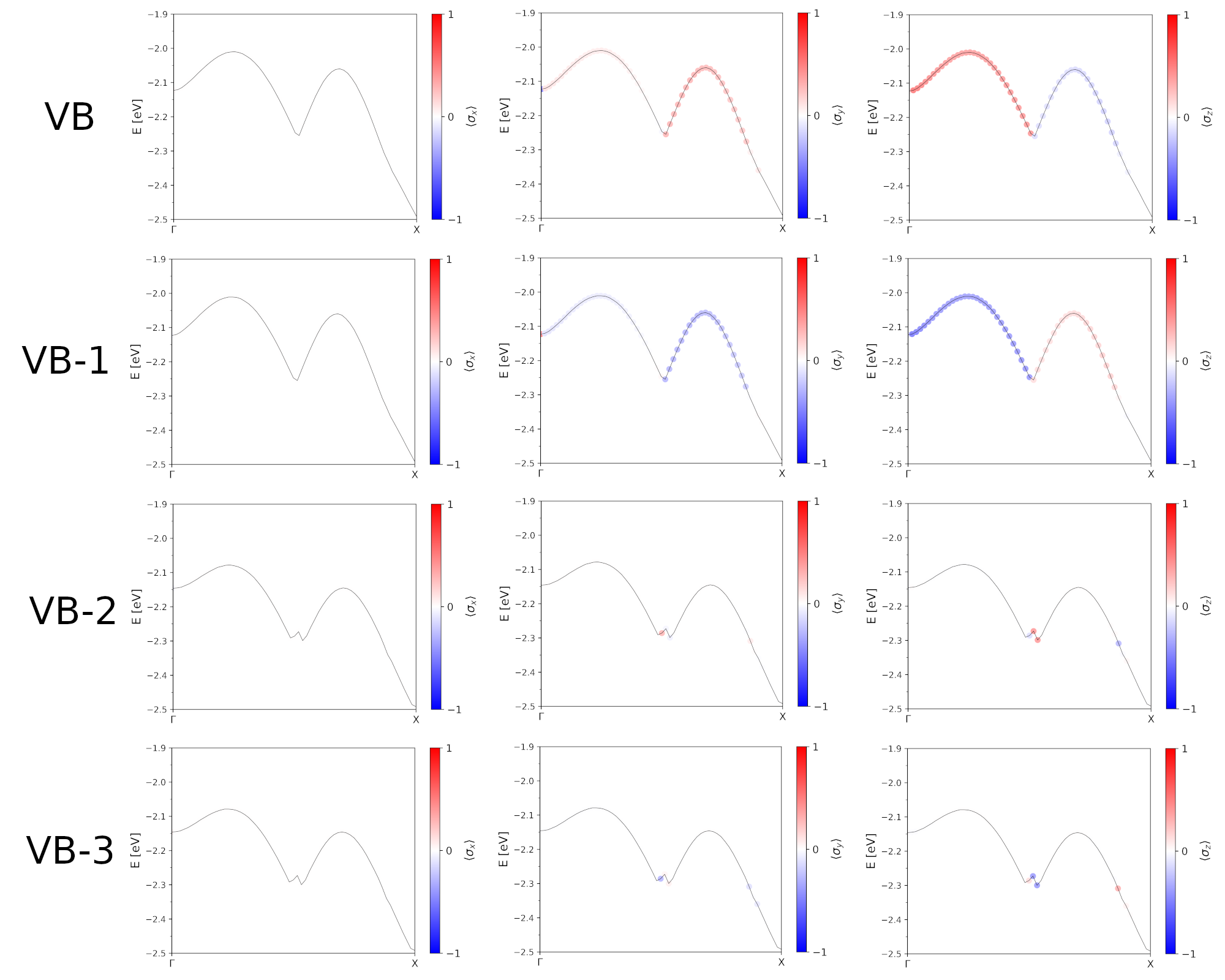}
\caption{Expectation values of the Pauli matrices $\langle\sigma_i\rangle$ for the four uppermost valence bands (VB) for the (11,11) NT as calculated with {\sc Quantum ESPRESSO}.\label{fig_spintubeVB_QE}}
\end{figure}

\begin{figure}
\centering
\includegraphics[width=\linewidth]{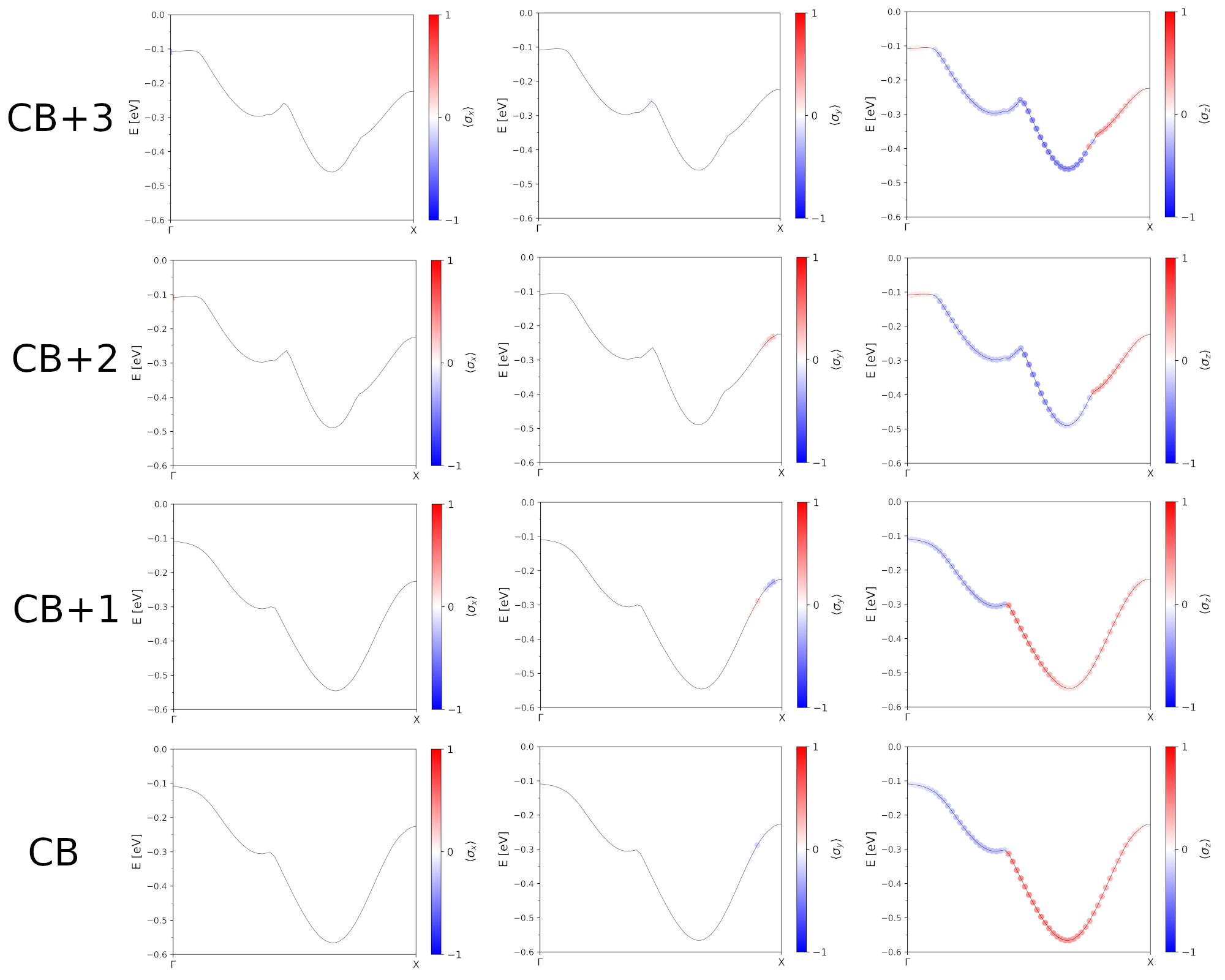}
\caption{Expectation values of the Pauli matrices $\langle\sigma_i\rangle$ for the four lowest conduction bands (CB) for the (11,11) wrinkle as calculated with {\sc Quantum ESPRESSO}.\label{fig_spinwrinkleCB_QE}}
\end{figure}

\begin{figure}
\centering
\includegraphics[width=\linewidth]{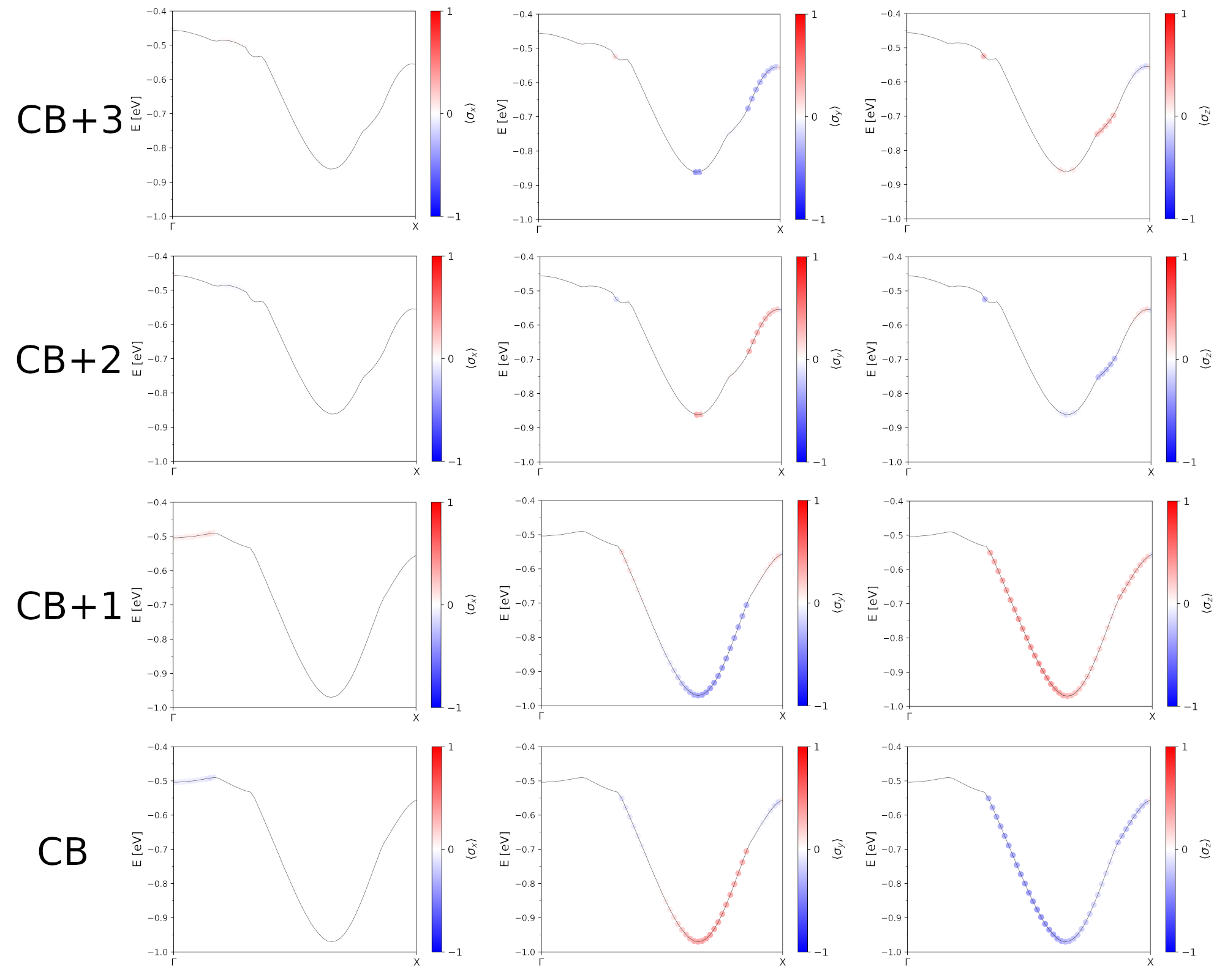}
\caption{Expectation values of the Pauli matrices $\langle\sigma_i\rangle$ for the four lowest conduction bands (CB) for the (11,11) NT as calculated with {\sc Quantum ESPRESSO}.\label{fig_spintubeCB_QE}}
\end{figure}

\begin{figure}
    \centering
    \includegraphics[width=0.95\columnwidth]{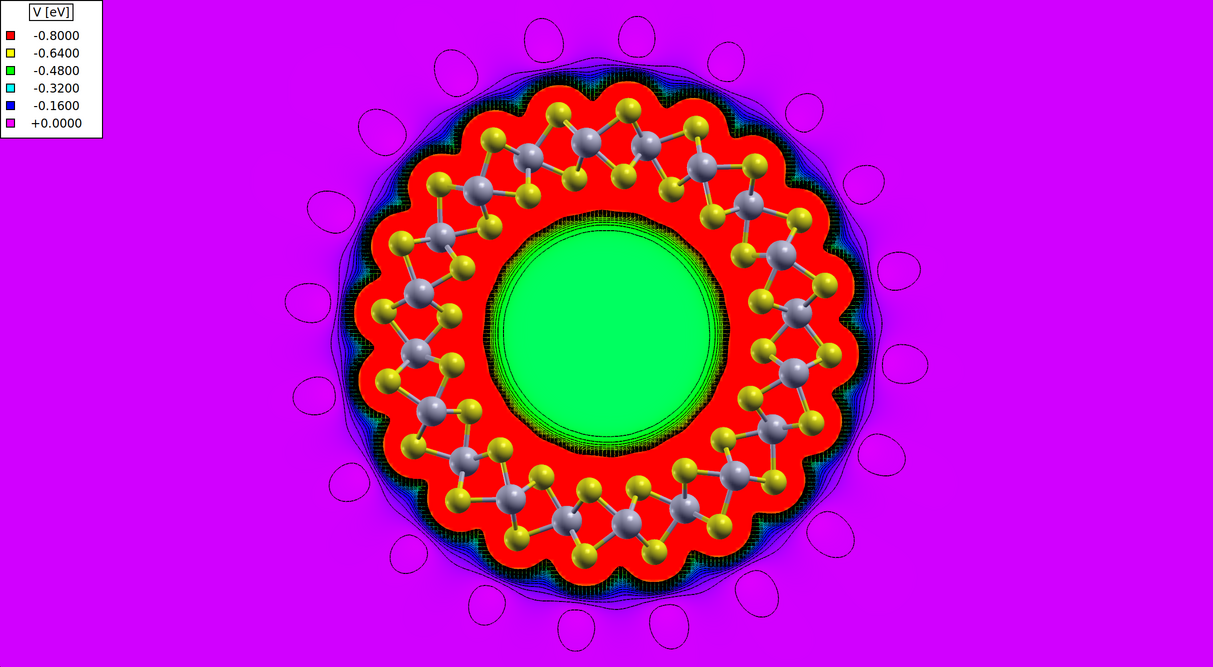}
    \includegraphics[width=0.95\columnwidth]{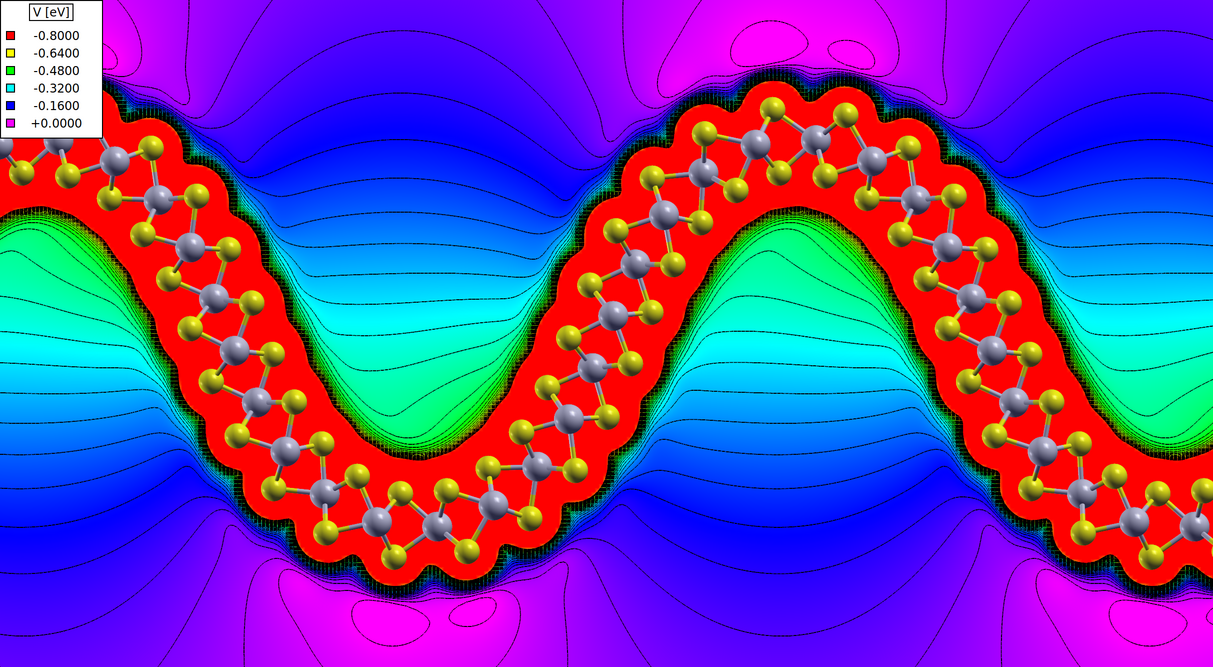}
    \caption{Contour plot of the total electrostatic potential for a (10,10) nanotube (upper panel) and wrinkle. The dipolar character of the corresponding electric field is clearly visible. The potential difference between the center of the NT and the outside is $\Delta V\approx-0.394\:\mathrm{eV}$.}\label{fig_tot_pot}
\end{figure}

The translational vector $\mathbf{T}$ in our NTs is parallel to the $\mathbf{\hat{x}}$ direction and accordingly the electric field of the induced dipoles is pointing in the $\mathbf{\hat{y}}$/$\mathbf{\hat{z}}$ (more specifically it is parallel to $\mathbf{\hat{\rho}}$, $E(\rho)\propto 1/\rho$, see also Figure~\ref{fig_tot_pot}). This leads to the following form of the spin-orbit Hamiltonian:
\begin{align}
  H_\mathrm{SOC} &\propto -\alpha \mathbf{\sigma} \cdot \left(\mathbf{E} \times \mathbf{k}\right)\\
  &= -\alpha \begin{pmatrix}\sigma_x\\\sigma_y\\\sigma_z\end{pmatrix} \cdot
      \left( \begin{pmatrix}0\\E_y\\E_z\end{pmatrix} \times 
             \begin{pmatrix}p_x\\0\\0\end{pmatrix} \right),
\end{align}
where $E_y=E(\rho) \cdot \cos \phi$ and $E_z=E(\rho) \cdot \sin \phi$ with $\phi$ being the azimuth in cylindrical coordinates. One can easily see that the resulting spin texture has only contributions in $\sigma_y$ and $\sigma_z$ and this can also be seen in Figure~\ref{fig_spinwrinkleVB} for the (11,11) wrinkle, however Figure \ref{fig_spintubeVB} also shows small contributions for $\langle\sigma_x\rangle$. Repeating the calculations with {\sc Quantum ESPRESSO}\cite{qe1,qe2} (using the structures as obtained with the FHI-aims code, Figures \ref{fig_spinwrinkleVB_QE}--\ref{fig_spintubeCB_QE}) the band structures and the spin textures look the same, except that the x-component of the spin is zero, $\langle\sigma_x\rangle=0$ (as expected) and the $\langle\sigma_i\rangle$ do not change the sign when going from one $\mathbf{k}$ point to the next. We thus attribute this small contribution seen in the FHI-aims calculations to a bug in the plotting routine since both states are degenerate.

There are 2 major differences between nanotubes and wrinkles. While the former always show 2-fold degenerate bands coming from the K and K' point of the 2D material, the degeneracy is slightly lifted in the wrinkle due to the different strain states along one period of the wrinkle -- in fact, a slight asymmetry is also visible in the contour plot of the total electrostatic potential shown in Figure \ref{fig_tot_pot}. This also leads to a different spin texture in which the VB does not automatically have the opposite spin expectation value of the VB-1.

\newpage
\begin{figure}
    \begin{subfigure}[b]{0.75\columnwidth}
        \includegraphics[width=0.75\columnwidth]{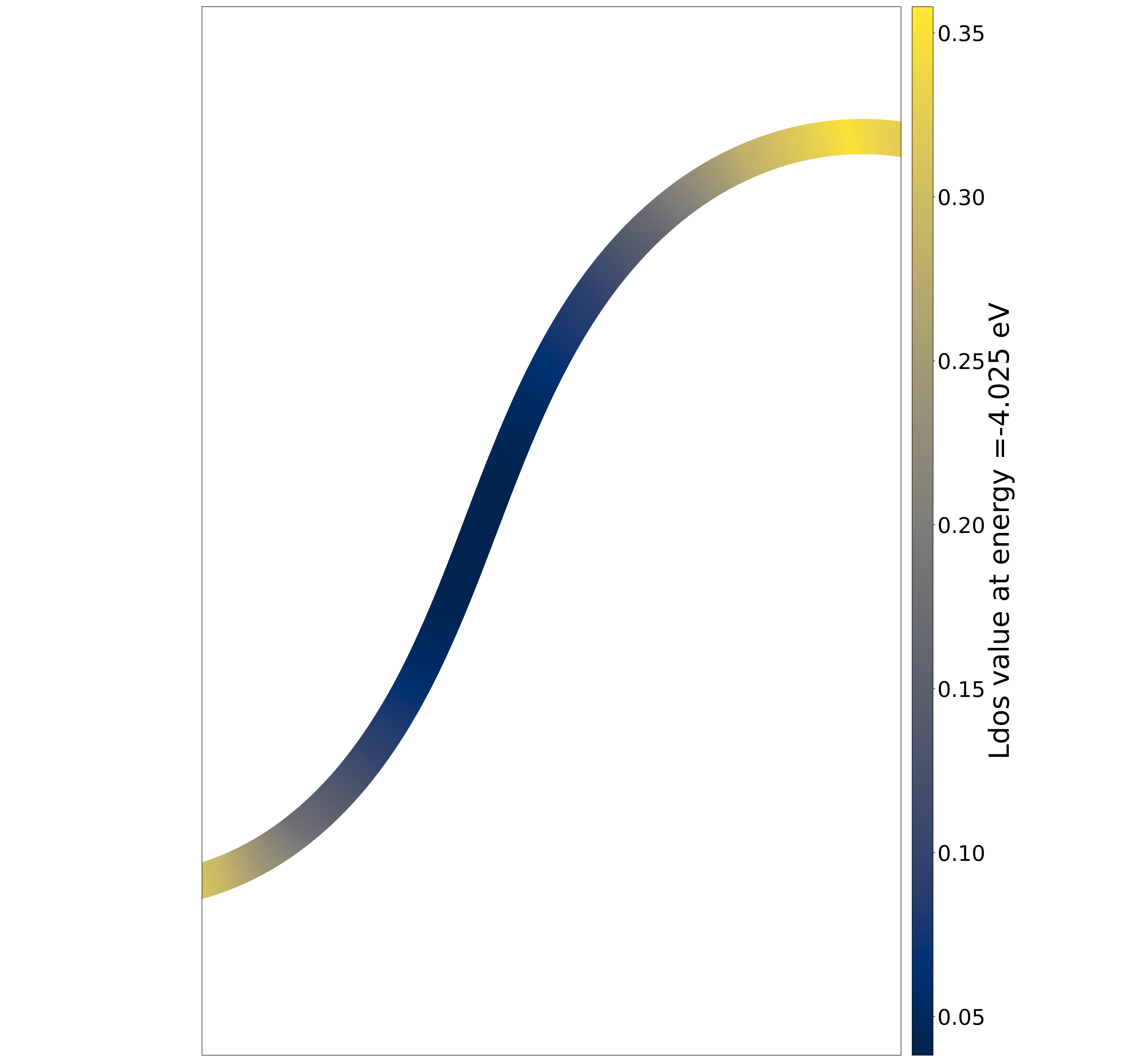}
        \caption{}
        \label{figure_ldosContributionCBM_n11}
    \end{subfigure}
    \begin{subfigure}[b]{0.75\columnwidth}
        \includegraphics[width=0.75\columnwidth]{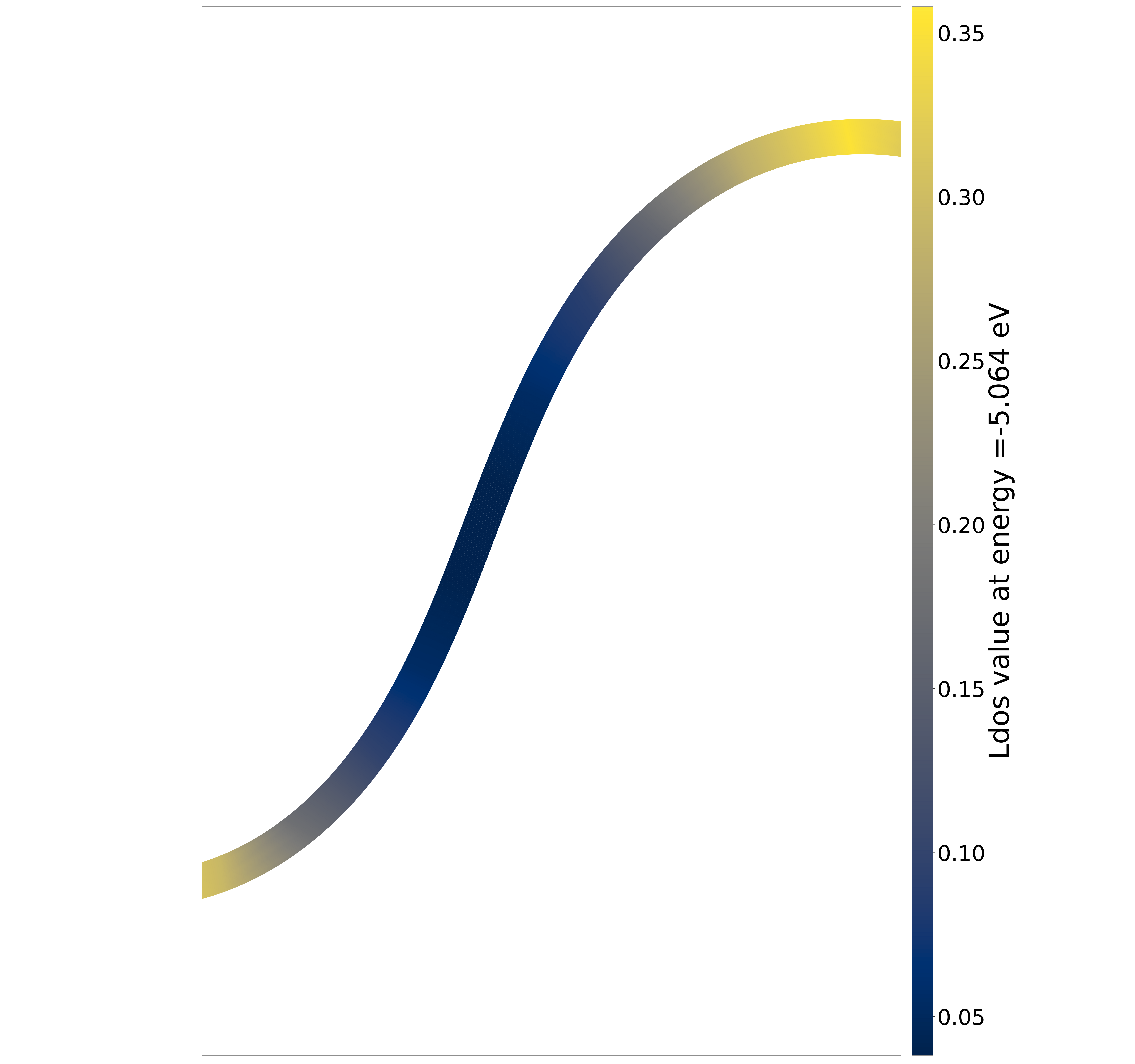}
        \caption{}
        \label{figure_ldosContributionVBM_n11}
    \end{subfigure}
    \caption{Variation of the contribution to the a) valence-band maximum and b) conduction-band minimum along the  (11,11) wrinkle.} \label{fig_ldosContribution}
\end{figure}

\begin{figure}
    \centering
    \includegraphics[width=.8\linewidth]{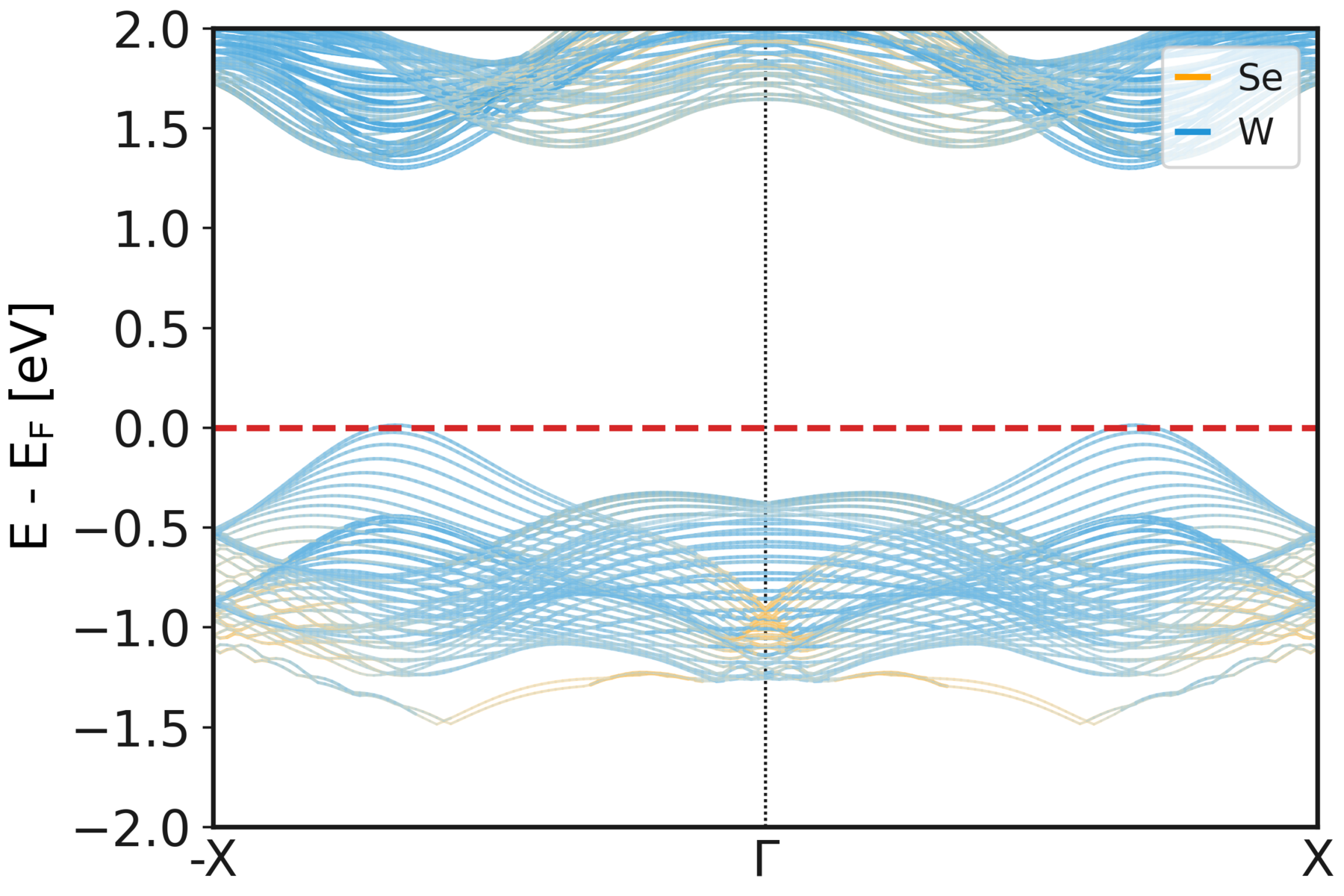}
    \includegraphics[width=.8\linewidth]{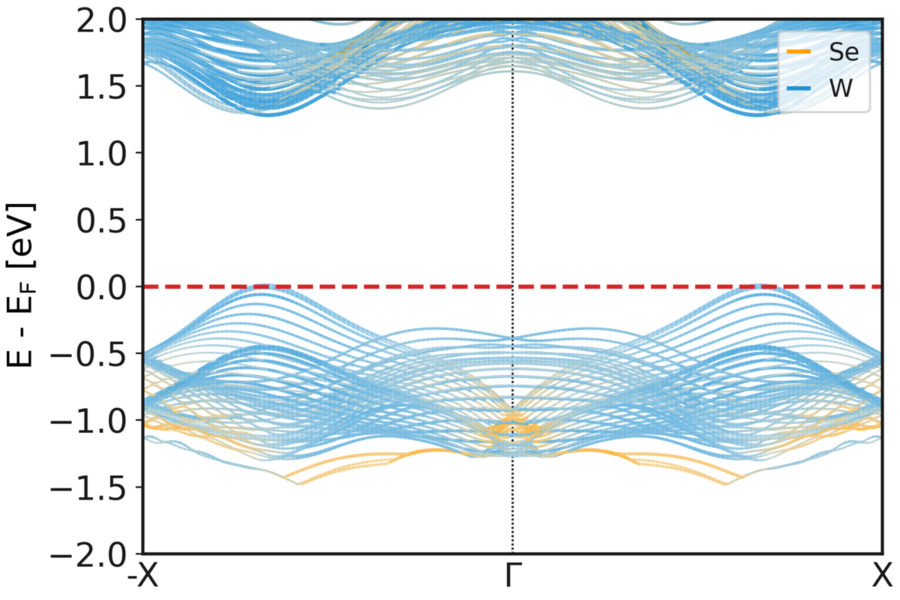}
    \caption{The comparison of the band dispersion of the (24,24) nanotube (upper panel) and wrinkle. The splitting in momentum direction of the VBs is visible in both cases.} \label{fig_nano_wrinkle2424_SI}
\end{figure}

\begin{figure}
\centering
\includegraphics[width=.8\linewidth]{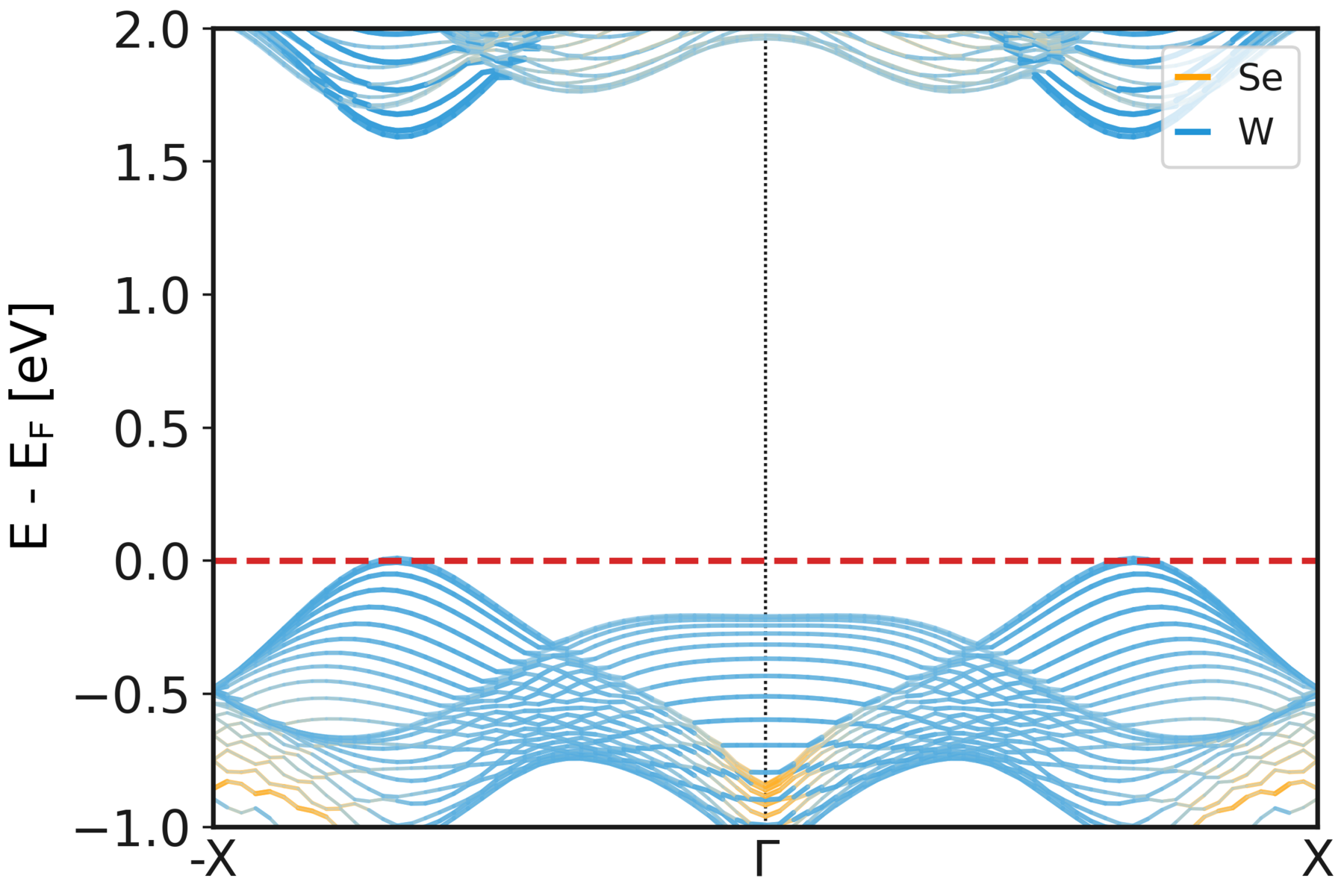}
\includegraphics[width=.8\linewidth]{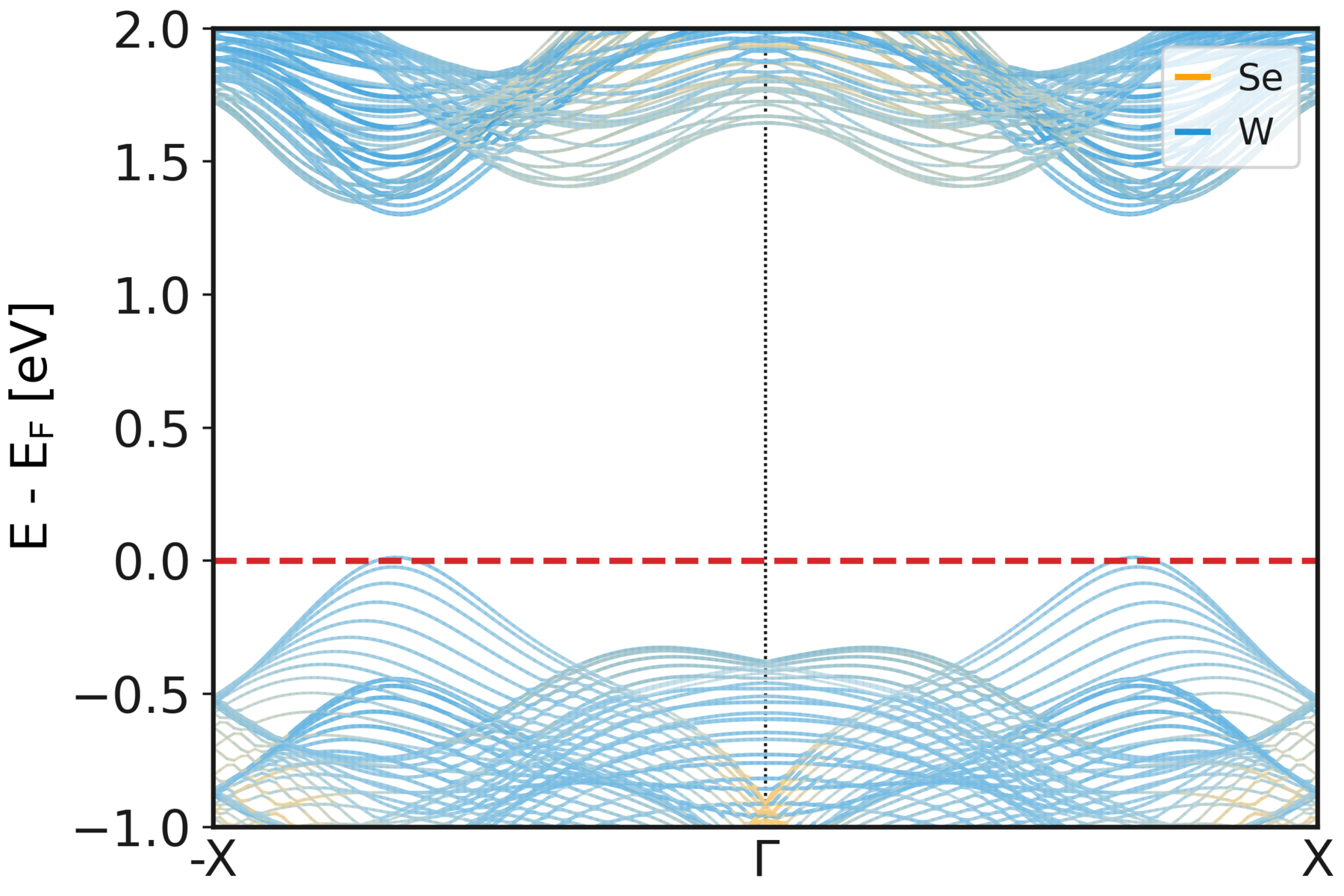}
\caption{The effect of spin-orbit coupling on the band dispersion for a  (24,24) armchair nanotube of \ce{WSe2} (upper panel) without spin-orbit coupling (lower panel) with spin-orbit coupling.}
\label{figure_results_armchair_n24}
\end{figure}

\begin{figure}
    \includegraphics[width=0.8\textwidth]{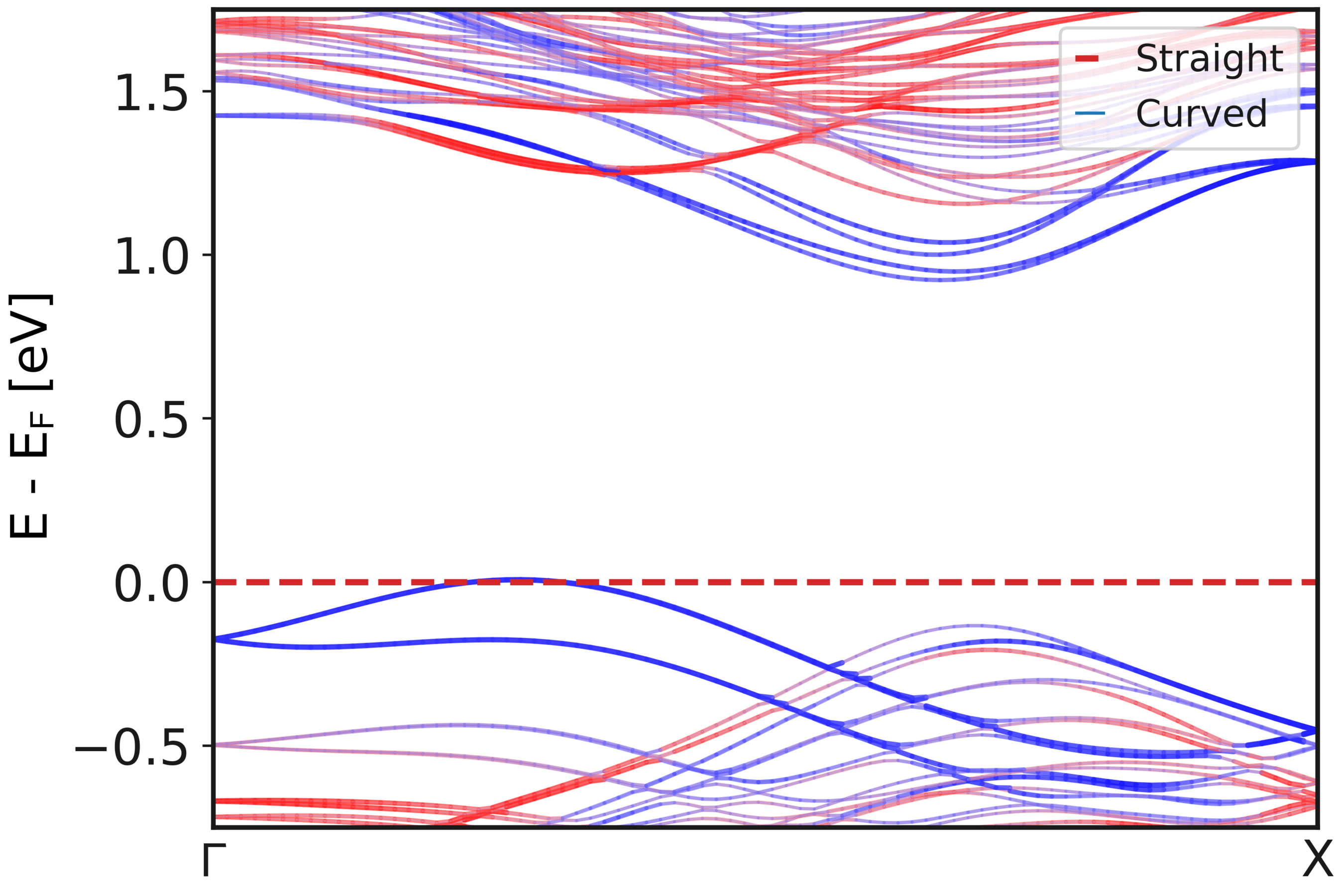}
    \caption{Contribution of curved and straight sections of the (10,10) wrinkle to its band structure.}
    \label{figure_regionContribution10}
\end{figure}
\begin{figure}
    \includegraphics[width=0.8\textwidth]{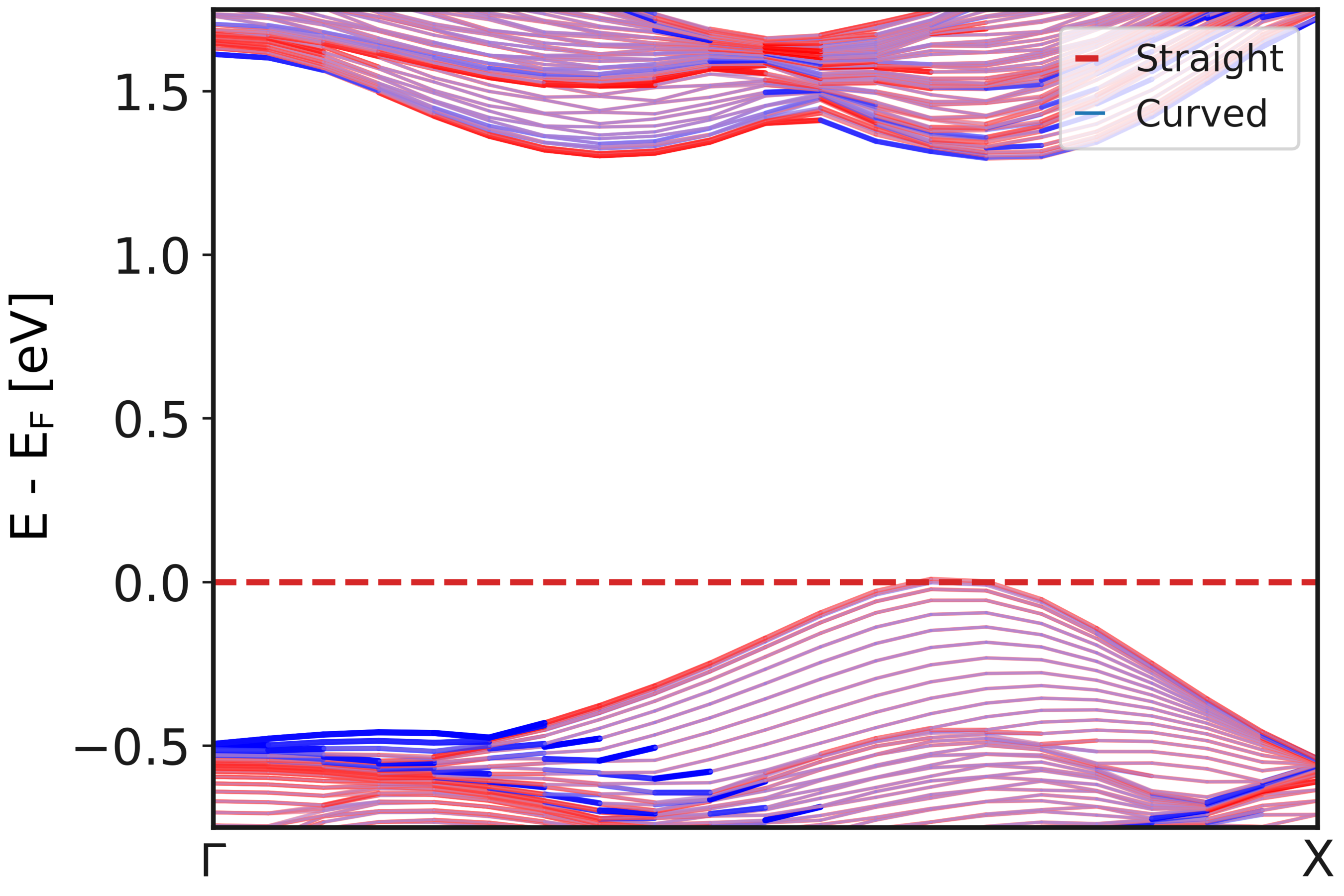}
     \caption{Contribution of curved and straight sections of the (41,41) wrinkle to its band structure.}
    \label{figure_regionContribution41}
\end{figure}

\newpage

\begin{figure}
    \centering
        \includegraphics[width=.5\textwidth]{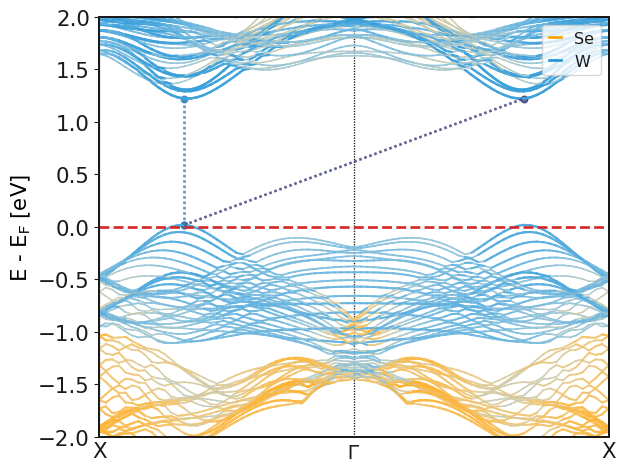} \\
        \includegraphics[width=.5\textwidth]{images/nano_armchair_n24_sp.png} \\
        \includegraphics[width=.5\textwidth]{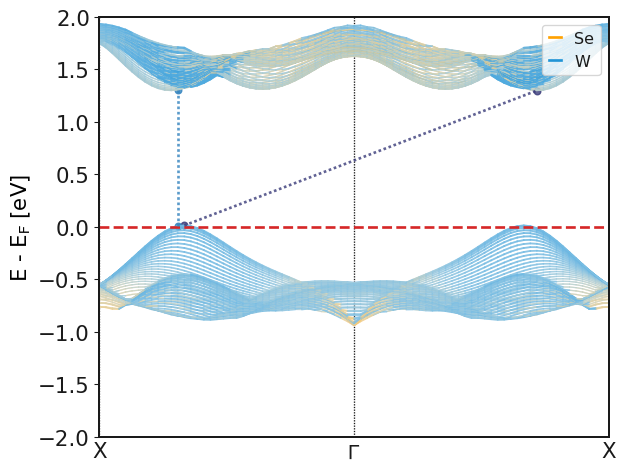}
    \caption{Variation of the band dispersion for armchair nanotubes with increasing diameter (from top to bottom: (15,15), (24,24), (56,56)) for 50  valence and 50 conduction bands.}\label{fig_nano_bg_comparison}
\end{figure}

\begin{figure}
    \centering
        \includegraphics[width=.5\textwidth]{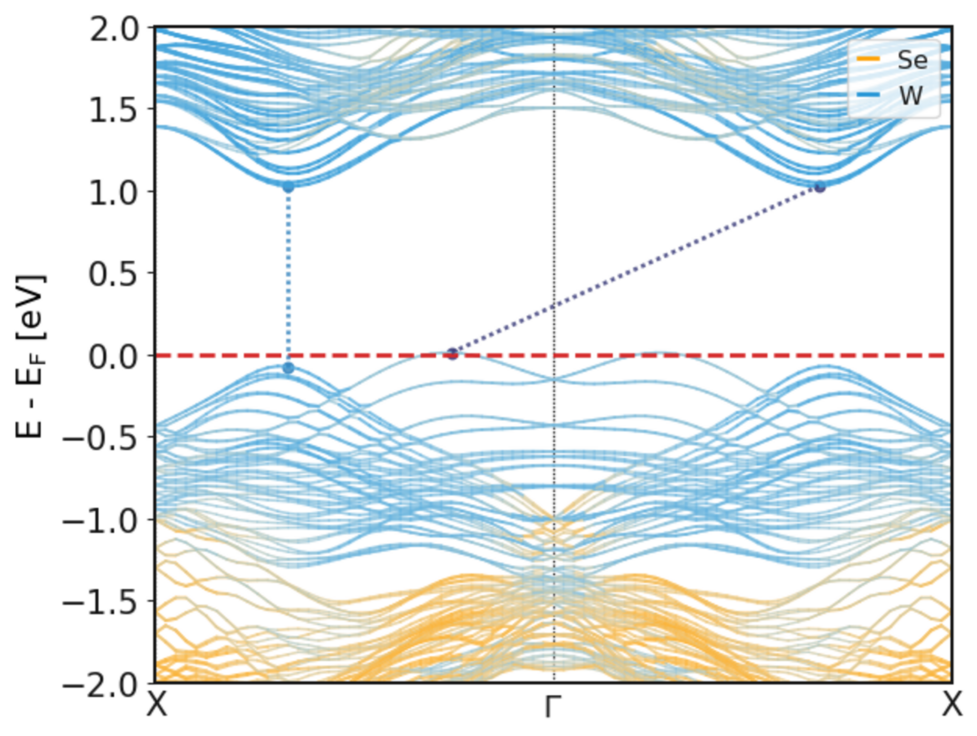}
        \includegraphics[width=.5\textwidth]{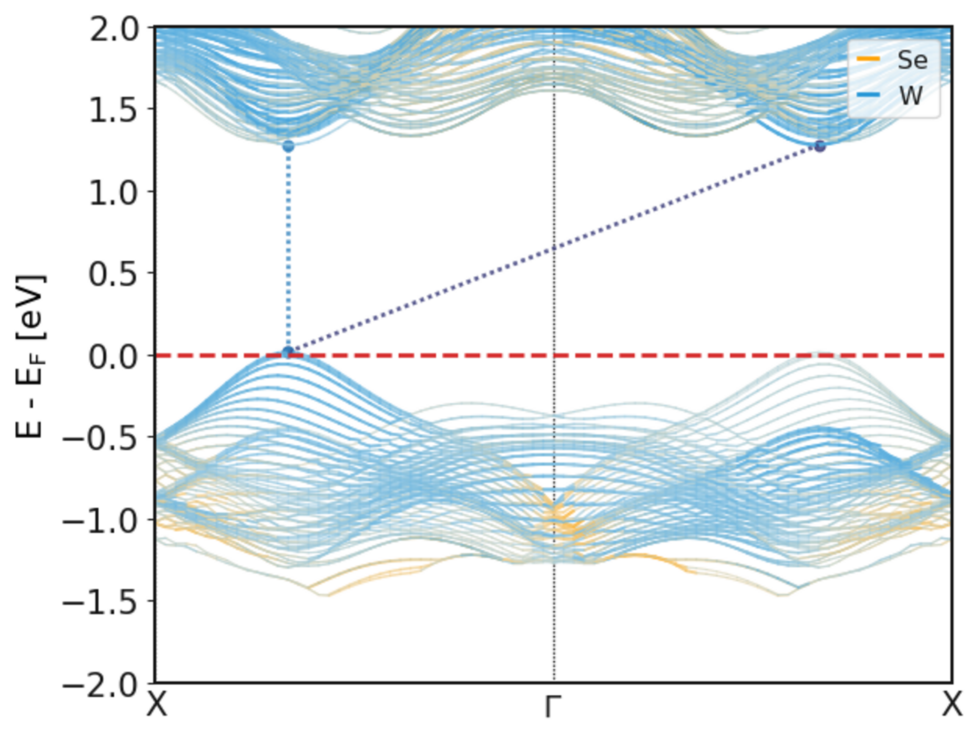}
        \includegraphics[width=.5\textwidth]{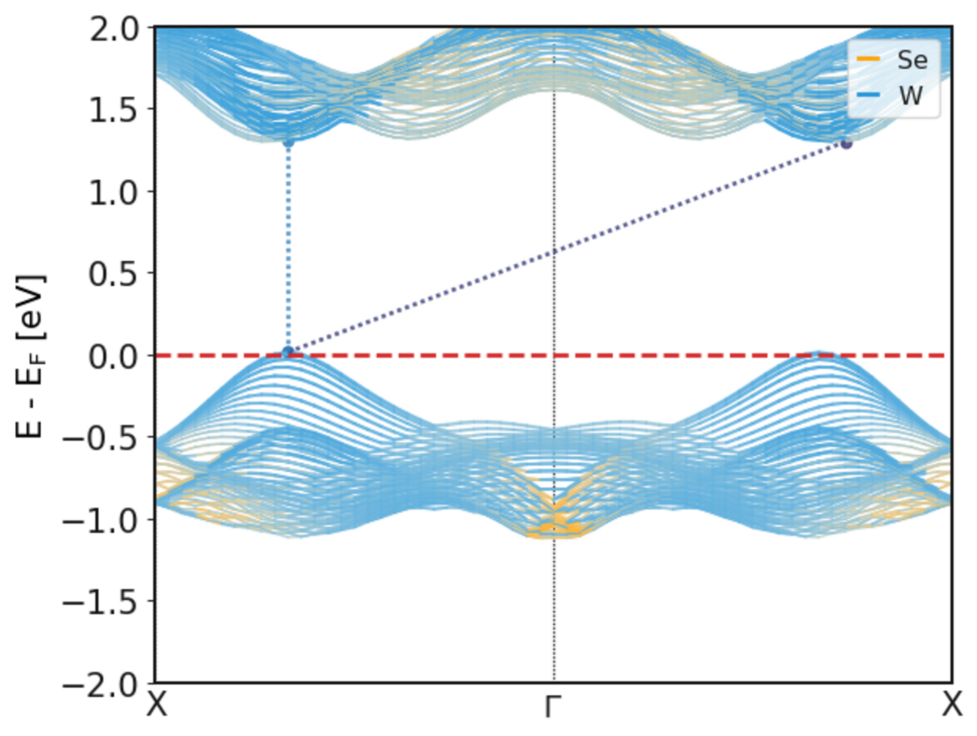}
    \caption{Variation of the band dispersion for armchair wrinkles with increasing wave length (from top to bottom: (11,11), (24,24), (33,33)) for 50 valence and 50 conduction bands.} \label{fig_wrinkle_bg_comparison}
\end{figure}

\newpage

\begin{figure}
    \centering
    \begin{subfigure}[b]{0.49\linewidth}
         \includegraphics[width=1\linewidth]{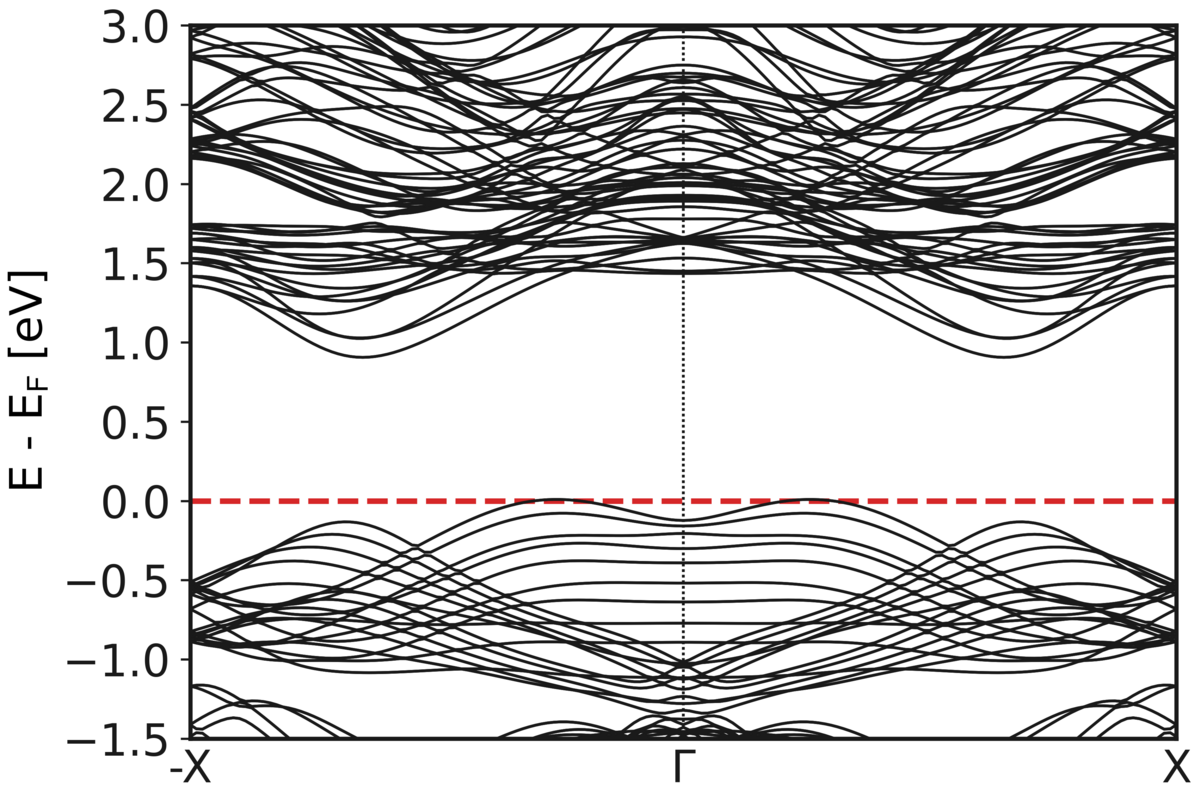}
         \caption*{(10,10)}
    \end{subfigure}
    \begin{subfigure}[b]{0.49\linewidth}
         \includegraphics[width=1\linewidth]{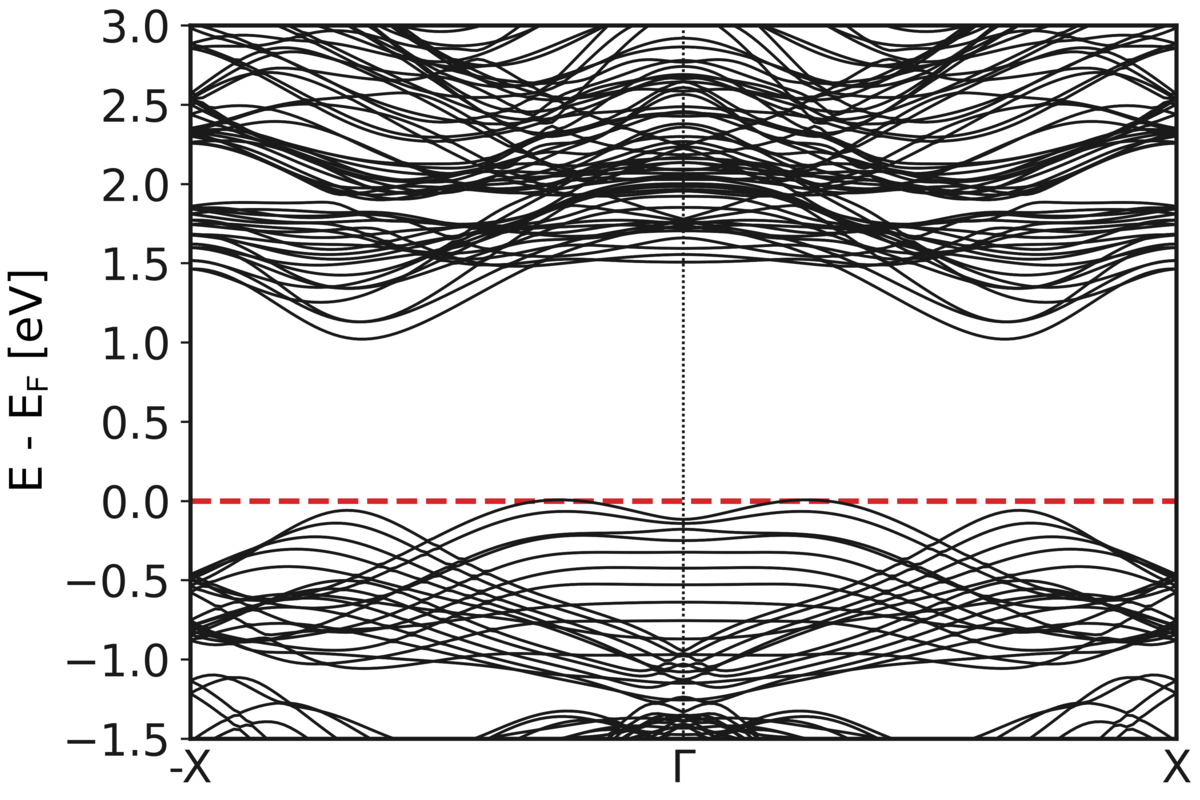}
         \caption*{(11,11)}
    \end{subfigure}
    \begin{subfigure}[b]{0.49\linewidth}
         \includegraphics[width=1\linewidth]{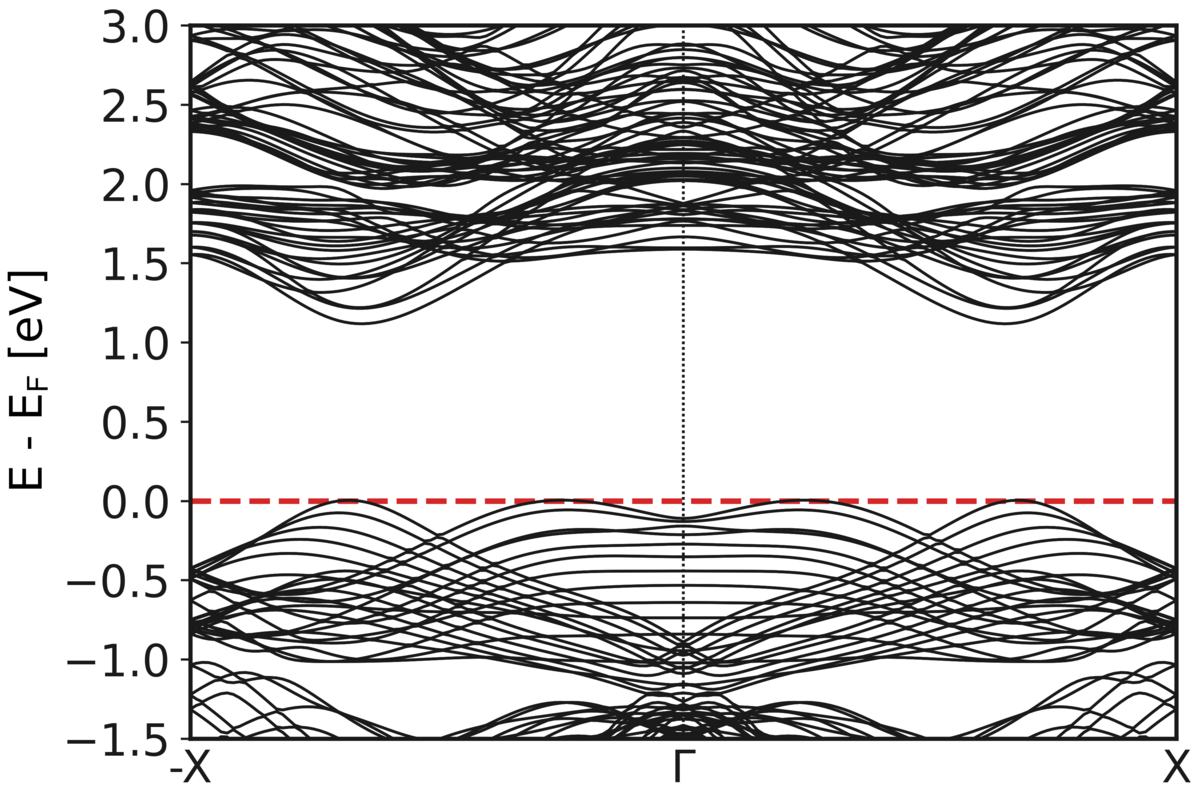}
         \caption*{(12,12)}
    \end{subfigure}
        \begin{subfigure}[b]{0.49\linewidth}
         \includegraphics[width=1\linewidth]{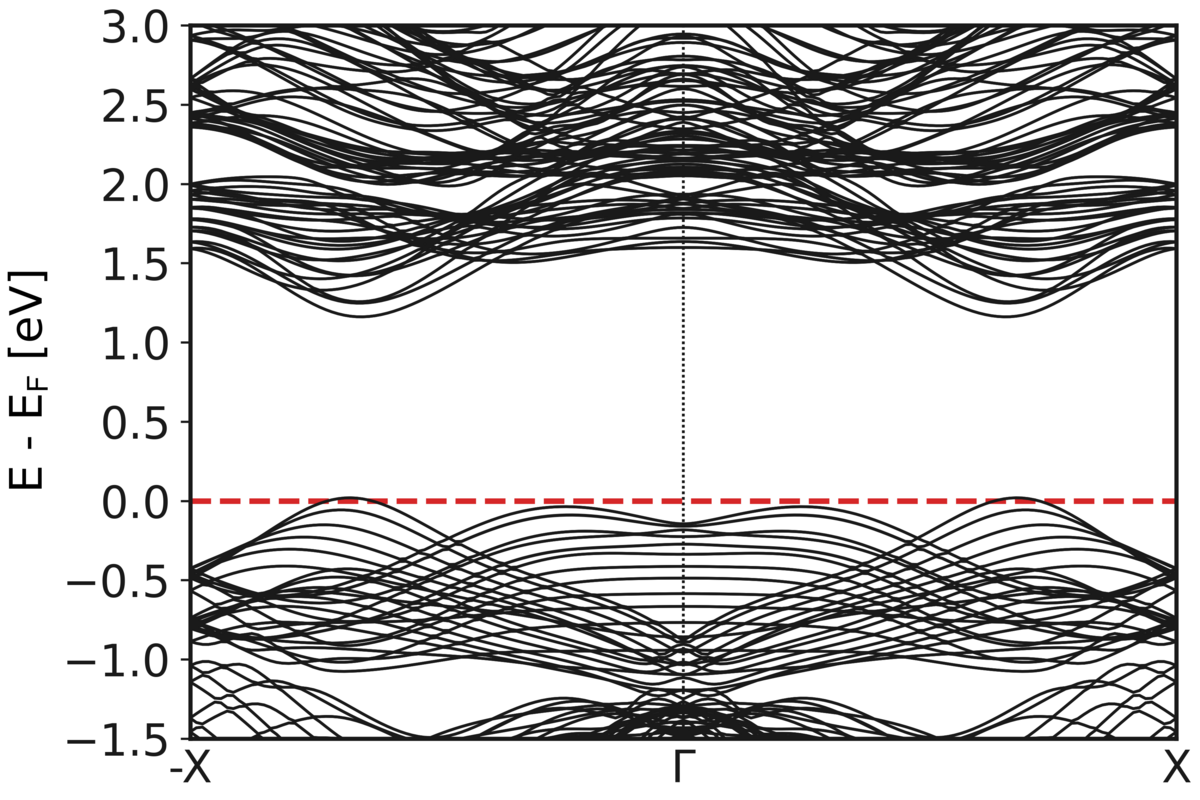}
         \caption*{(13,13)}
    \end{subfigure}
           \begin{subfigure}[b]{0.49\linewidth}
         \includegraphics[width=1\linewidth]{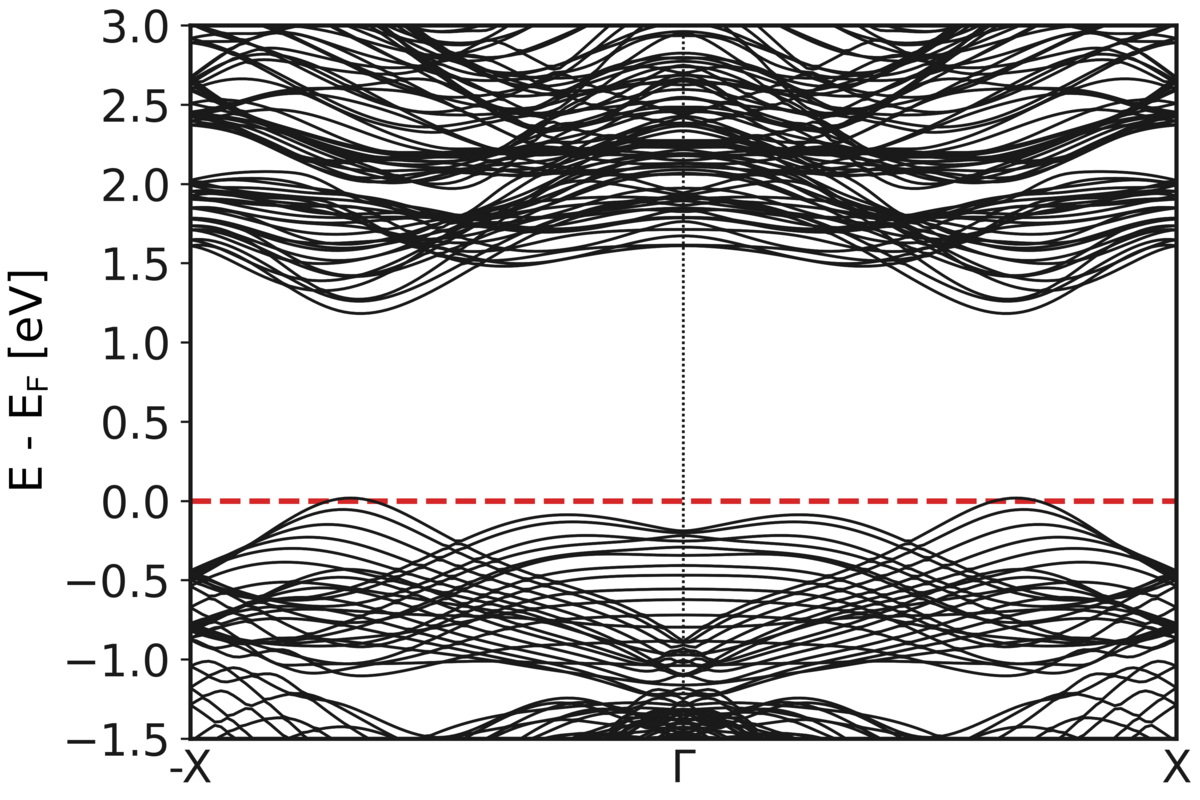}
         \caption*{(14,14)}
    \end{subfigure}
    \begin{subfigure}[b]{0.49\linewidth}
        \includegraphics[width=1\linewidth]{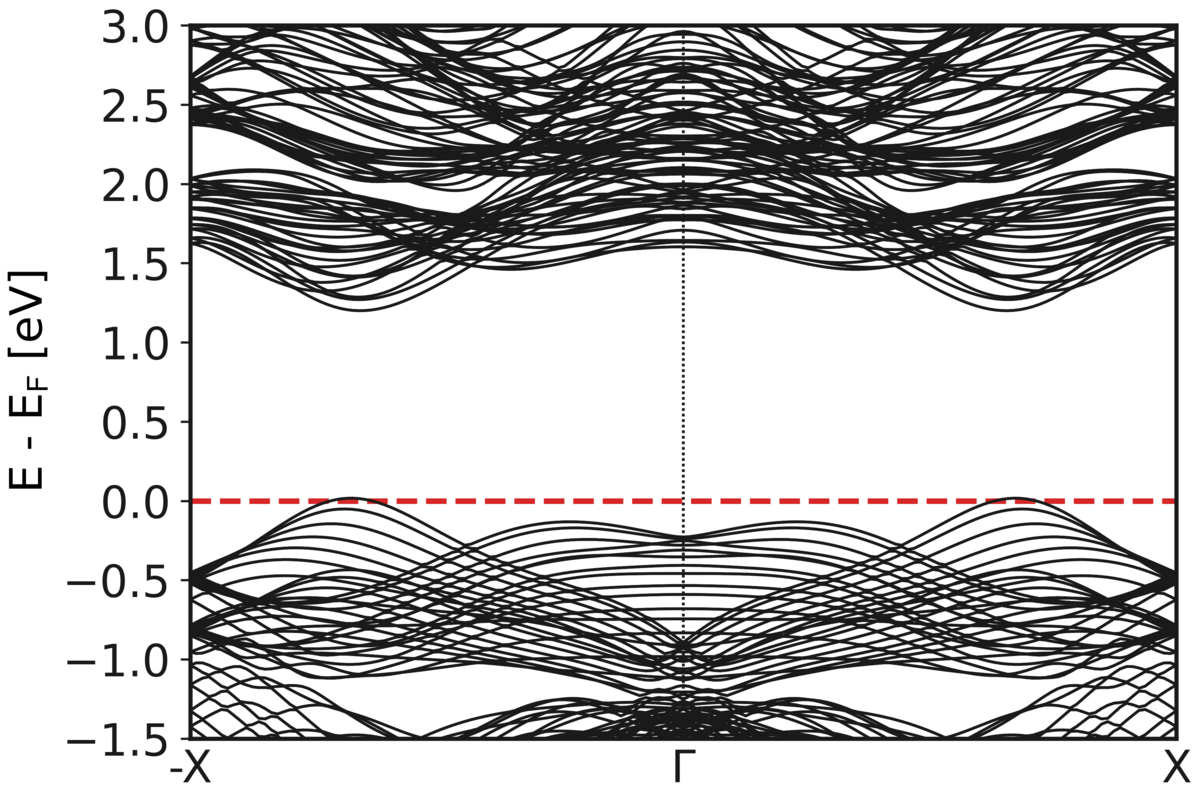}
        \caption*{(15,15)}
    \end{subfigure}

\end{figure}
\begin{figure}
\ContinuedFloat.

    \begin{subfigure}[b]{0.49\linewidth}
        \includegraphics[width=1\linewidth]{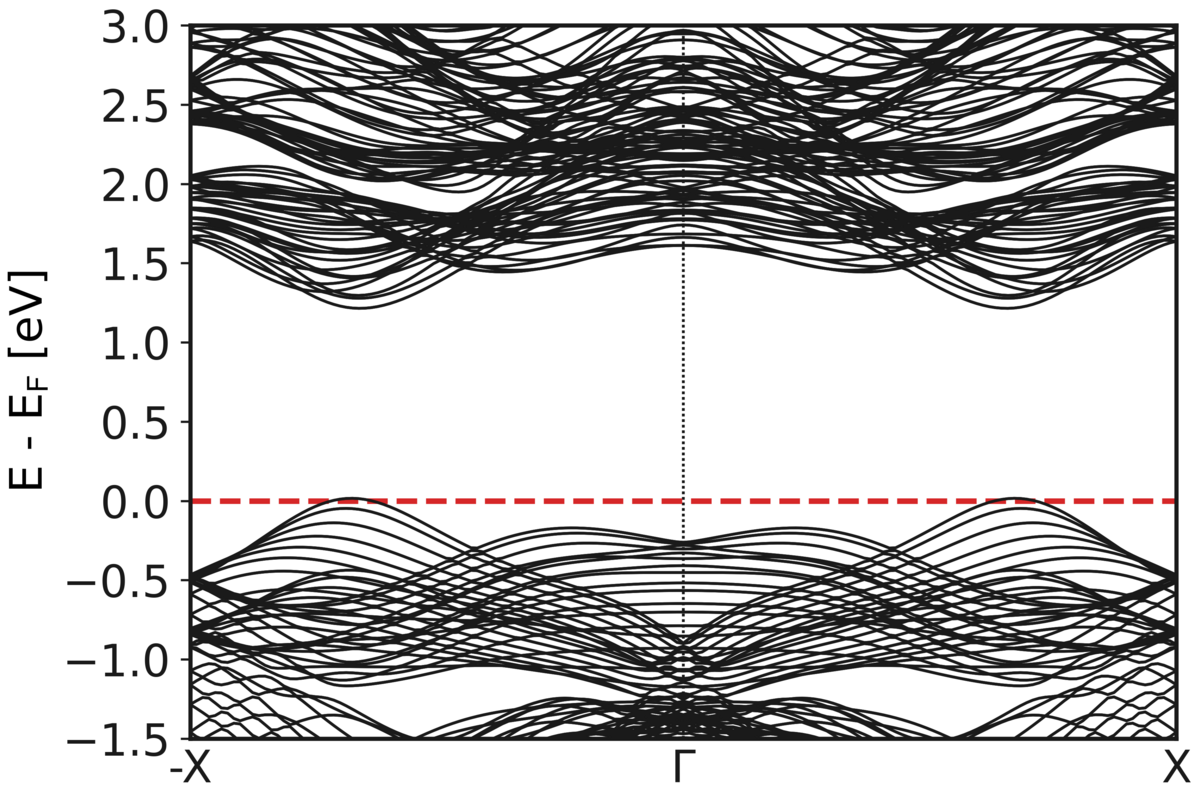}
        \caption*{(16,16)}
    \end{subfigure}
    \begin{subfigure}[b]{0.49\linewidth}
        \includegraphics[width=1\linewidth]{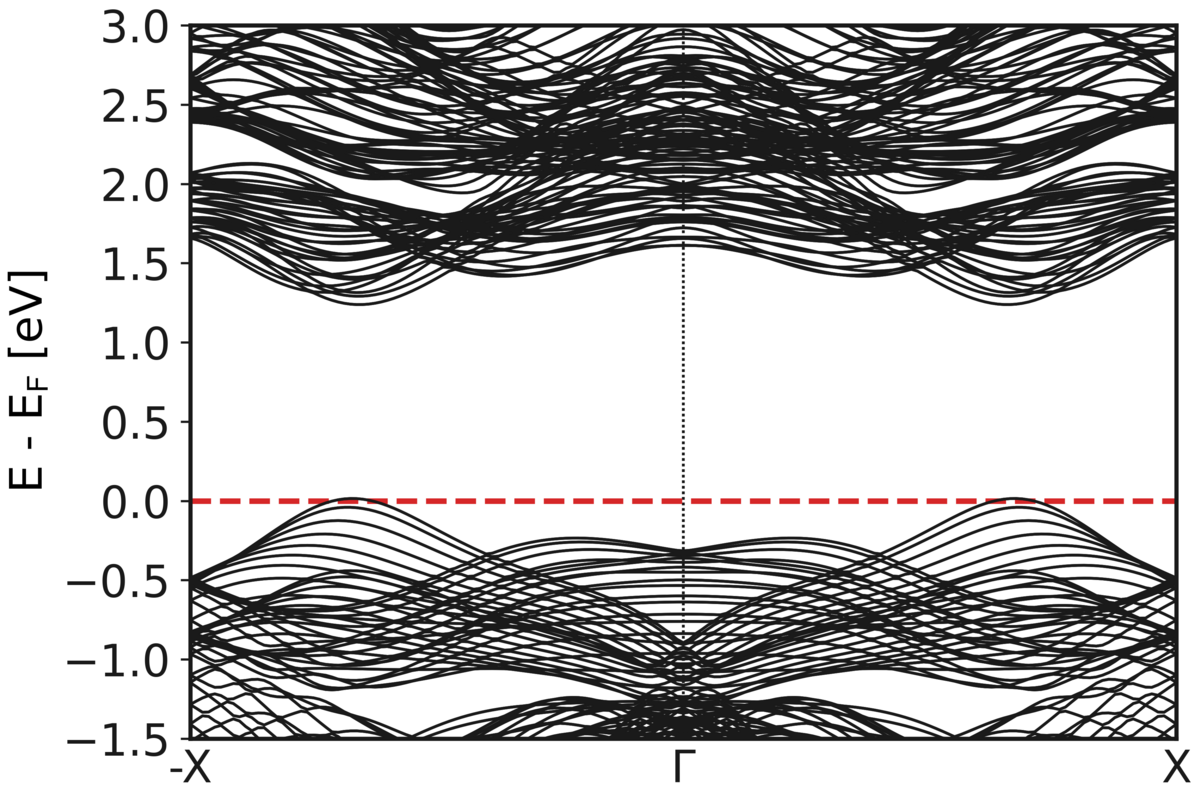}
        \caption*{(18,18)}
    \end{subfigure}
        \begin{subfigure}[b]{0.49\linewidth}
        \includegraphics[width=1\linewidth]{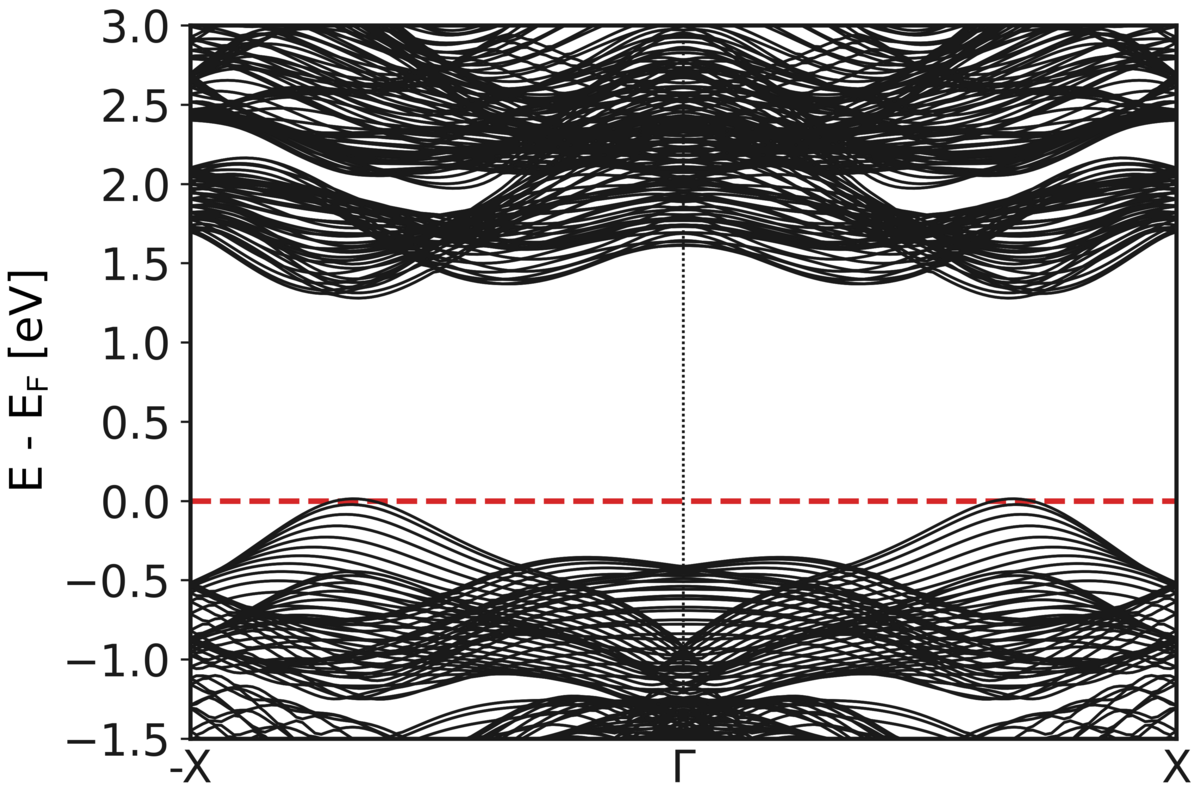}
        \caption*{(24,24)}
    \end{subfigure}
        \begin{subfigure}[b]{0.49\linewidth}
            \includegraphics[width=1\linewidth]{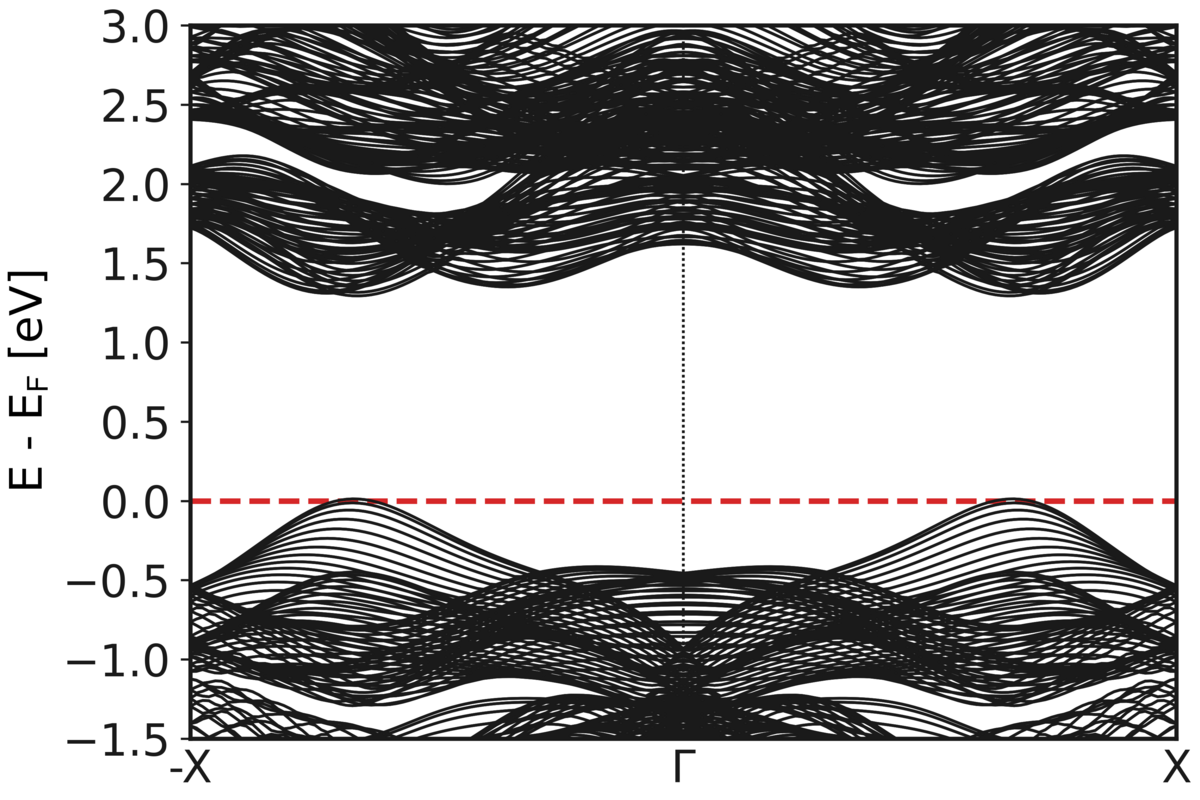}
            \caption*{(30,30)}
    \end{subfigure}
    \begin{subfigure}[b]{0.49\linewidth}
            \includegraphics[width=1\linewidth]{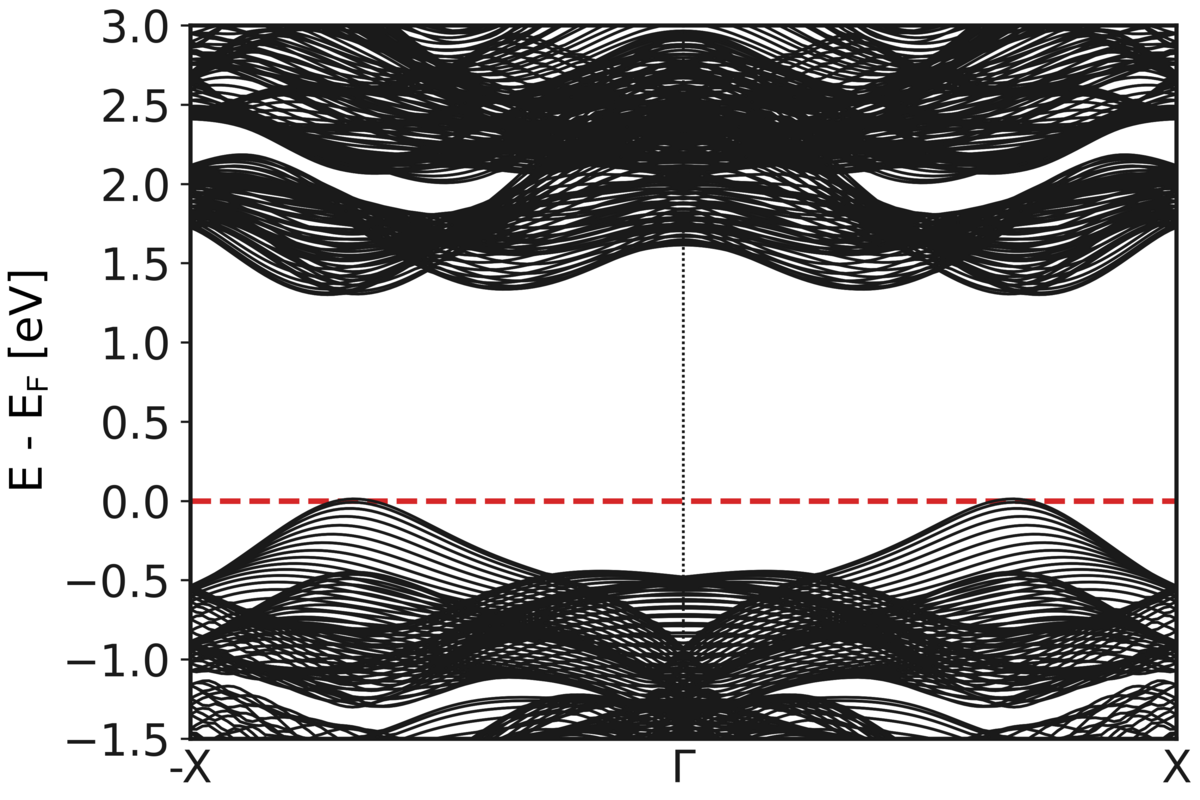}
            \caption*{(33,33)}
    \end{subfigure}
    \begin{subfigure}[b]{0.49\linewidth}
            \includegraphics[width=1\linewidth]{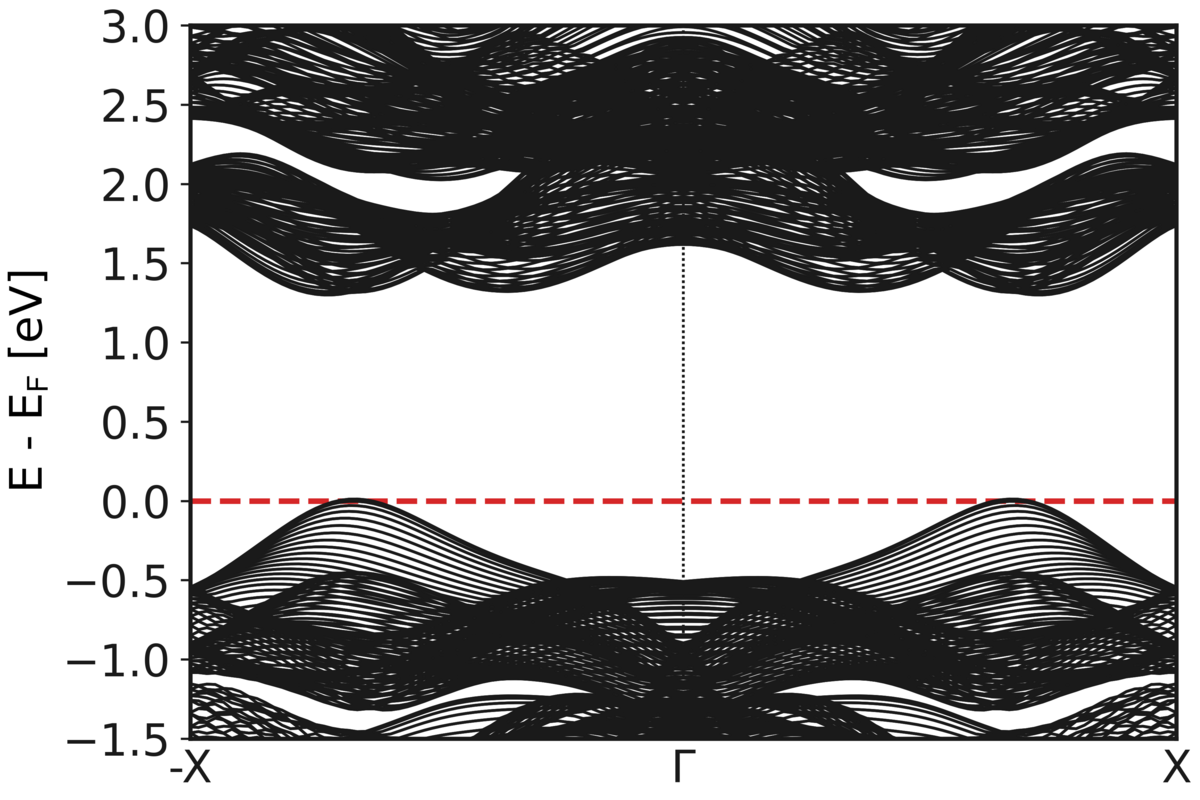}
            \caption*{(41,41)}
    \end{subfigure}
\end{figure}

\begin{figure}
\ContinuedFloat.
    \begin{subfigure}[b]{0.49\linewidth}
            \includegraphics[width=1\linewidth]{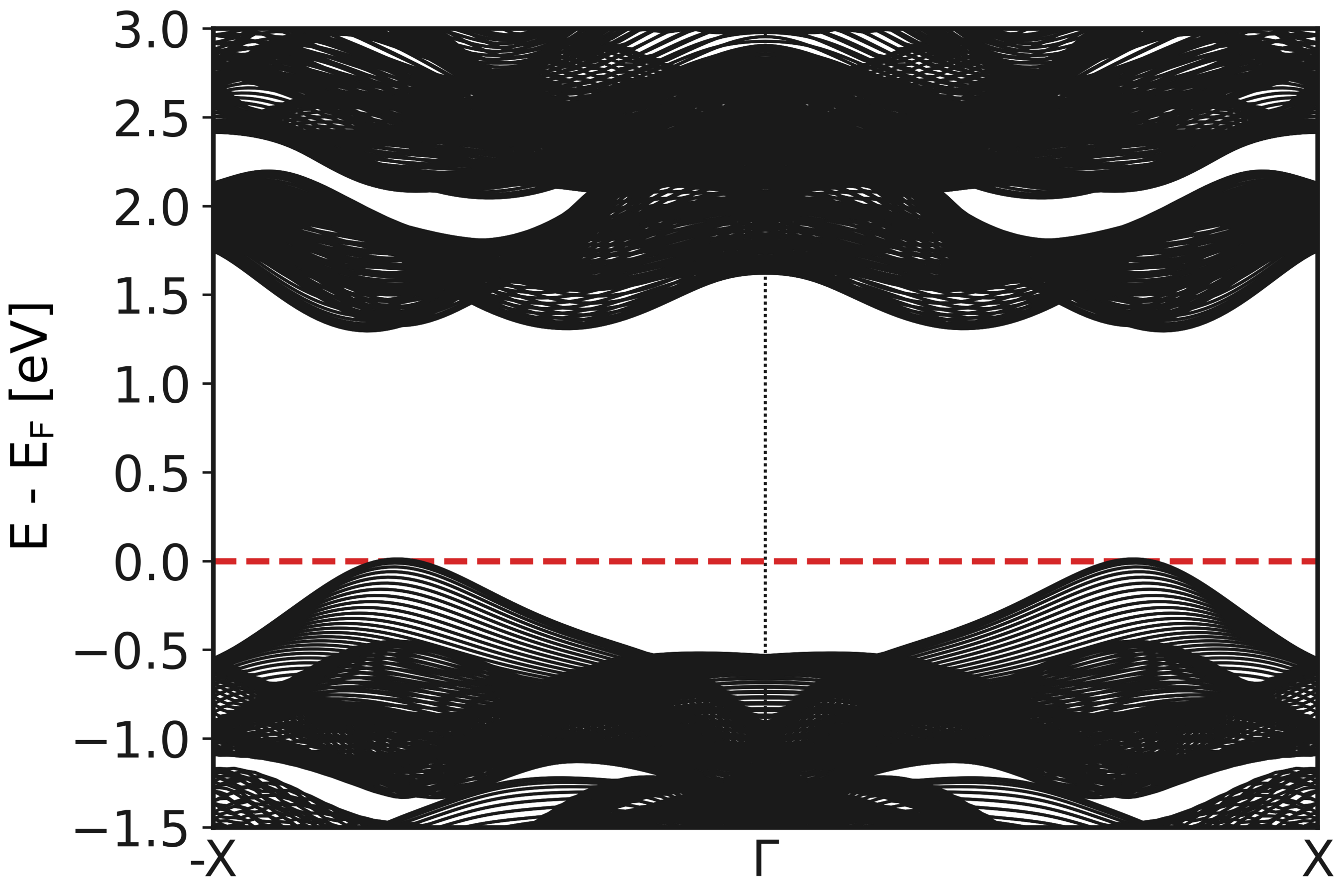}
            \caption*{(56,56)}
    \end{subfigure}
    \caption{Band structure of investigated nanotubes in armchair direction}
\end{figure}

\begin{figure}
    \centering
    \begin{subfigure}[b]{0.49\linewidth}
         \includegraphics[width=1\linewidth]{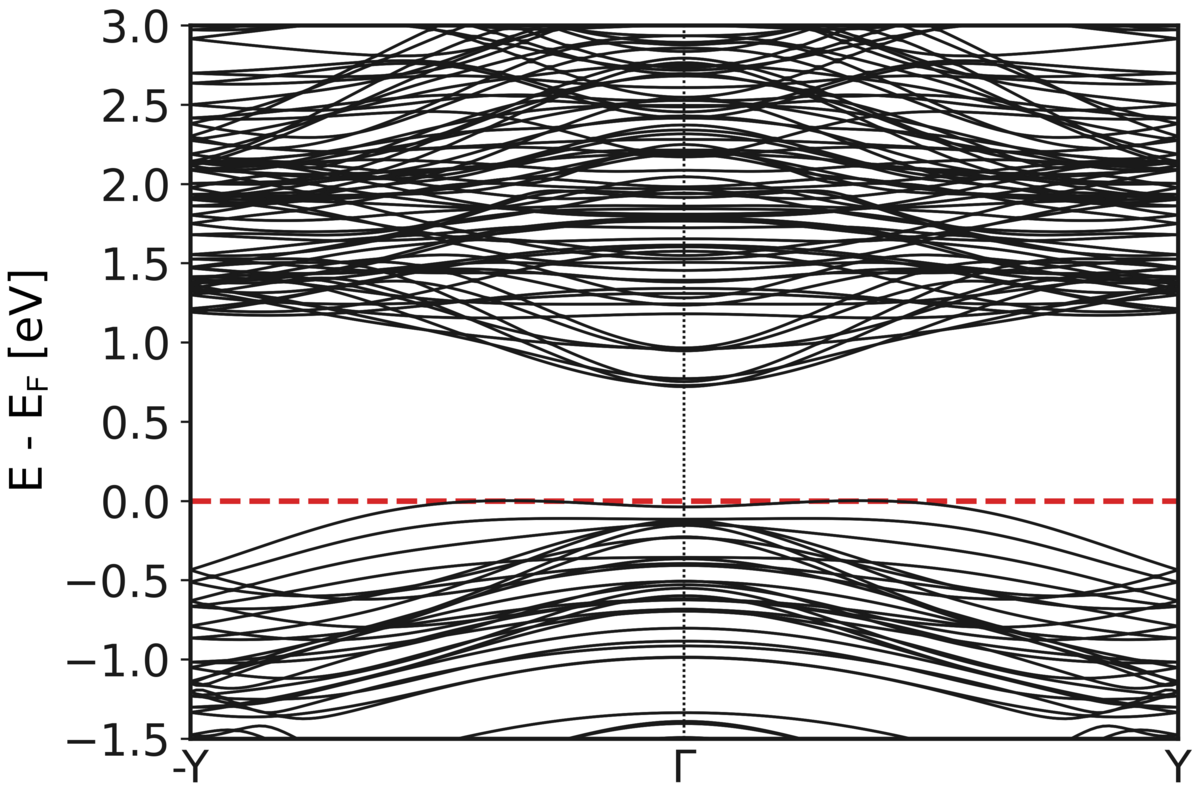}
         \caption*{(14,0)}
    \end{subfigure}
    \begin{subfigure}[b]{0.49\linewidth}
        \includegraphics[width=1\linewidth]{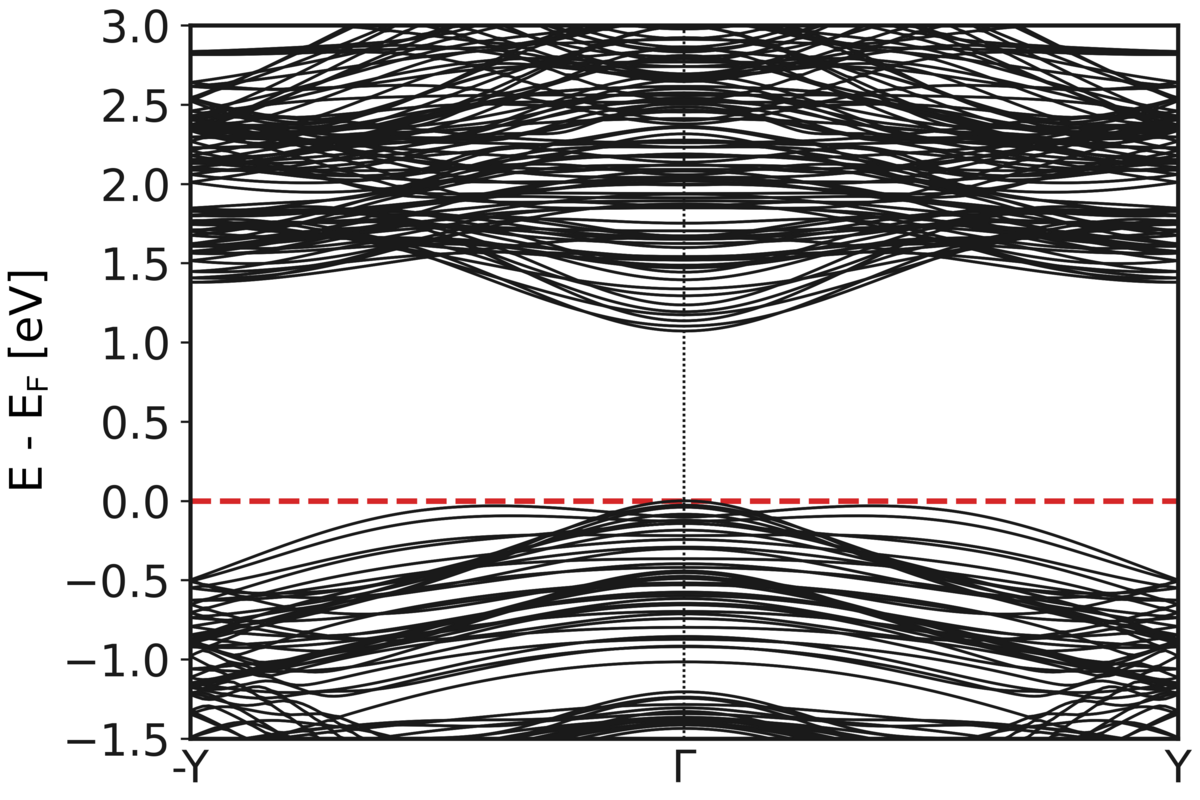}
        \caption*{(20,0)}
    \end{subfigure}
    \begin{subfigure}[b]{0.49\linewidth}
        \includegraphics[width=1\linewidth]{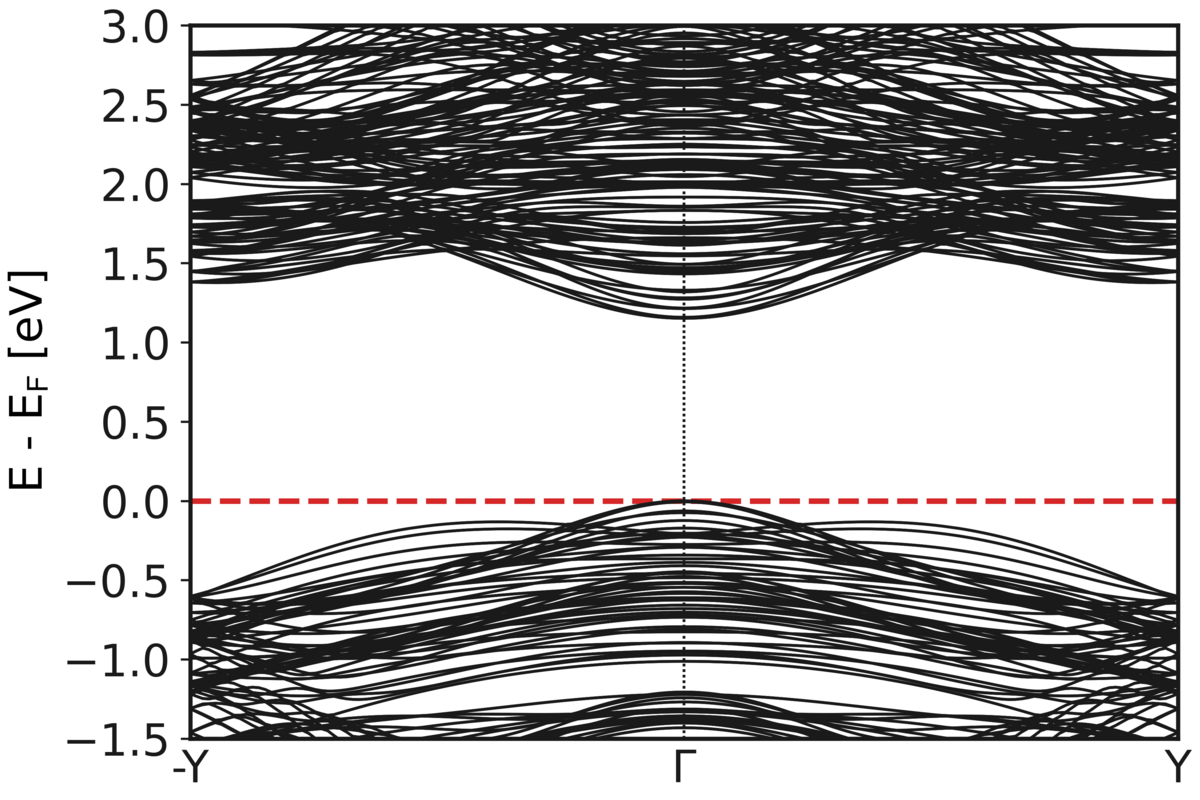}
        \caption*{(24,0)}
    \end{subfigure}
        \begin{subfigure}[b]{0.49\linewidth}
        \includegraphics[width=1\linewidth]{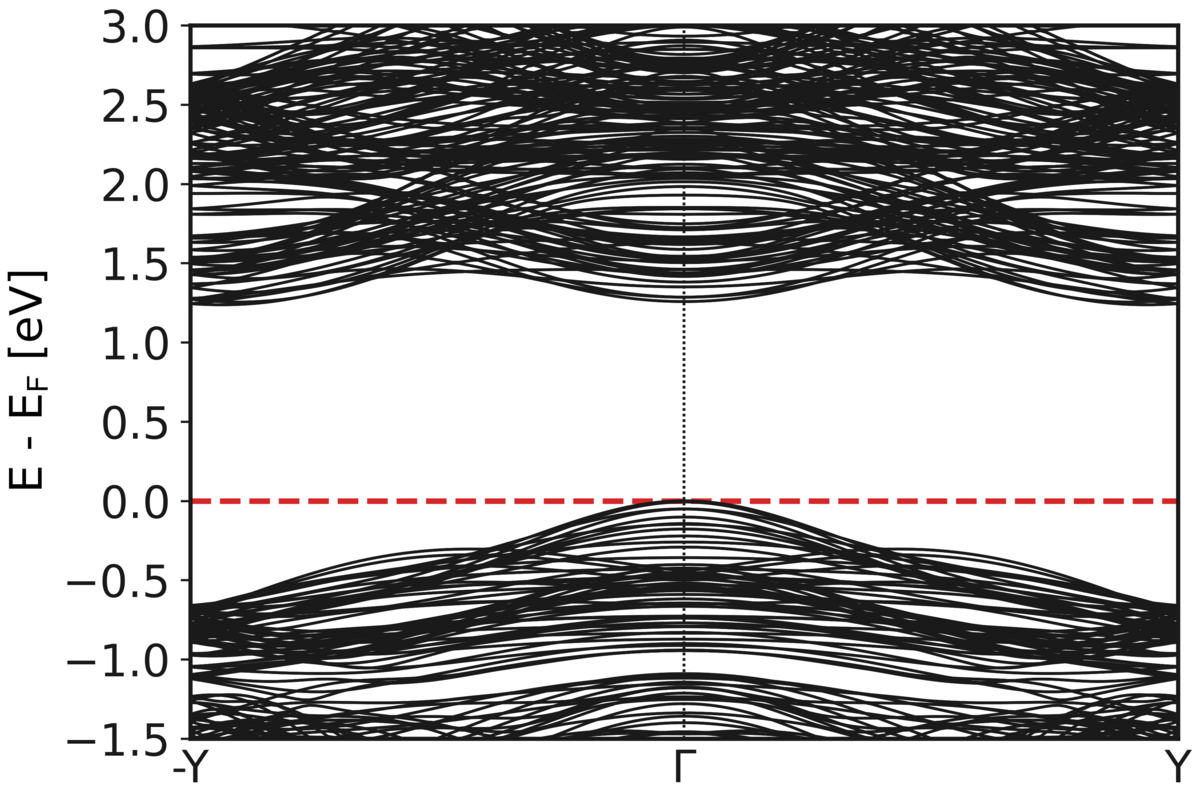}
        \caption*{(27,0)}
    \end{subfigure}
        \begin{subfigure}[b]{0.49\linewidth}
            \includegraphics[width=1\linewidth]{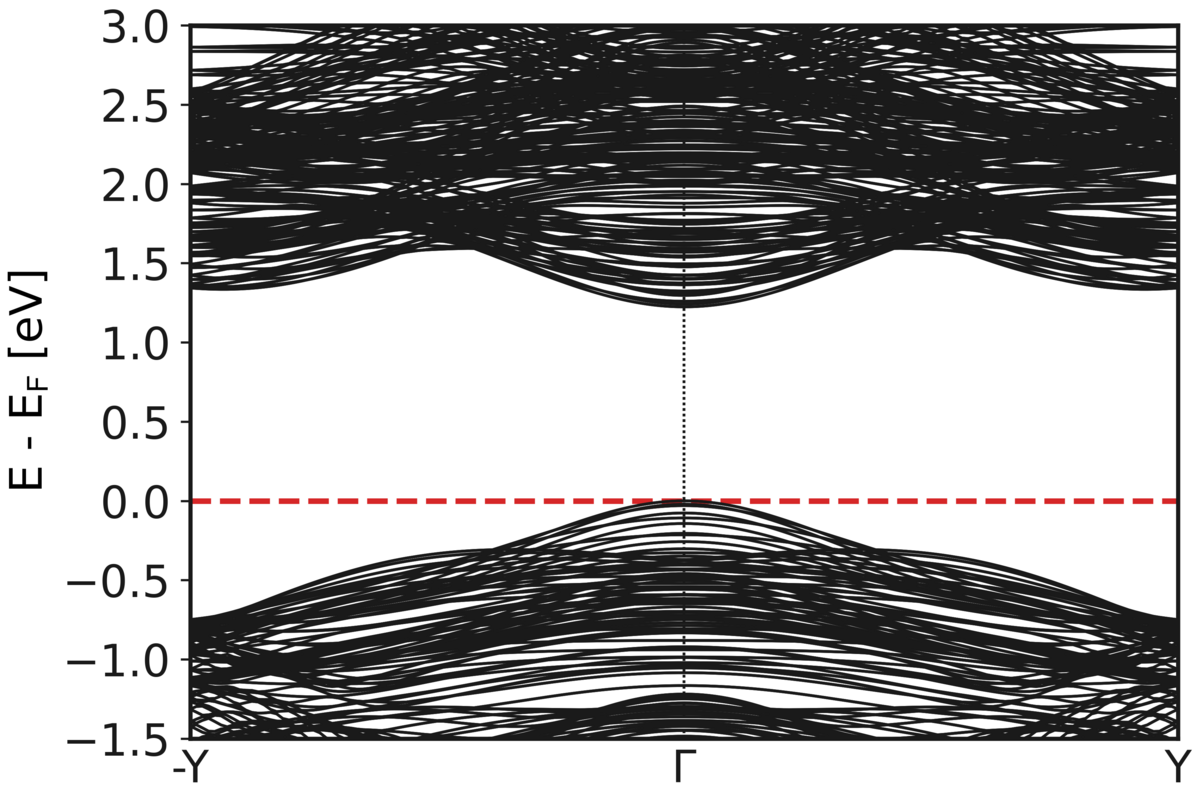}
            \caption*{(34,0)}
    \end{subfigure}
    \begin{subfigure}[b]{0.49\linewidth}
            \includegraphics[width=1\linewidth]{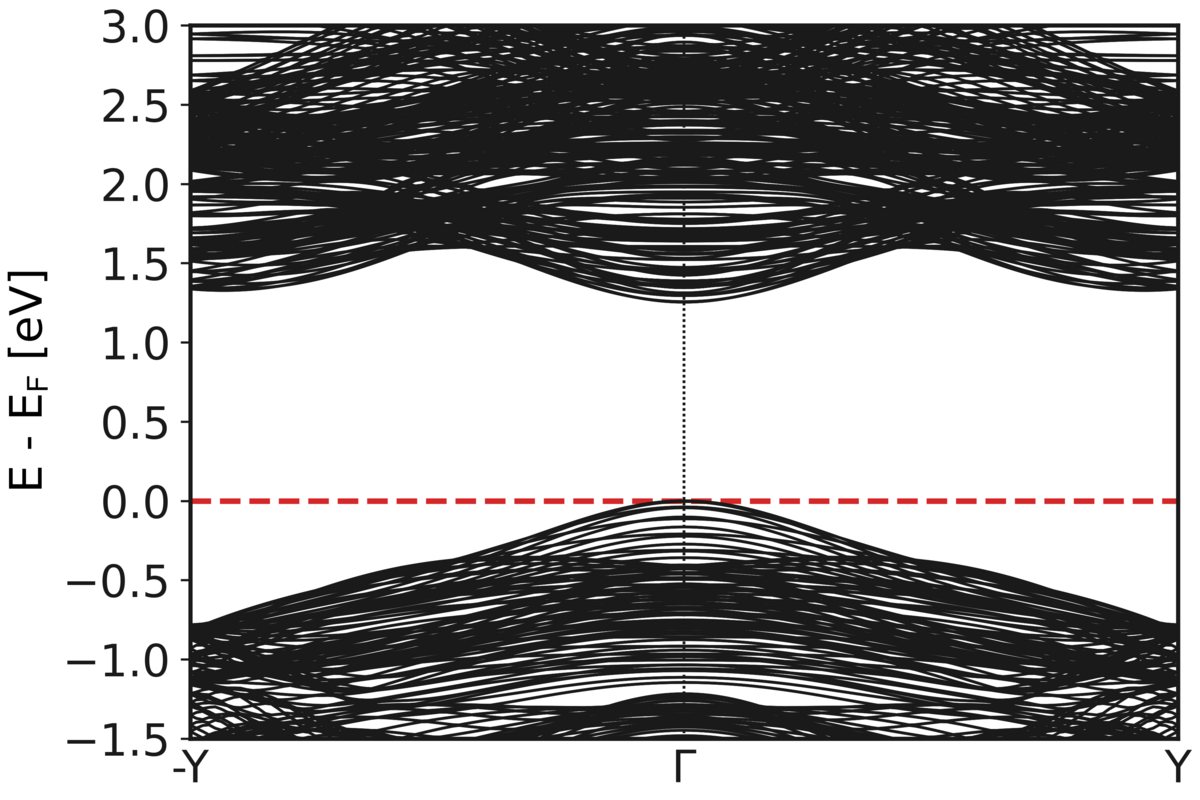}
            \caption*{(39,0)}
    \end{subfigure}
\end{figure}
\begin{figure}
\ContinuedFloat.

    \begin{subfigure}[b]{0.49\linewidth}
            \includegraphics[width=1\linewidth]{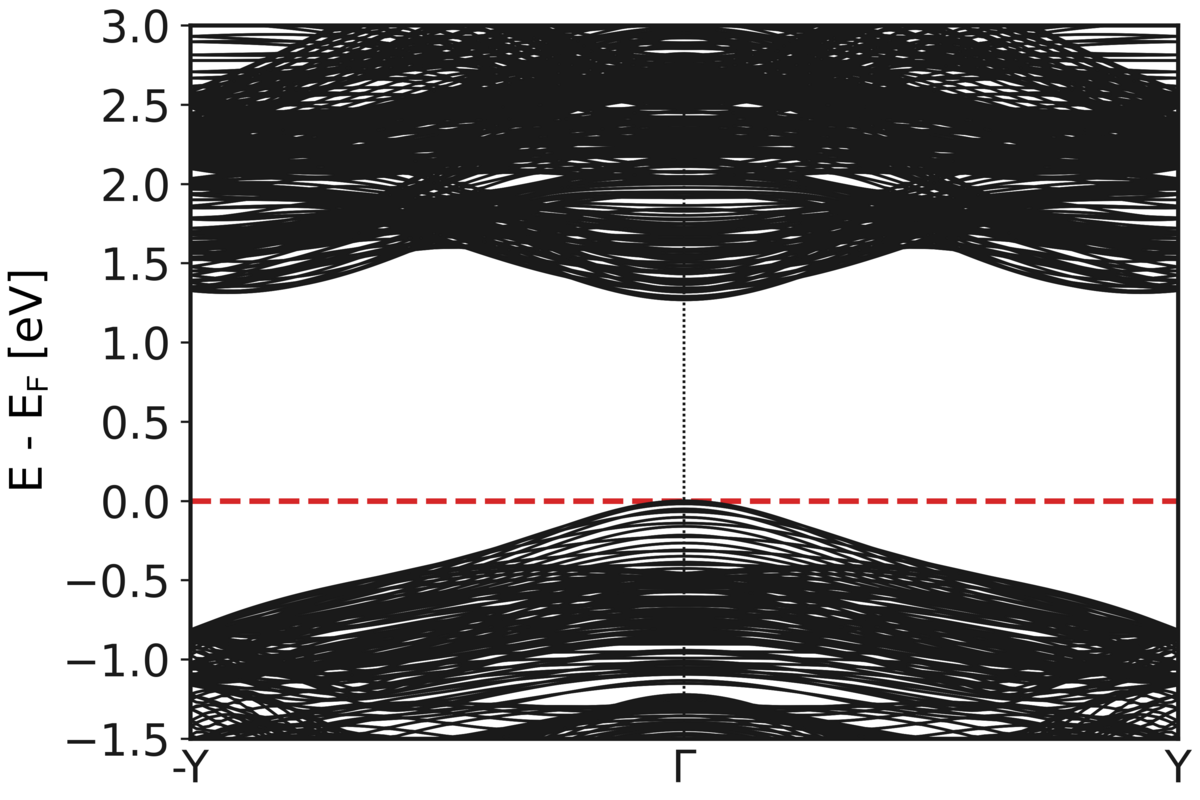}
            \caption*{(46,0)}
    \end{subfigure}
    \begin{subfigure}[b]{0.49\linewidth}
            \includegraphics[width=1\linewidth]{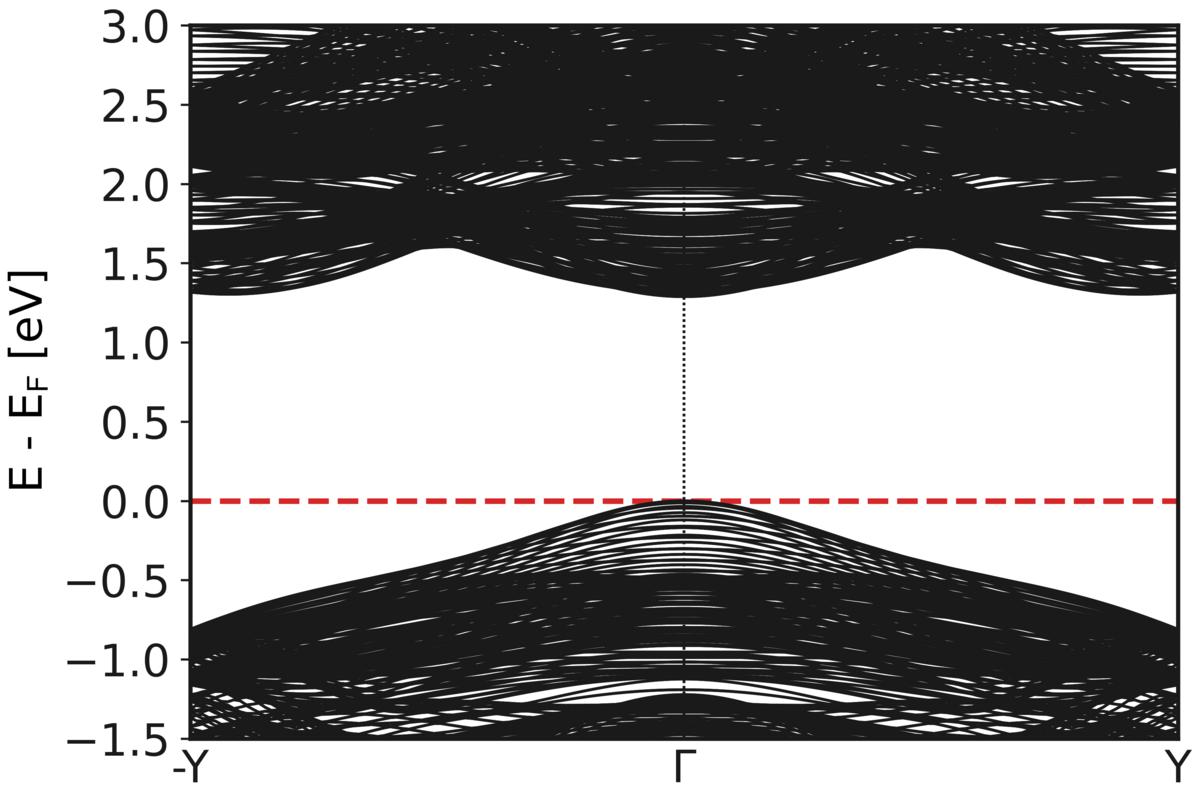}
            \caption*{(58,0)}
    \end{subfigure}
        \begin{subfigure}[b]{0.49\linewidth}
            \includegraphics[width=1\linewidth]{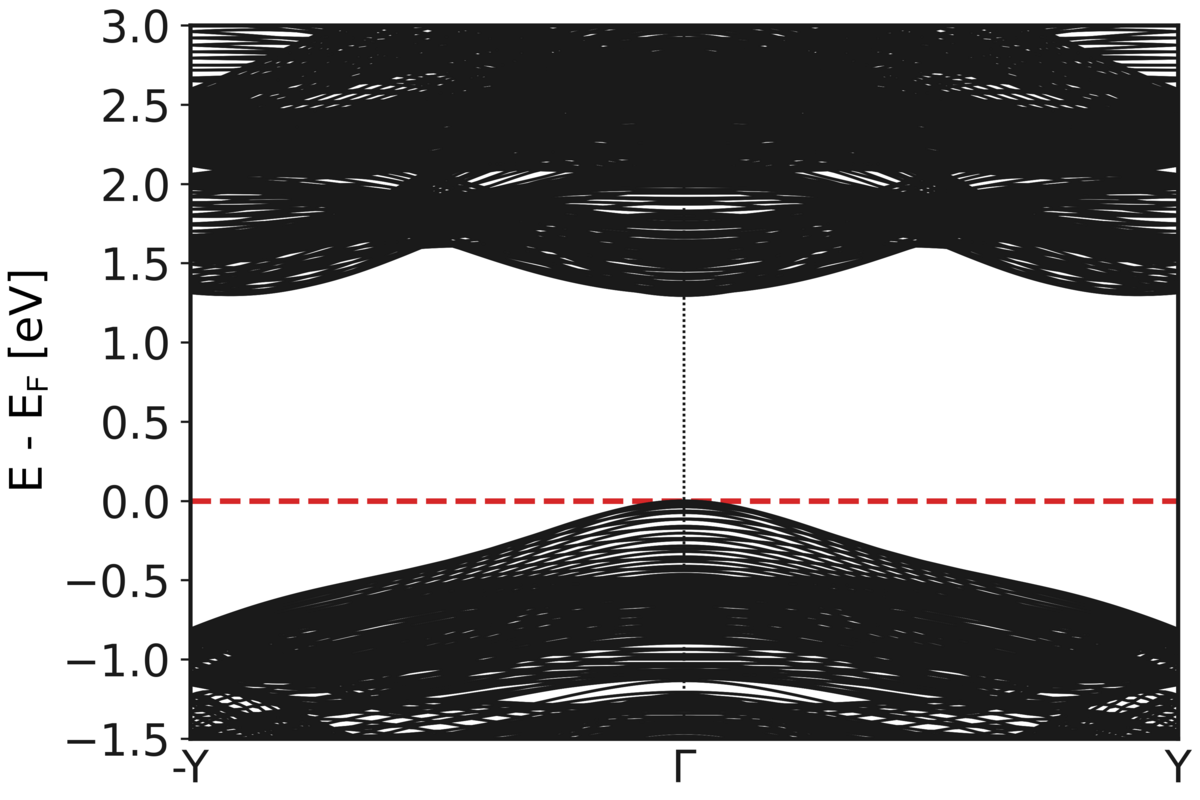}
            \caption*{(65,0)}
    \end{subfigure}
        \caption{Band structure of investigated nanotubes in zigzag direction}
\end{figure}
\begin{figure}
    \centering
        \begin{subfigure}[b]{0.49\linewidth}
         \includegraphics[width=1\linewidth]{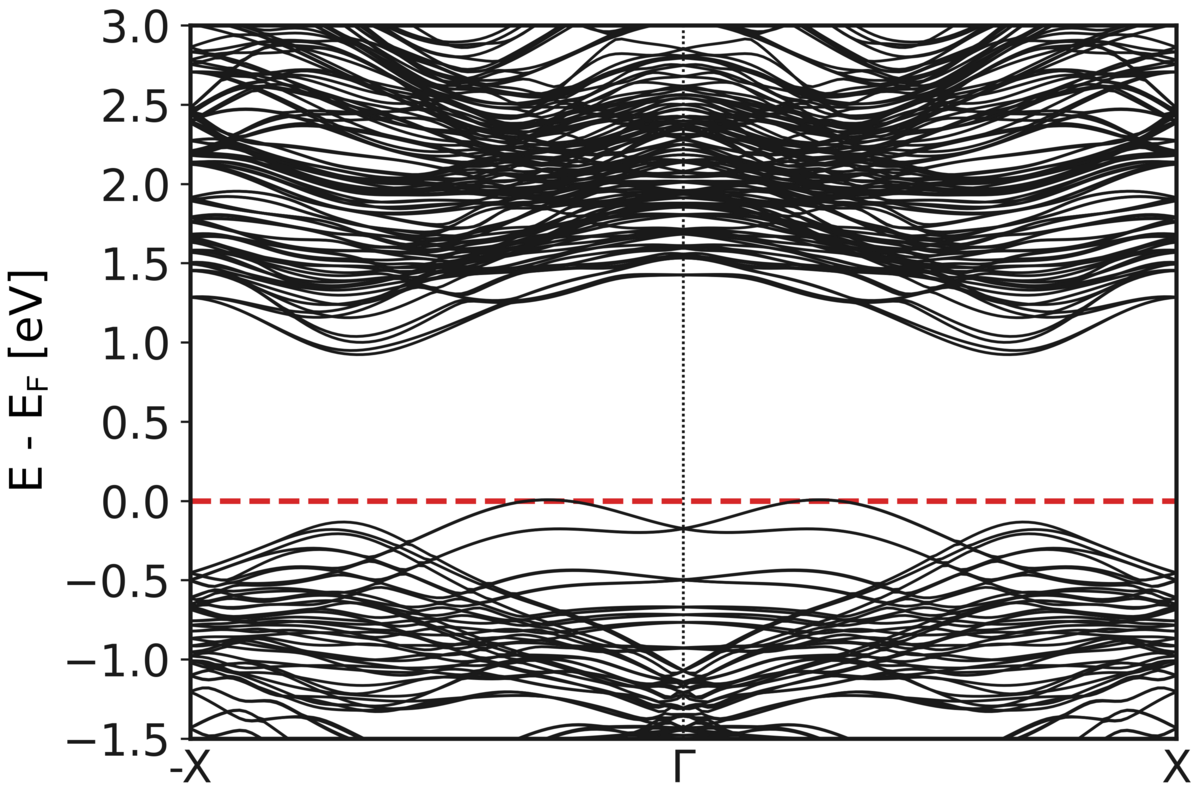}
         \caption*{(10,10)}
    \end{subfigure}
    \begin{subfigure}[b]{0.49\linewidth}
         \includegraphics[width=1\linewidth]{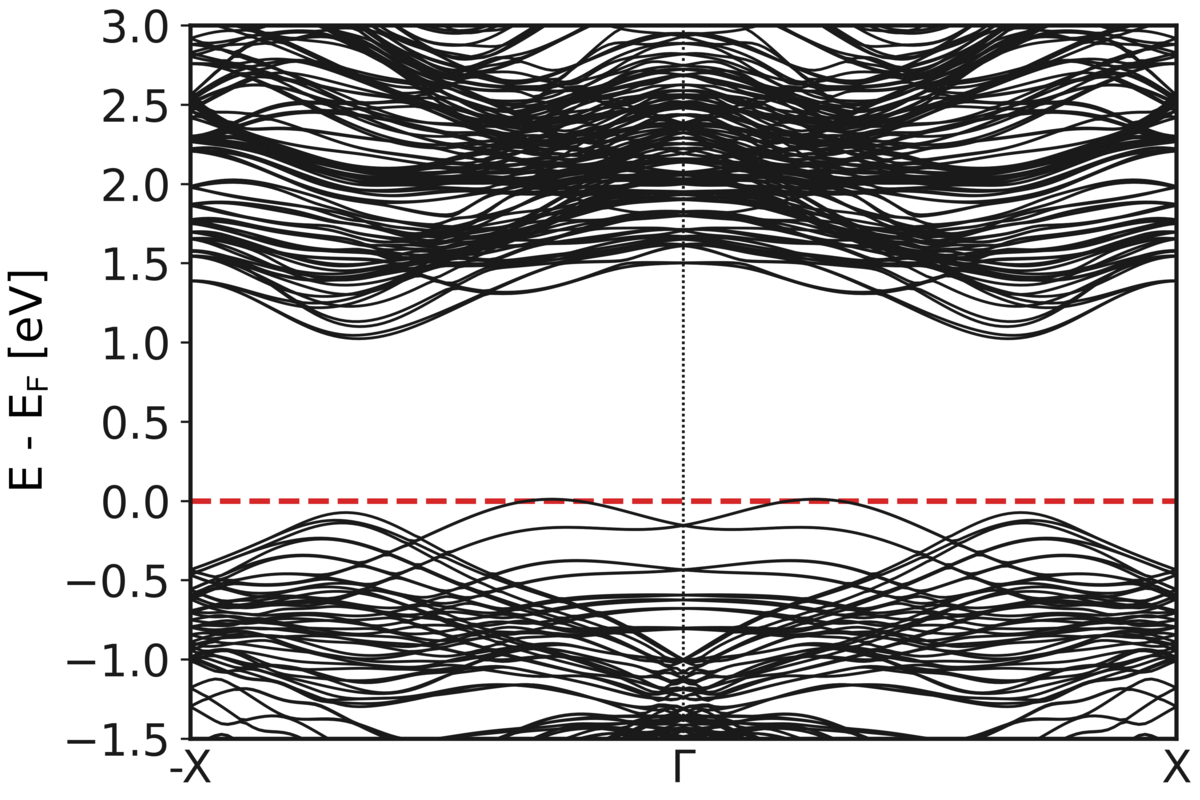}
         \caption*{(11,11)}
    \end{subfigure}
        \begin{subfigure}[b]{0.49\linewidth}
         \includegraphics[width=1\linewidth]{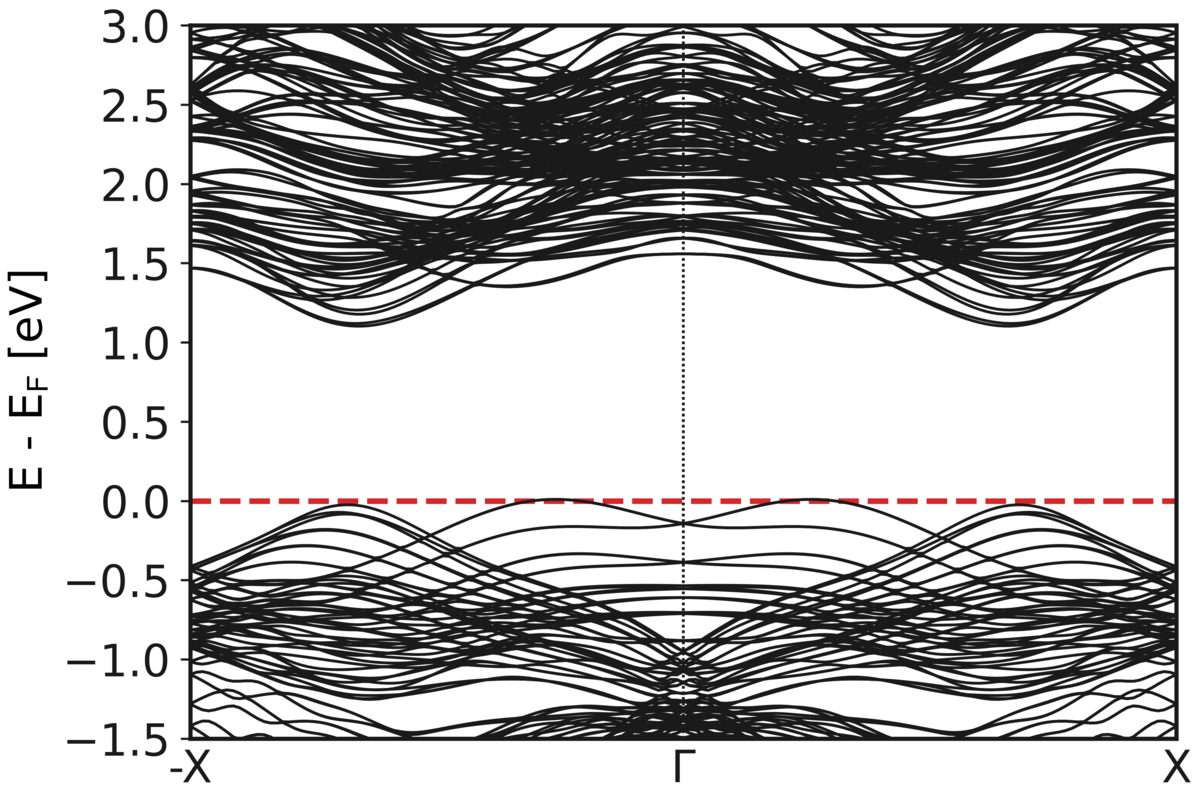}
         \caption*{(12,12)}
    \end{subfigure}
        \begin{subfigure}[b]{0.49\linewidth}
         \includegraphics[width=1\linewidth]{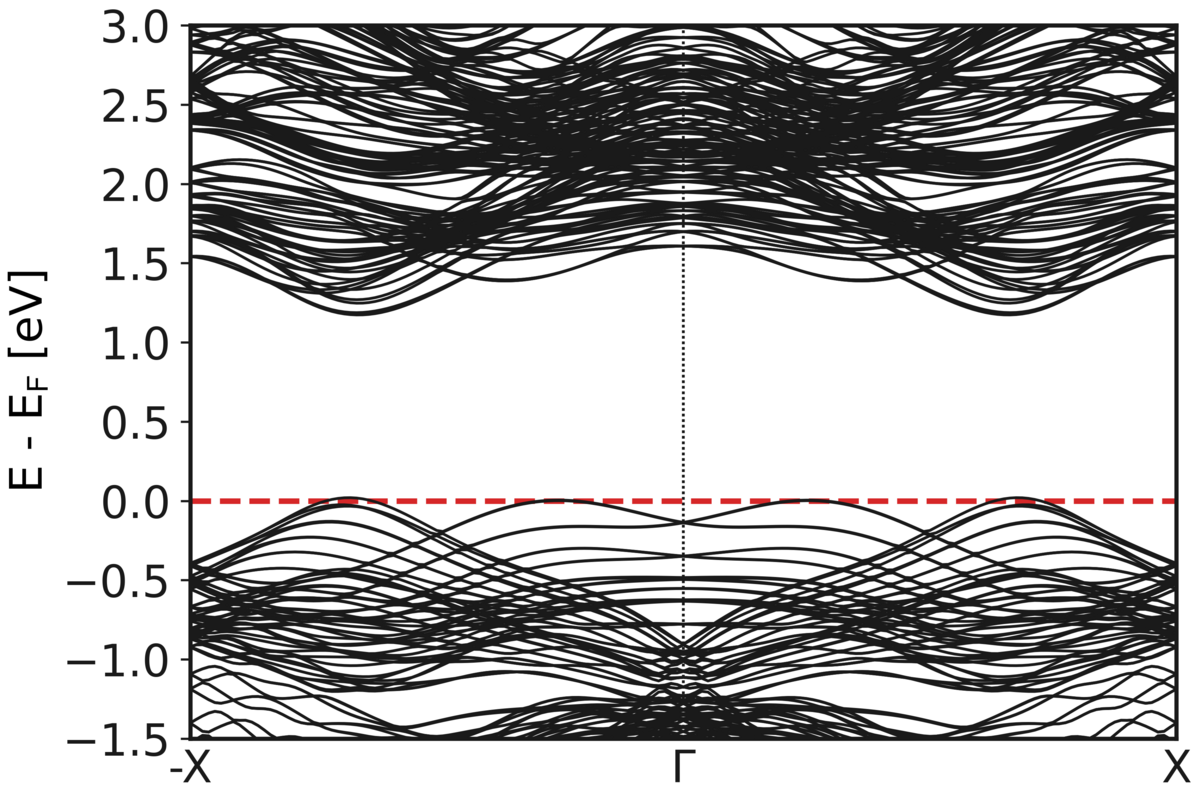}
         \caption*{(13,13)}
    \end{subfigure}
        \begin{subfigure}[b]{0.49\linewidth}
         \includegraphics[width=1\linewidth]{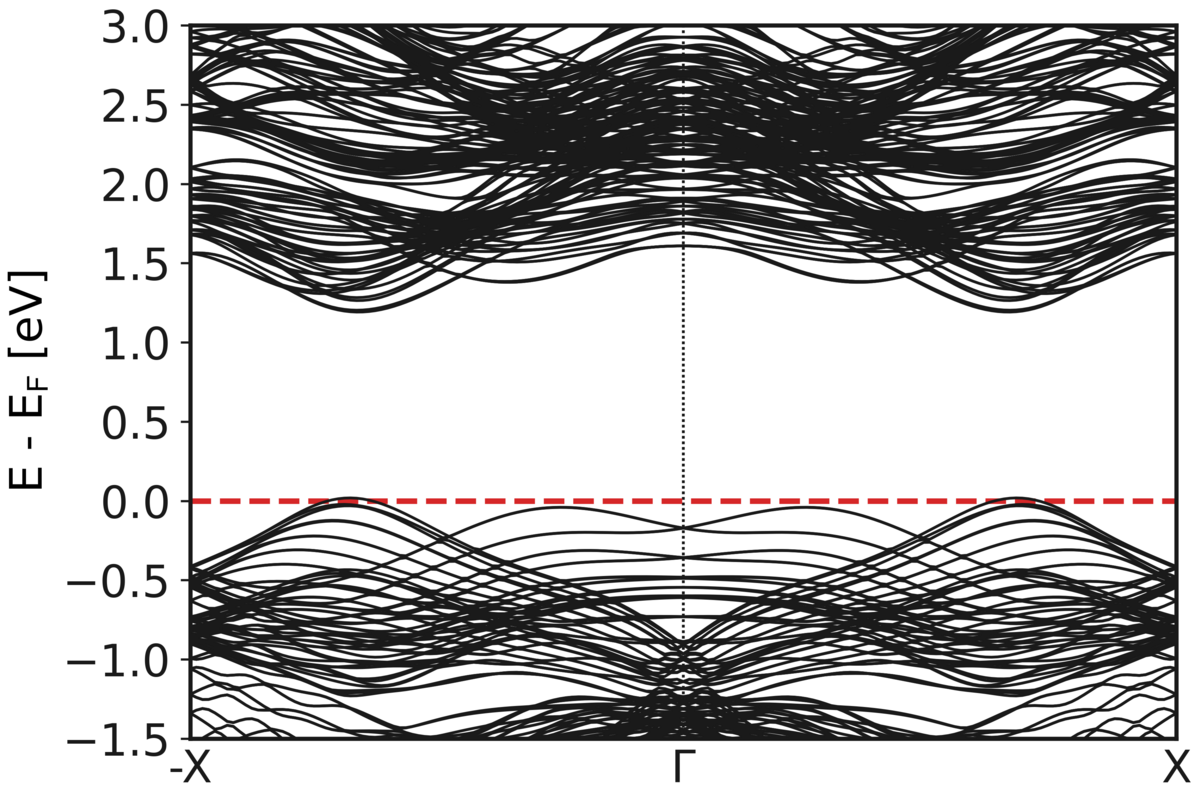}
         \caption*{(14,14)}
    \end{subfigure}
    \begin{subfigure}[b]{0.49\linewidth}
        \includegraphics[width=1\linewidth]{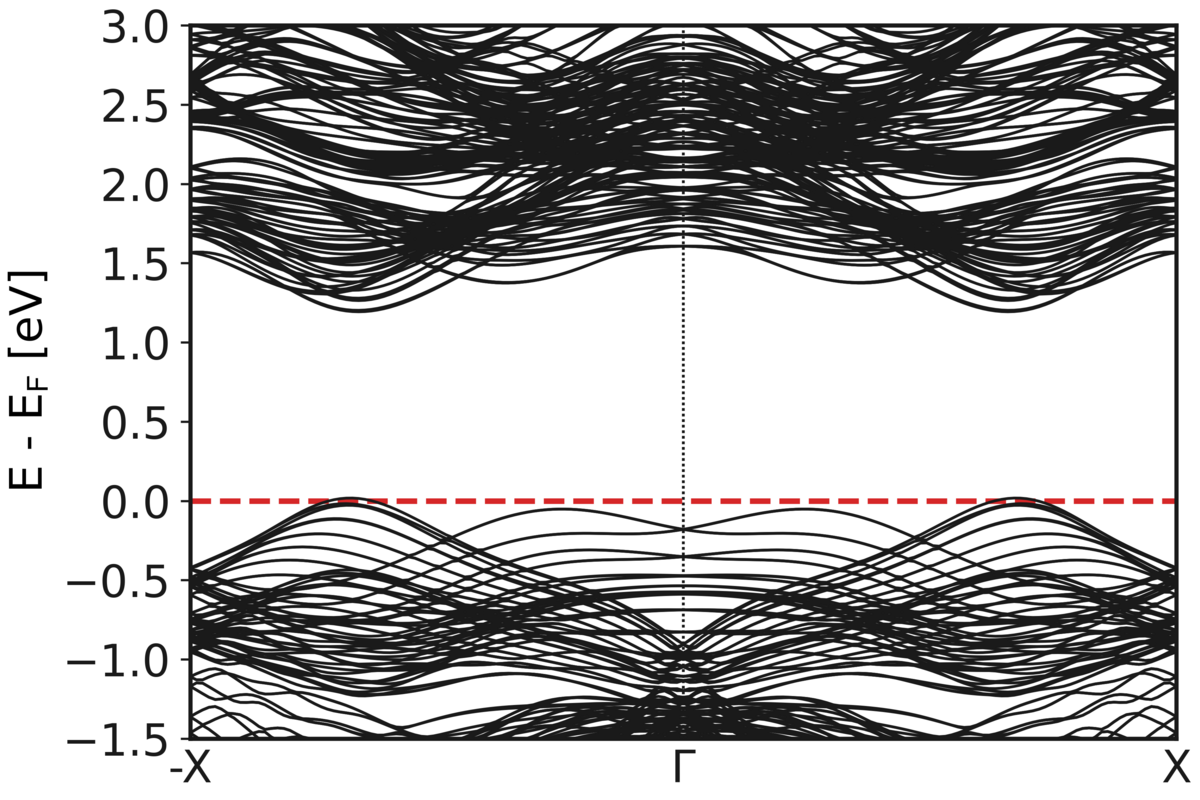}
        \caption*{(15,15)}
    \end{subfigure}
  
\end{figure}
\begin{figure}
\ContinuedFloat.

        \begin{subfigure}[b]{0.49\linewidth}
        \includegraphics[width=1\linewidth]{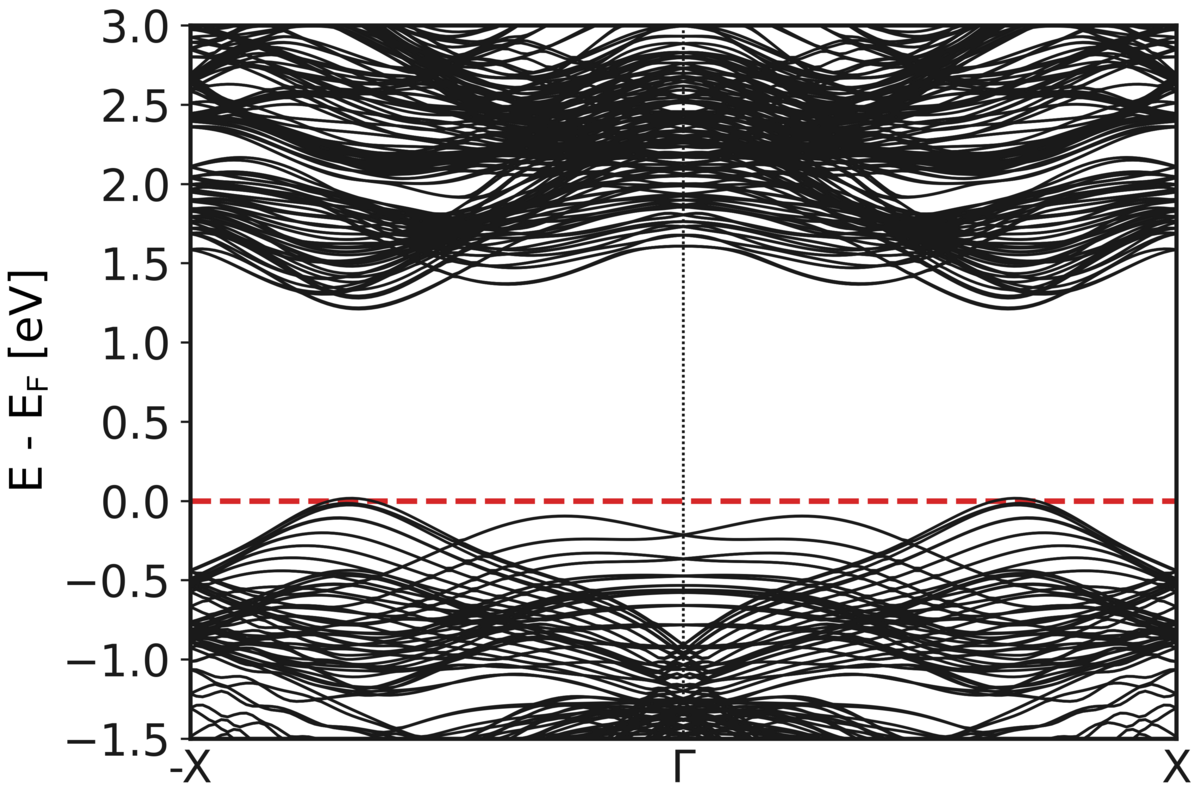}
        \caption*{(16,16)}
    \end{subfigure}
        \begin{subfigure}[b]{0.49\linewidth}
        \includegraphics[width=1\linewidth]{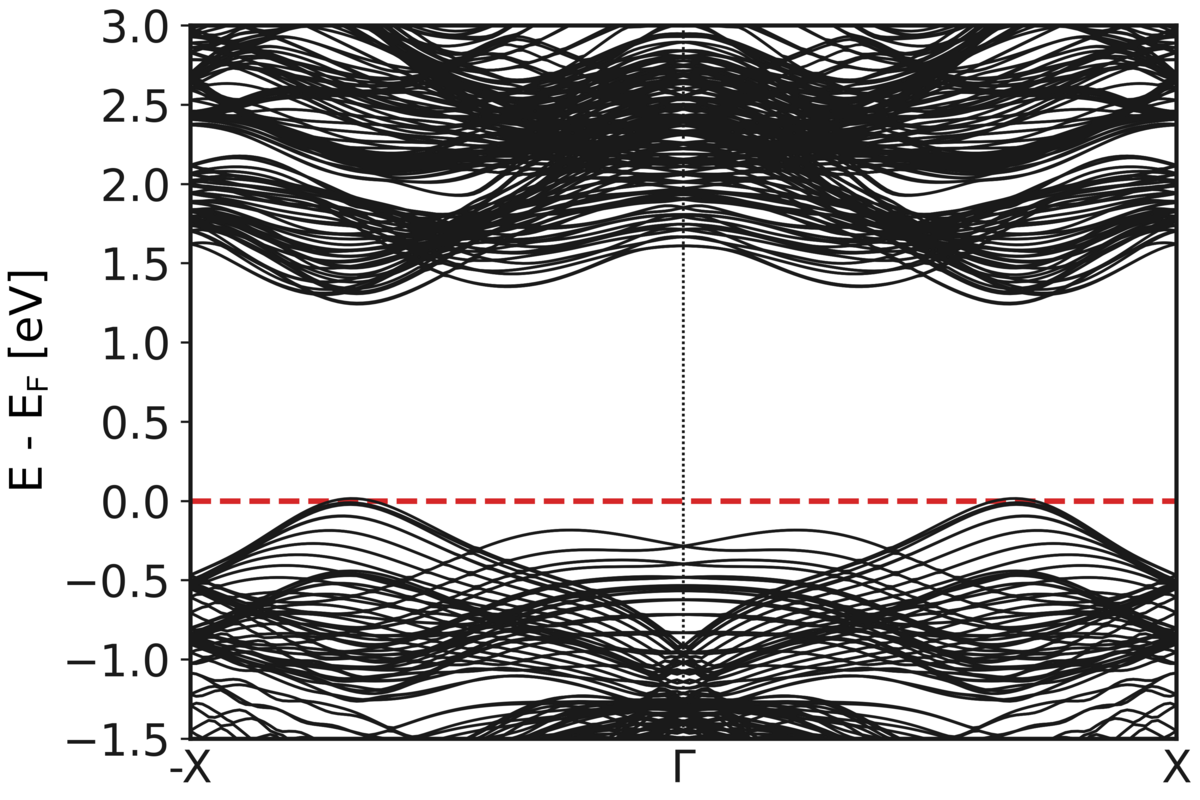}
        \caption*{(18,18)}
    \end{subfigure}
        \begin{subfigure}[b]{0.49\linewidth}
        \includegraphics[width=1\linewidth]{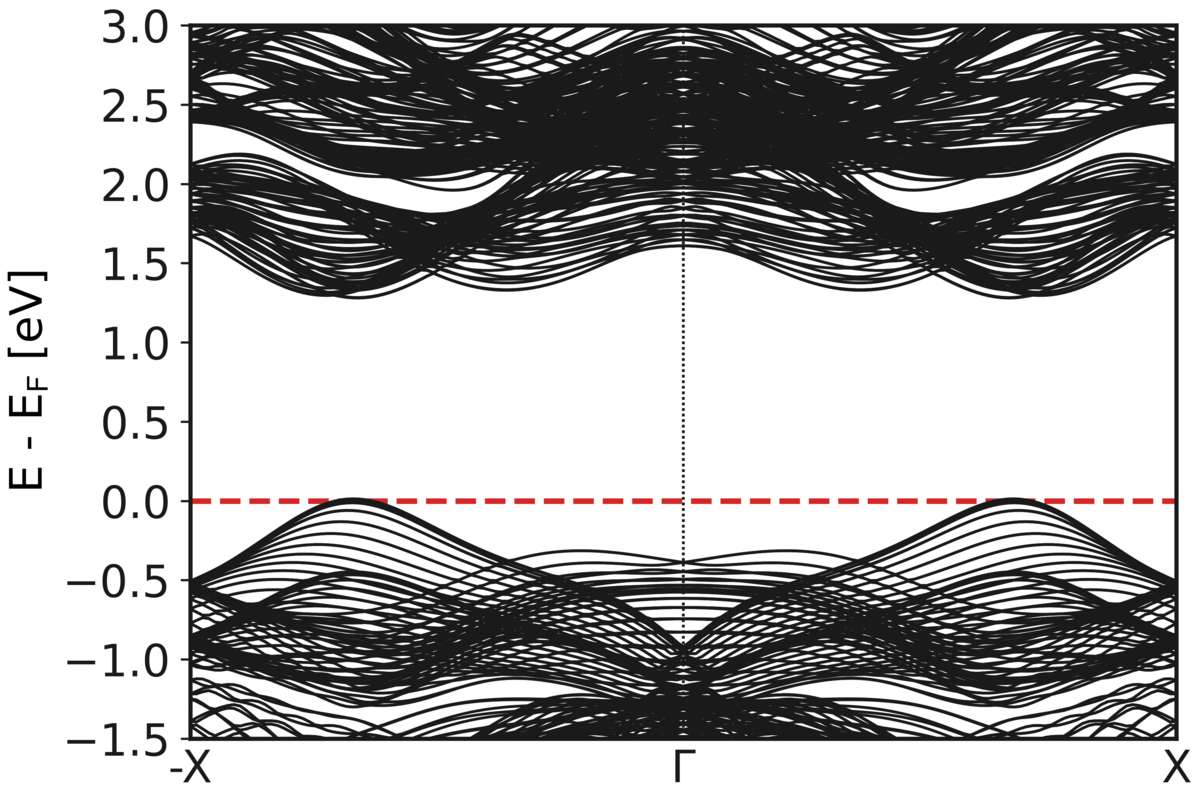}
        \caption*{(24,24)}
    \end{subfigure}
        \begin{subfigure}[b]{0.49\linewidth}
            \includegraphics[width=1\linewidth]{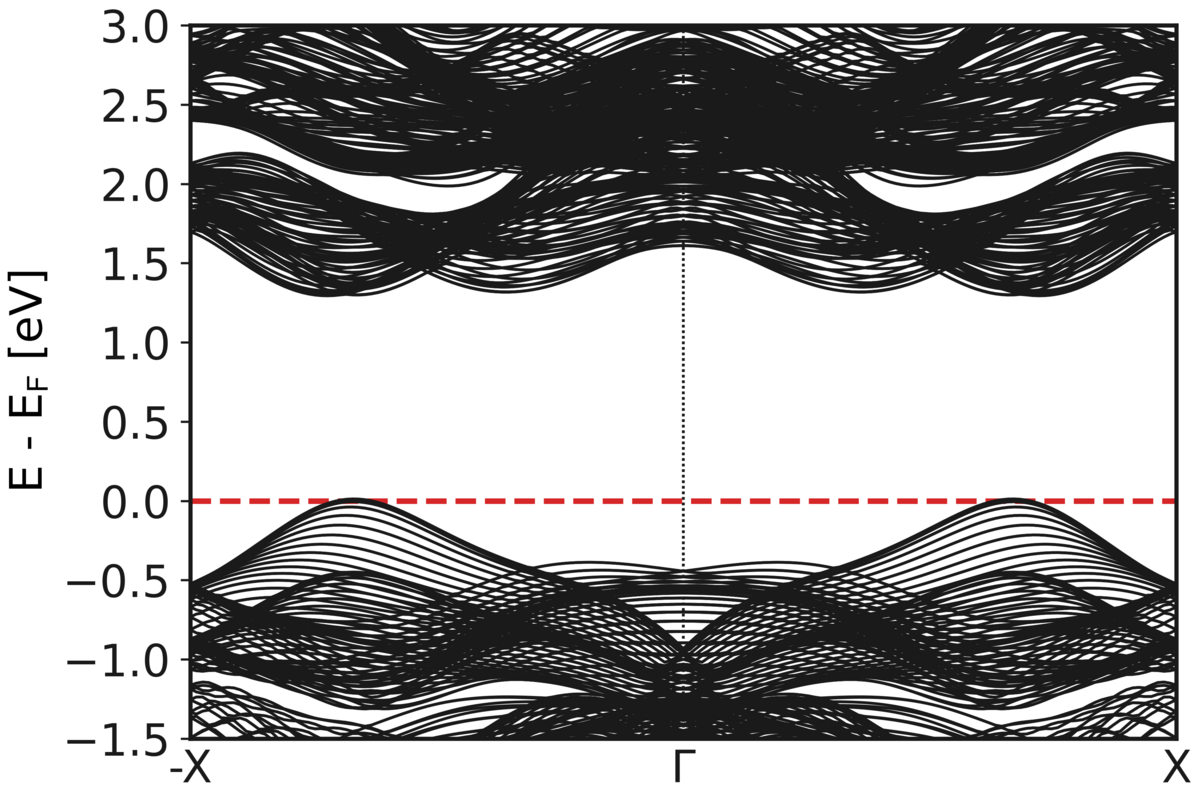}
            \caption*{(30,30)}
    \end{subfigure}
    \begin{subfigure}[b]{0.49\linewidth}
            \includegraphics[width=1\linewidth]{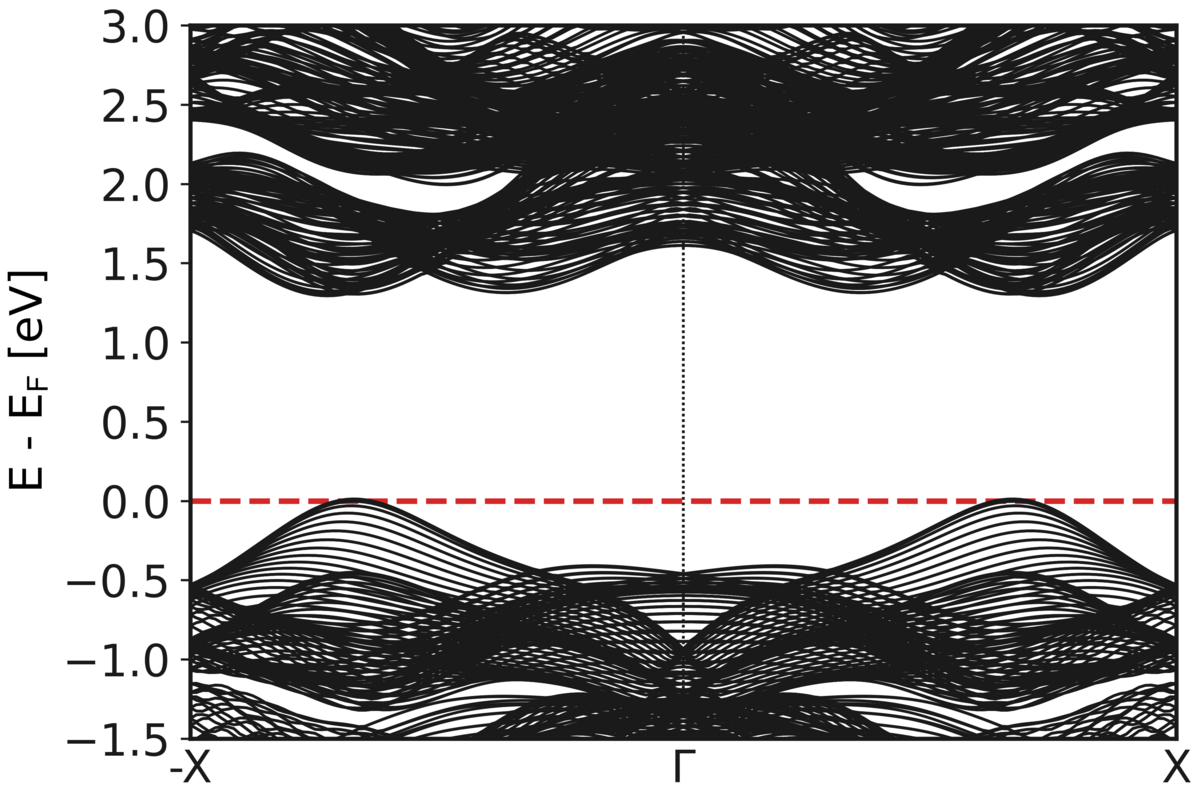}
            \caption*{(33,33)}
    \end{subfigure}
    \begin{subfigure}[b]{0.49\linewidth}
            \includegraphics[width=1\linewidth]{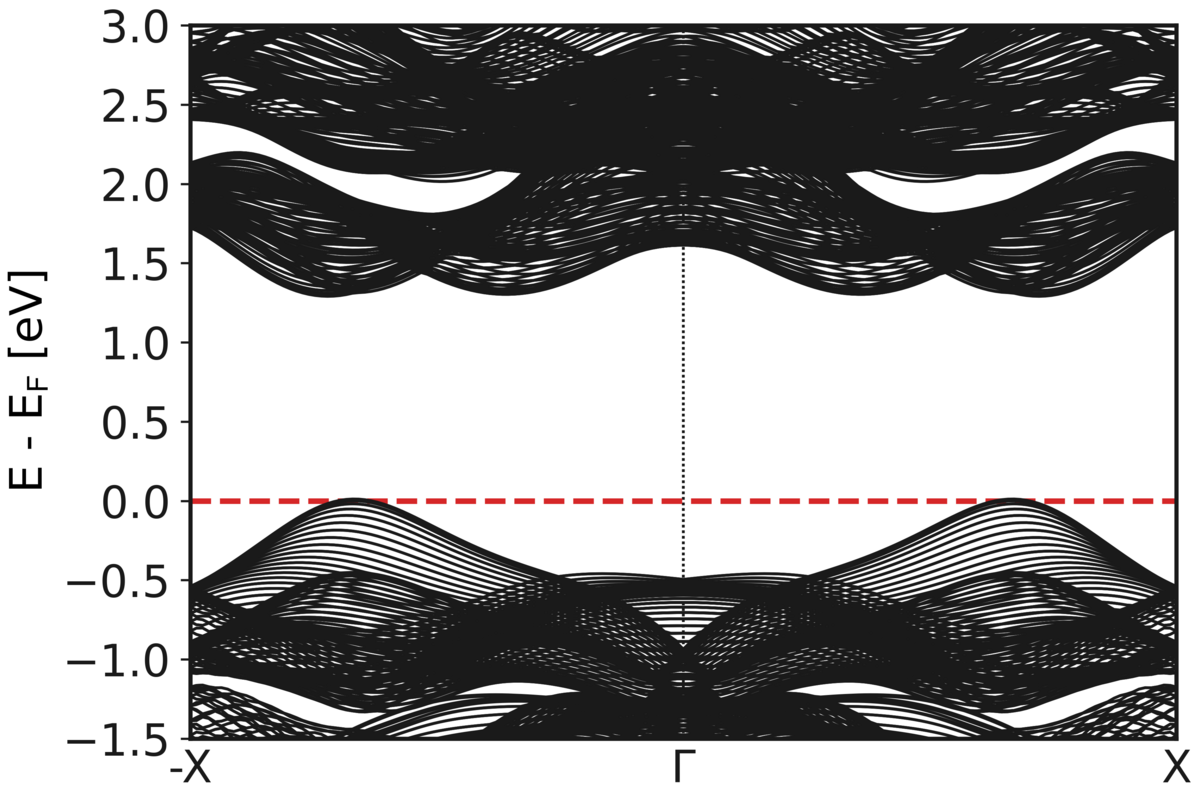}
            \caption*{(41,41)}
    \end{subfigure}
    \caption{Band structure of investigated wrinkles in armchair direction}
\end{figure}

\begin{figure}
    \centering
    \begin{subfigure}[b]{0.49\linewidth}
         \includegraphics[width=1\linewidth]{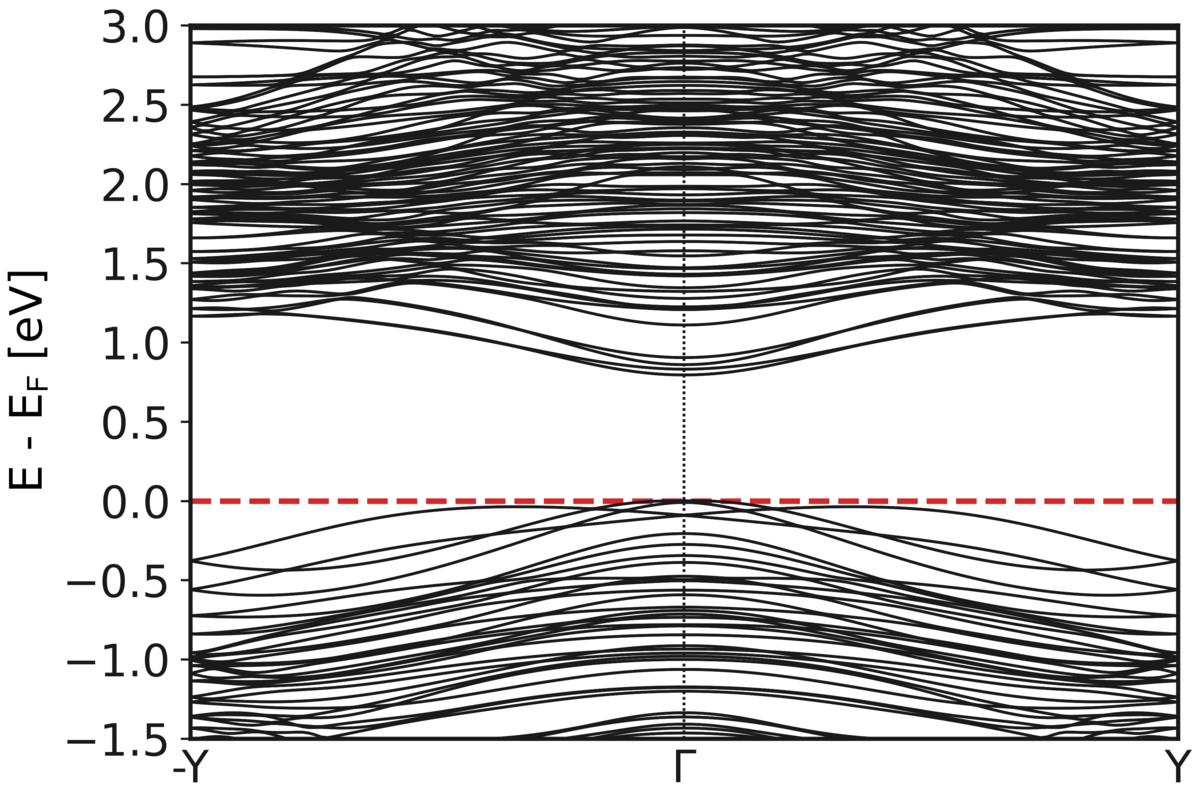}
         \caption*{(14,0)}
    \end{subfigure}
    \begin{subfigure}[b]{0.49\linewidth}
        \includegraphics[width=1\linewidth]{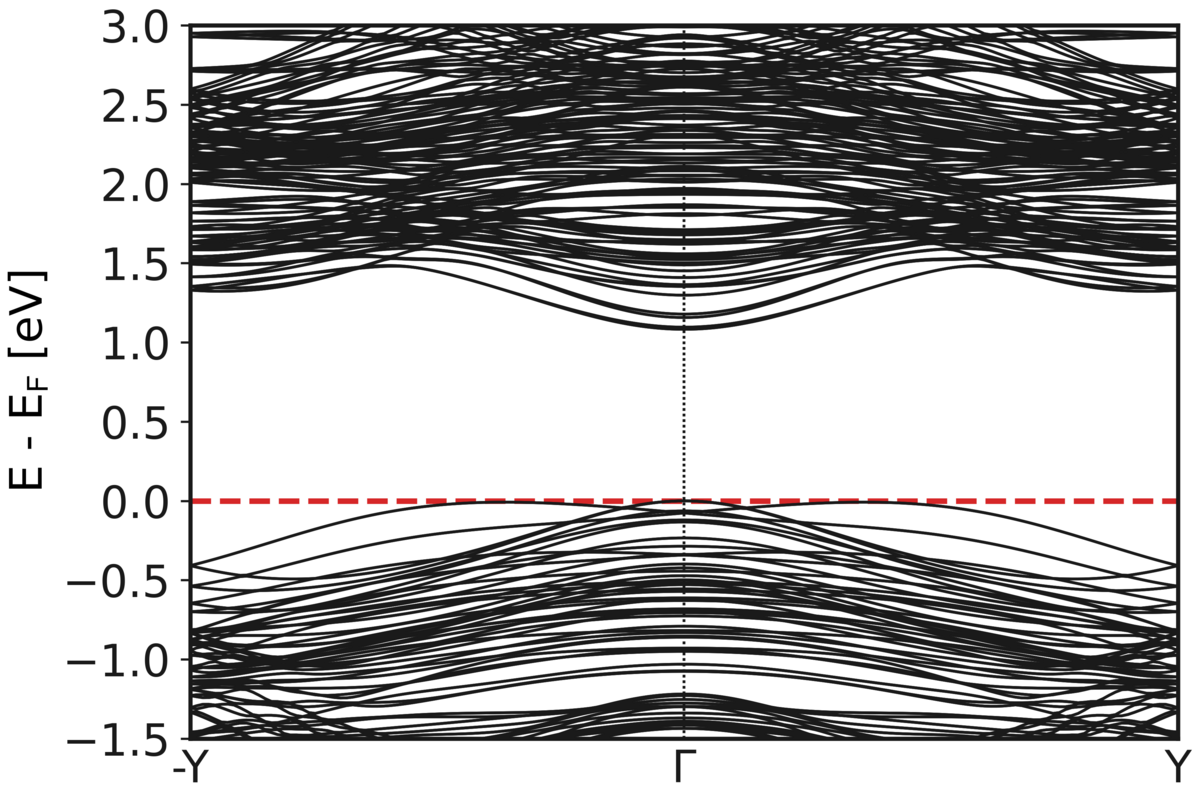}
        \caption*{(20,0)}
    \end{subfigure}
    \begin{subfigure}[b]{0.49\linewidth}
        \includegraphics[width=1\linewidth]{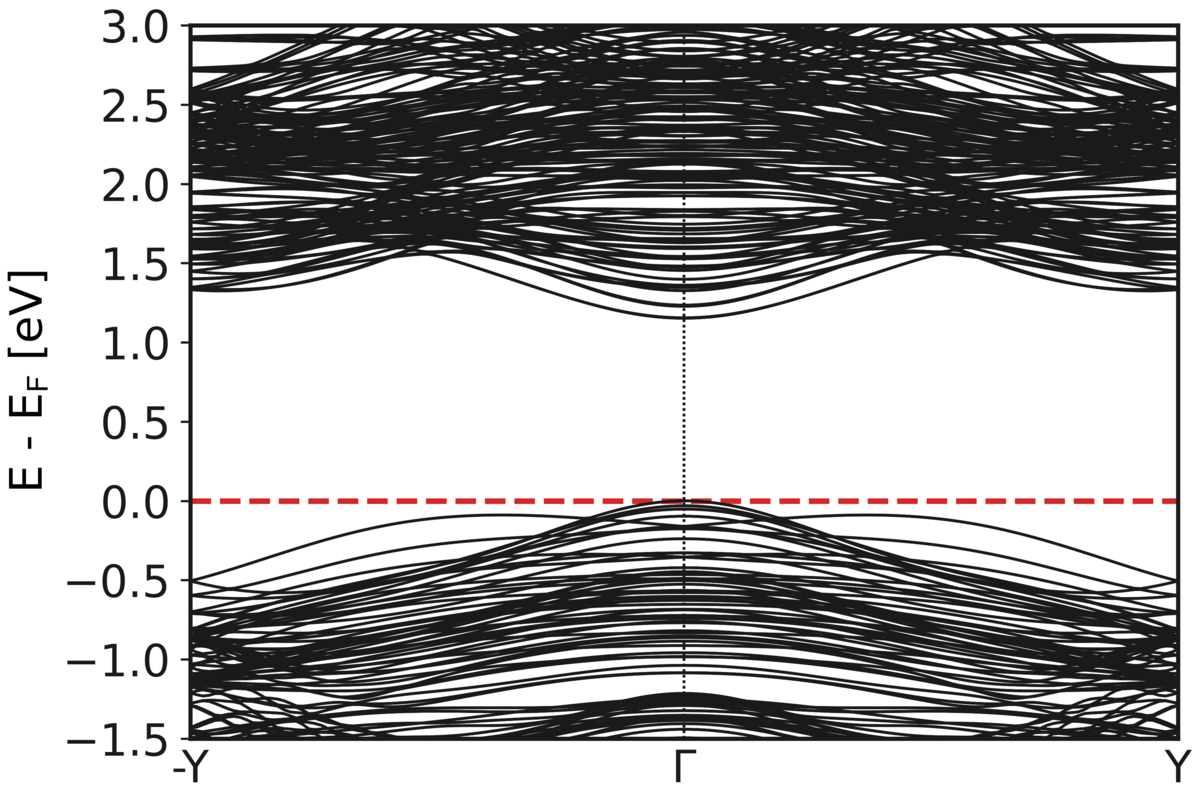}
        \caption*{(24,0)}
    \end{subfigure}
        \begin{subfigure}[b]{0.49\linewidth}
        \includegraphics[width=1\linewidth]{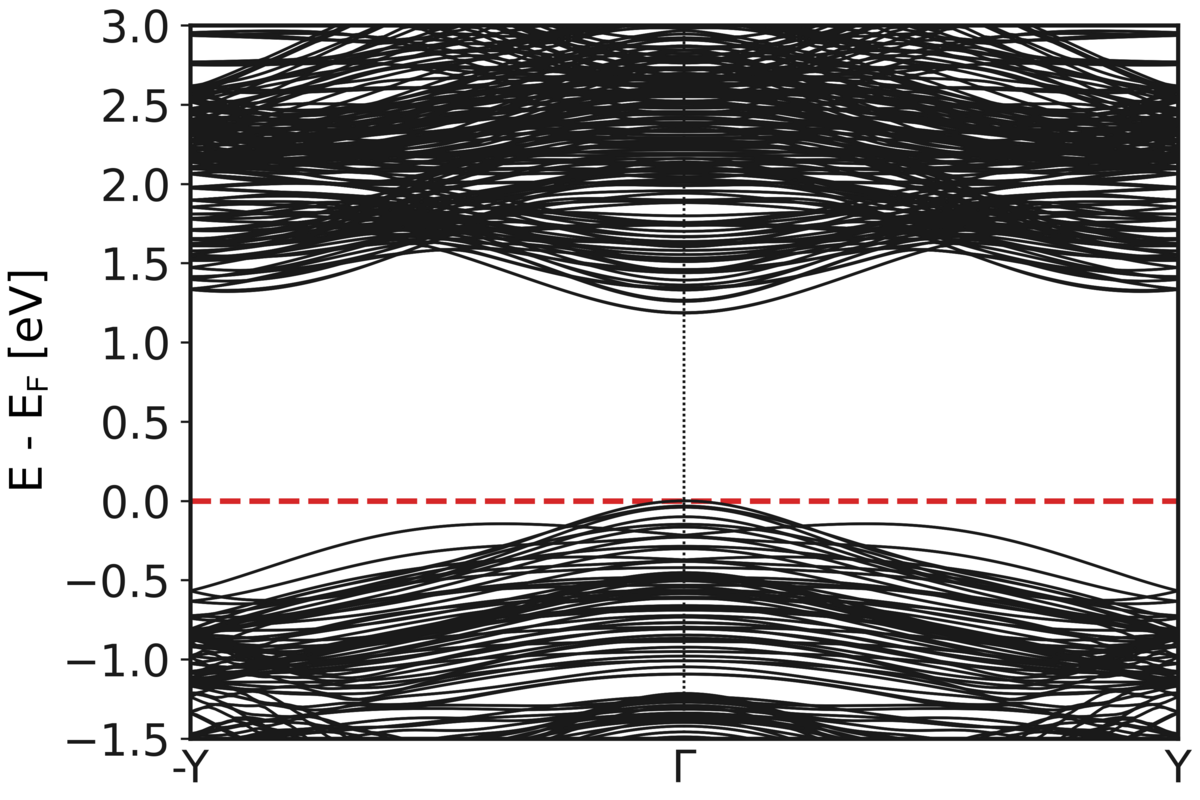}
        \caption*{(27,0)}
    \end{subfigure}
        \begin{subfigure}[b]{0.49\linewidth}
            \includegraphics[width=1\linewidth]{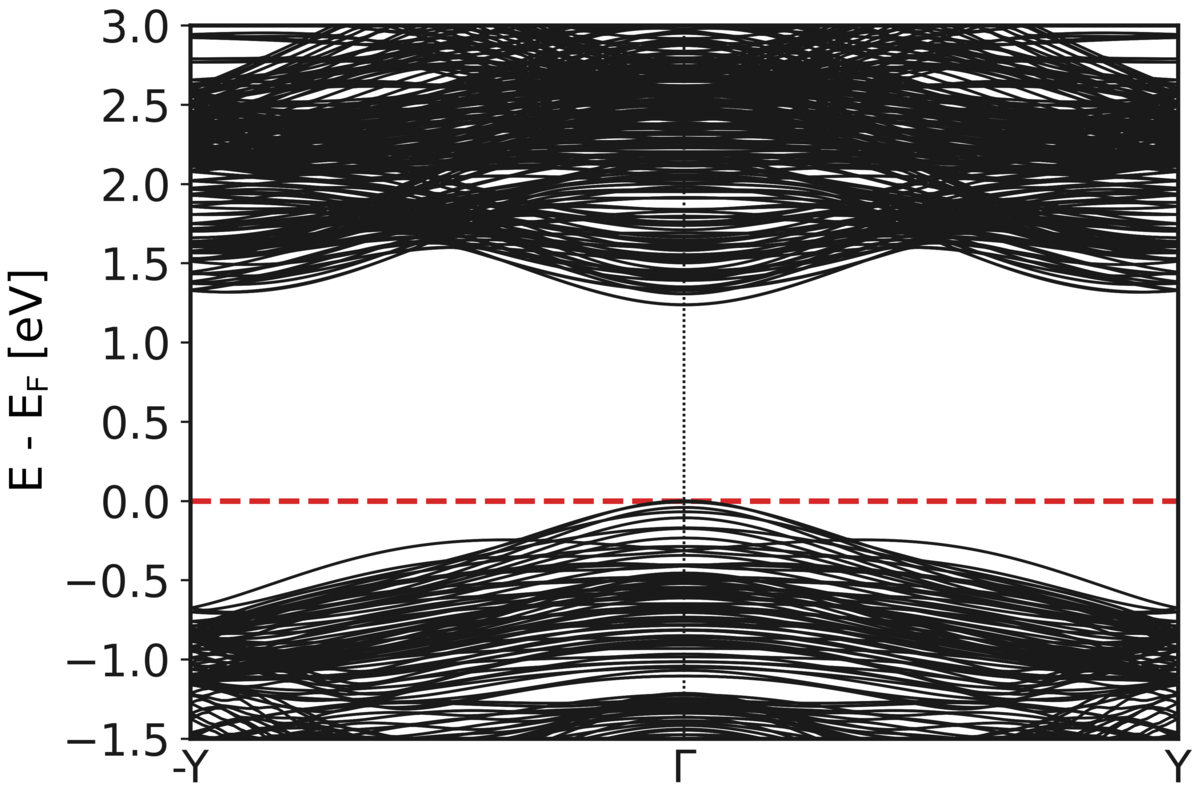}
            \caption*{(34,0)}
    \end{subfigure}
    \begin{subfigure}[b]{0.49\linewidth}
            \includegraphics[width=1\linewidth]{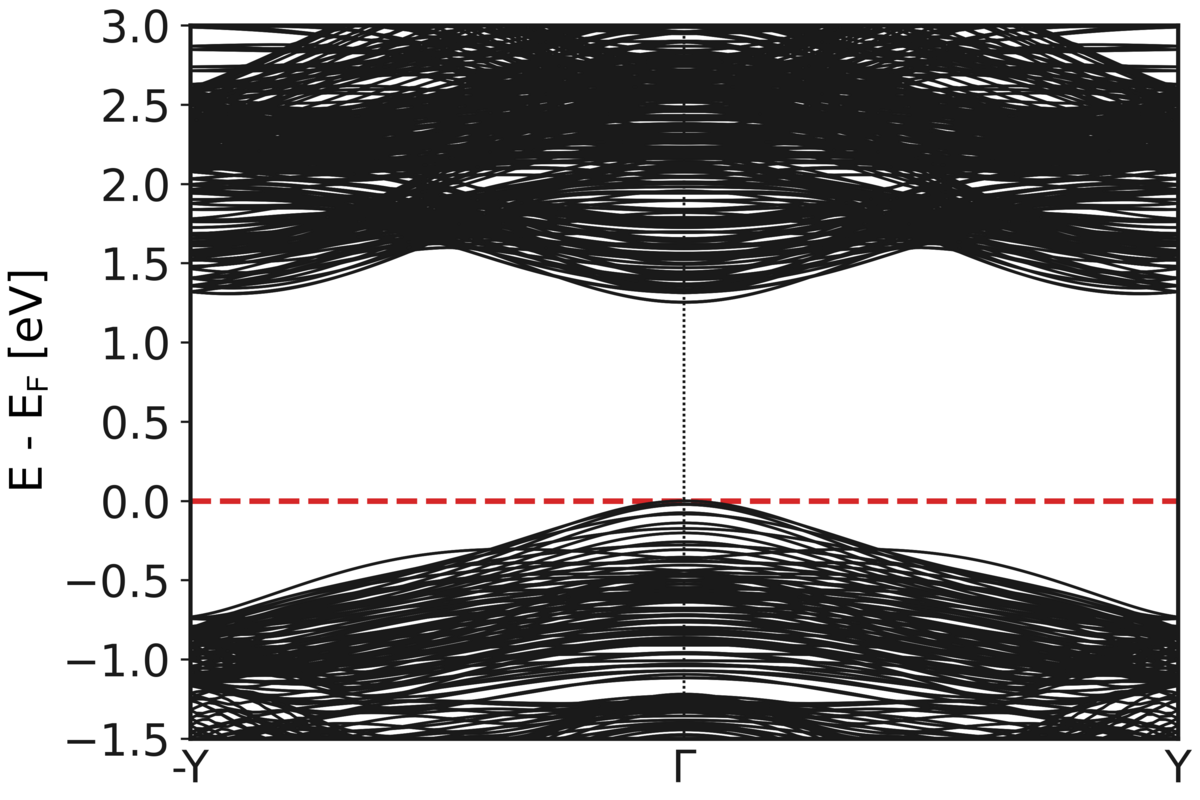}
            \caption*{(39,0)}
    \end{subfigure}
\end{figure}
\begin{figure}
\ContinuedFloat.

    \begin{subfigure}[b]{0.49\linewidth}
            \includegraphics[width=1\linewidth]{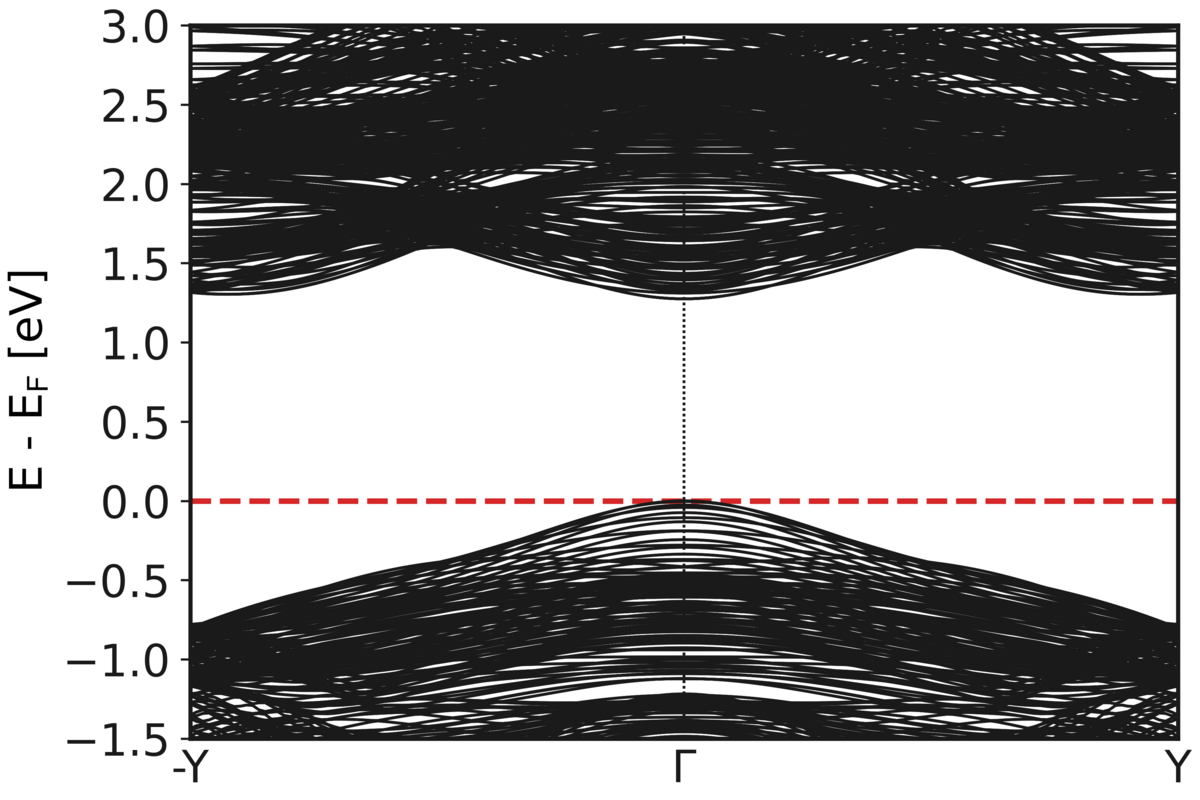}
            \caption*{(46,0)}
    \end{subfigure}
    \begin{subfigure}[b]{0.49\linewidth}
            \includegraphics[width=1\linewidth]{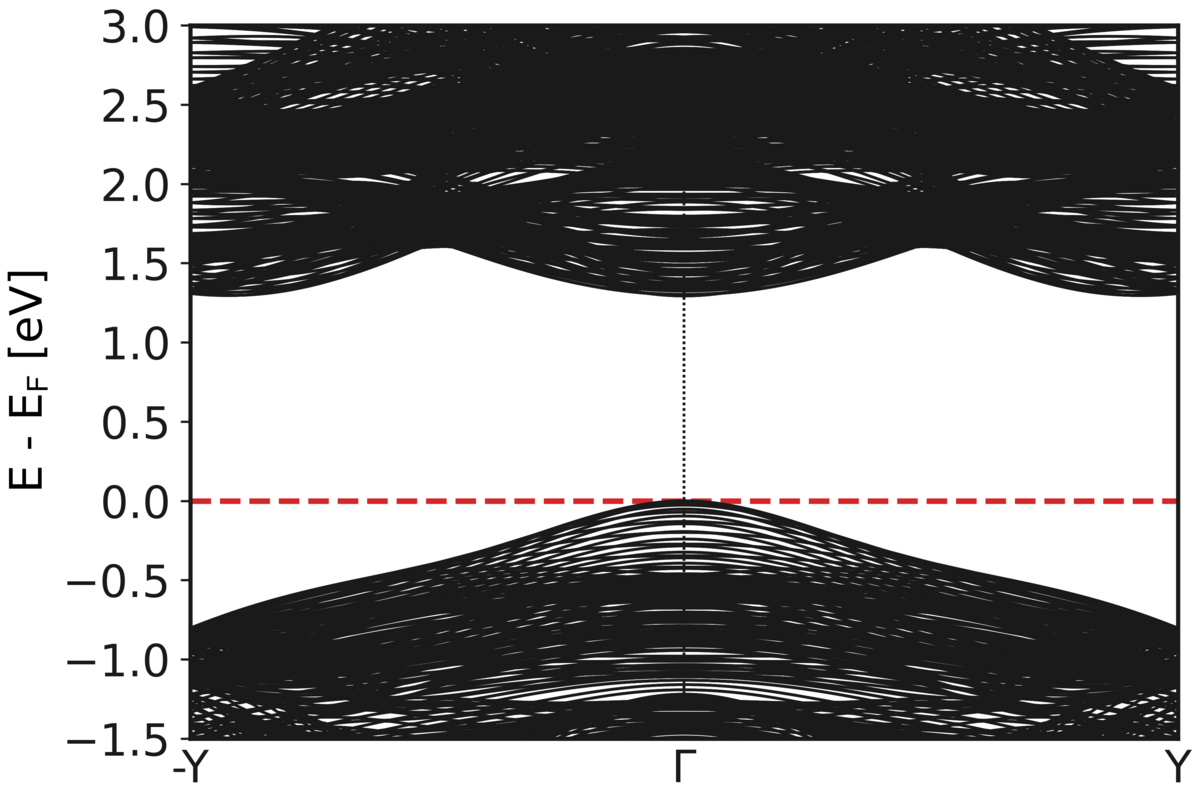}
            \caption*{(58,0)}
    \end{subfigure}
        \caption{Band structure of investigated nanotubes in zigzag direction}
\end{figure}

\end{document}